\pdfoutput=1
\documentclass[iop]{emulateapj}

\usepackage{graphicx, subfigure}
\usepackage{amsmath}
\usepackage{listings}

\usepackage{times}
\usepackage[english]{babel}
\usepackage{color}
\usepackage{url}
\usepackage[colorlinks=true,citecolor=blue,urlcolor=green]{hyperref}%

\renewcommand{\S}[1]{Section~}
\newcommand{\ie}{i.\,e.}

\newcommand{\eg}{e.\,g.}

\numberwithin{equation}{section}

\newcommand{\M}{\text{MC12}} 
\newcommand{\B}{BL11} 

\shorttitle{Search for $\ell=4$ and $\ell=5$ modes}
\shortauthors{Lund et al.}

\begin{document}

\title{\large  DETECTION OF $\ell=\, $4 AND $\ell=\, $5 MODES IN 12 YEARS OF SOLAR VIRGO-SPM DATA \hspace*{\fill}\\ --- TESTS ON \textit{KEPLER} OBSERVATIONS OF 16 Cyg A AND B ---  \vspace*{0.3cm}}
\author{Mikkel~N\o rup~Lund$^{1\bigstar}$}
\author{Hans~Kjeldsen$^1$}
\author{J\o rgen~Christensen-Dalsgaard$^1$}
\author{\hspace*{\fill}\\Rasmus~Handberg$^{2,1}$}
\author{Victor~Silva~Aguirre$^1$\vspace*{0.2cm}}

\affil{$^1$Stellar Astrophysics Centre, Department of Physics and Astronomy, Aarhus University,\\ Ny Munkegade 120, DK-8000 Aarhus C, Denmark\\
$^2$School of Physics and Astronomy, University of Birmingham, Edgbaston, Birmingham, B15 2TT, UK}

\email{$^{\bigstar}$mikkelnl@phys.au.dk}


\begin{abstract}
\centering
We present the detection of $\ell=4$ and $\ell=5$ modes in power spectra of the Sun, constructed from 12 yr full-disk VIRGO-SPM data sets. A method for enhancing the detectability of these modes in asteroseismic targets is presented and applied to \textit{Kepler} data of the two solar analogues 16 Cyg A and B. For these targets we see indications of a signal from $\ell=4$ modes, while nothing is yet seen for $\ell=5$ modes. We further simulate the power spectra of these stars and from this we estimate that it should indeed be possible to see such indications of $\ell=4$ modes at the present length of the data sets. In the simulation process we briefly look into the apparent misfit between observed and calculated mode visibilities. We predict that firm detections of at least $\ell=4$ should be possible in any case at the end of the \textit{Kepler} mission. For $\ell=5$ we do not predict any firm detections from \textit{Kepler} data.  
\vspace{0.4cm}
\end{abstract}



\section{Introduction}
\label{sec:intro}
Stars observed for the purpose of asteroseismology are measured in full-disk integrated light.
In such observations there will be an unavoidable geometrical cancellation effect between bright and dark patches on the stellar surface from the standing harmonic oscillations excited in the star.
The NASA \textit{Kepler} satellite, dedicated to finding planets using the transit method \citep{2010ApJ...713L..79K}, delivers very high-quality photometric data which are ideal for asteroseismic studies \citep[][]{2010PASP..122..131G}. However, even with such exquisite data it has so far only been possible to detect modes of degree $\ell=3$ (\textit{octupole}) in two main sequence (MS) stars, 16~Cyg~A~and~B \citep[][]{2012ApJ...748L..10M}, while detections in subgiants and red giants have been seen for some time \citep[see, \eg,][]{2010ApJ...713L.176B, 2010ApJ...723.1607H, 2012A&A...540A.143M}.

Turning to our own star, the Sun has been studied extensively using helioseismology. Here the surface can be resolved whereby the concern of cancellation effects is avoided and very high degree modes can be studied. As mentioned above, we do not have this luxury when studying other stars using asteroseismology; here instead the global low-$\ell$ modes can be studied. To learn about other stars from our Sun, the unresolved surface has been studied in so-called Sun-as-a-star observations, and this has been ongoing for more than a decade with velocity observations from ground \citep[\eg, BiSON\footnote{Birmingham Solar Oscillation Network.}; see][]{1996SoPh..168....1C}, and space \citep[\eg, GOLF\footnote{Global Oscillations at Low Frequencies.}; see][]{1995SoPh..162...61G} and space-born photometric observations \citep[\eg, VIRGO\footnote{Variability of Solar Irradiance and Gravity Oscillations.}; see][]{1995SoPh..162..101F,1997SoPh..170....1F}. These observations suffer from the same cancellation effects as experienced for distant stars.
Owing to the difference in the stellar noise properties between velocity and photometric measurements, the highest $\ell=4$ modes (\textit{hexadecapole}) have been seen in \eg\ BiSON \citep{1996MNRAS.280.1162C} and GOLF data \citep{1998ESASP.418..323R} for a long time - with the detection of these already predicted by \citet*{1989SoPh..119....5C}. 
In full irradiance observations of the Sun on the other hand only modes up to $\ell=3$ have so far been directly observed \citep[see, \eg,][]{2009AIPC.1170..560G}.

It is clear that the detection of higher-degree modes would be of great importance for stellar modeling with the extra constraints added. It would to a greater extent become possible to perform stellar structure inversions \citep[\eg,][]{1997MNRAS.292..243B}, albeit not with very high precision. In the ideal case where these higher degree modes could actually be resolved, a wealth of information could be extracted on for instance the rotational properties of the star \citep[\eg,][]{2004SoPh..220..169G}.

In this study, we take a closer look at the photometric data from VIRGO as these data sets hint at what might be possible with long time series from the \textit{Kepler} satellite, and the results of this will be presented in \S~\ref{sec:sol}.
In \S~\ref{sec:col}, we describe the method we propose to use for other stars in the search for these higher degree modes. We test in \S~\ref{sec:kep} this method on two of the most promising targets in the \textit{Kepler} field, viz. 16 Cyg A and B, with results presented in \S~\ref{sec:results1}. Furthermore, we simulate the power spectra of these two targets, described in \S~\ref{sec:simulation}, and present the results from these in \S~\ref{sec:results2}. In \S~\ref{sec:solardat}, we test the signal from $\ell=4$ in the solar data when using data lengths corresponding to the amount of data available for 16 Cyg A and B. Finally, we will discuss our findings in \S~\ref{sec:discuss}.

\clearpage

\section{Solar analysis}
\label{sec:sol}

\subsection{Data}

\begin{figure}
\centering
\includegraphics[scale=0.45]{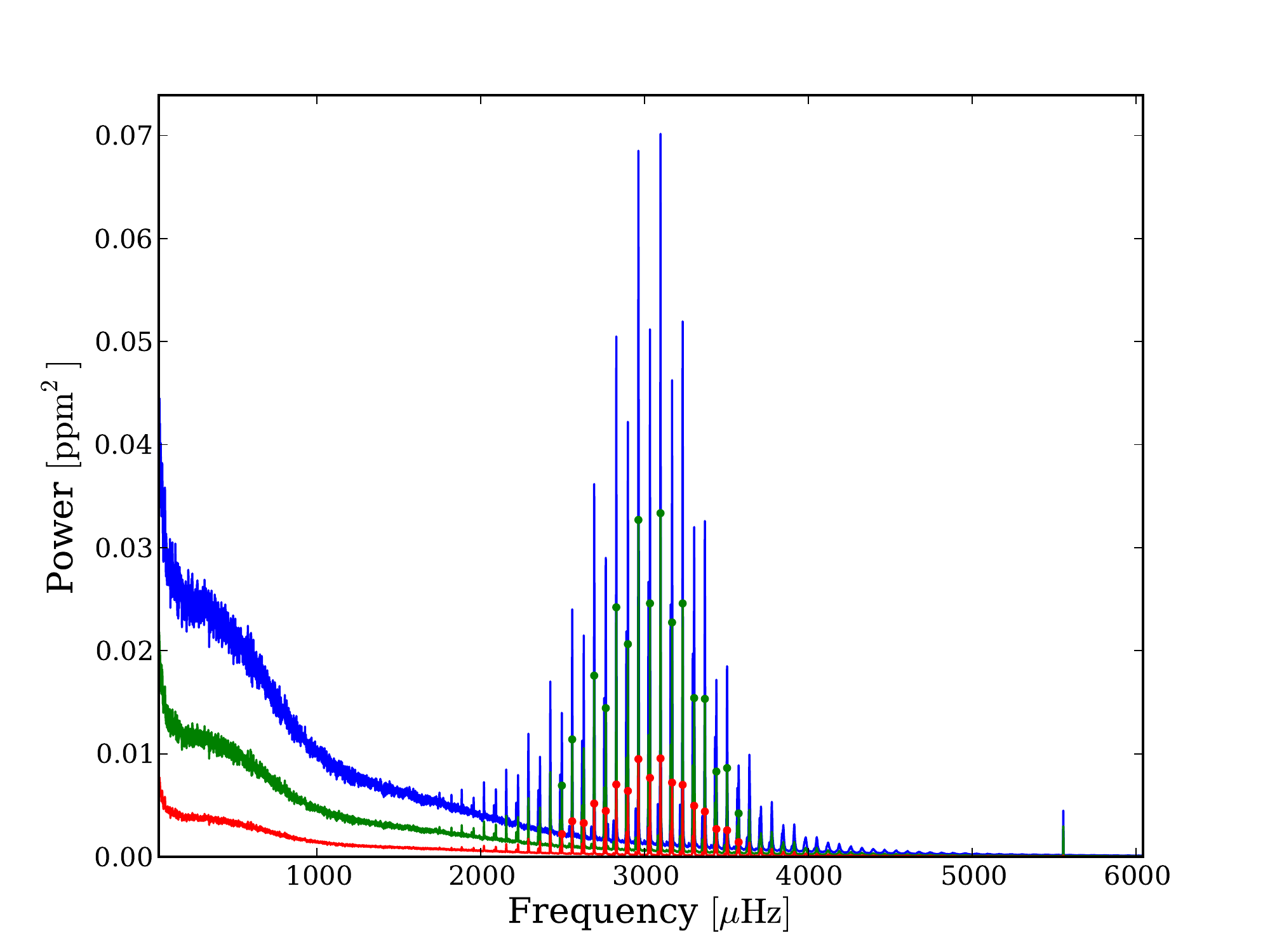}
\caption{\footnotesize Power spectra of VIRGO-SPM data, smoothed with a 1.8 $\mu$Hz boxcar filter. The color of the respective power spectra corresponds to color of the SPM-filter used. As seen the power levels are clearly highest in the blue band, follow by the green band and lastly the red band. In order to make the distinction between the power spectra easier circles have been added on the peaks of the central $\ell=0,1$ modes for the red and the green power spectra. See also Figure~\ref{fig:filters} for the position in wavelength of the different filters.}
\label{fig:powerspecs}
\end{figure}

For the solar analysis we used the data from \cite{2009A&A...501L..27F} with the corrections described therein.
More specifically we use 12 yr data sets from the three Sun photometers (SPM) of the VIRGO instrument on-board the ESA/NASA \textit{SoHO}\footnote{\textit{Solar and Heliospheric Observatory}.} spacecraft. The mid wavelengths of these photometers are at 402 nm (blue), 500 nm (green), and 862 nm (red).
The power spectra computed from the corrected time series are shown in Figure~\ref{fig:powerspecs}. These were calculated using a sine-wave fitting method \citep[see,\eg,][]{1992PhDT.......208K,1995A&A...301..123F}, normalized according to the amplitude-scaled version of Parseval's theorem \citep[see][]{1992PASP..104..413K}, in which a sine wave of peak amplitude, A, will have a corresponding peak in the power spectrum of $\rm A^2$.

The peak seen at $\rm 5555 \, \rm \mu Hz$ is an artefact and stems from the Data Acquisition System (DAS) of VIRGO which have a corresponding reference period of 3 minutes \citep{2005ApJ...623.1215J}.

\subsection{$\ell=\,$4 and $\ell=\,$5 Modes}
\label{sec:l4l5}

\begin{figure*}
\centering
\includegraphics[scale=0.35]{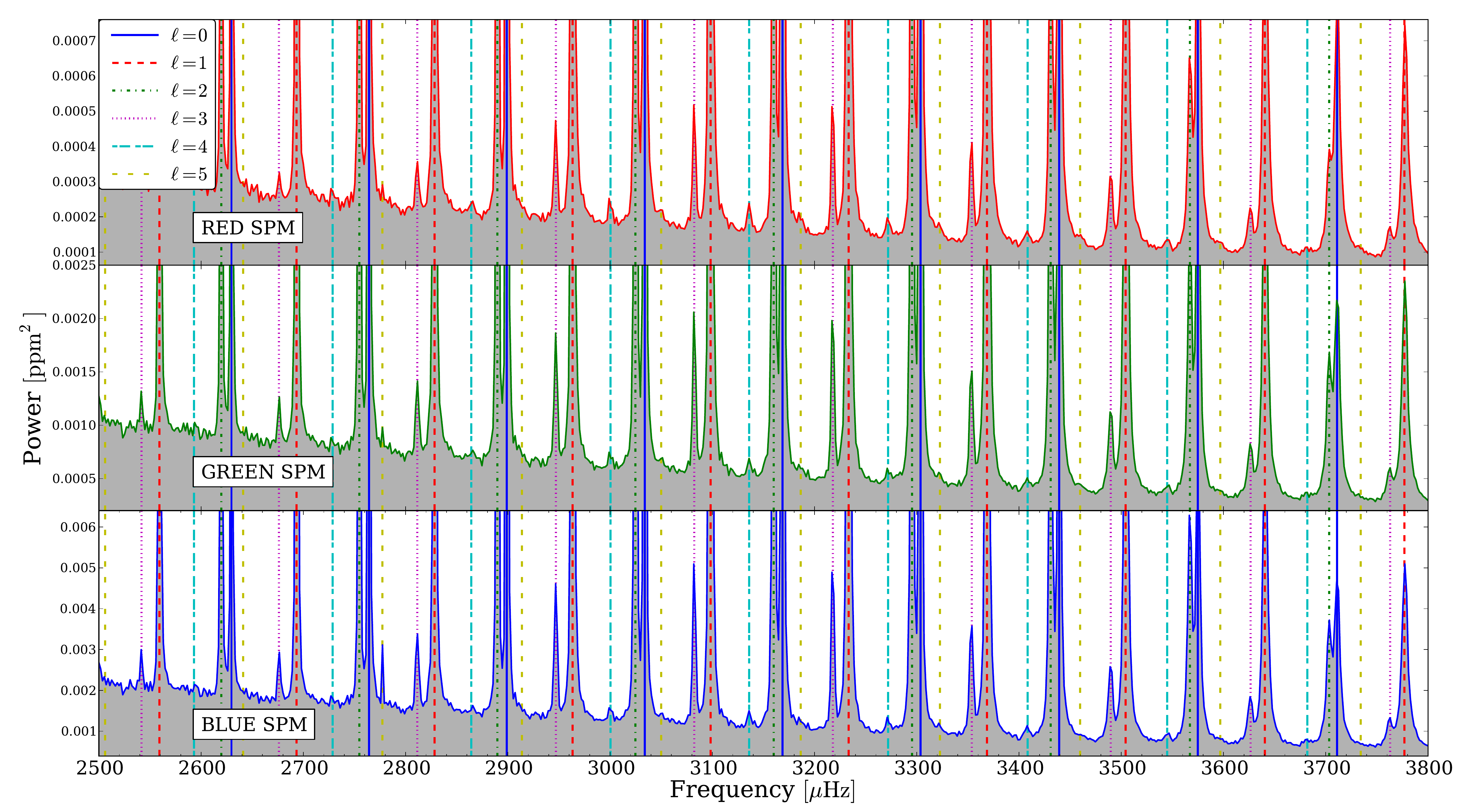}
\caption{\footnotesize Zoom-in on the power spectra of the VIRGO-SPM data, binned in segments covering 1.8 $\mu$Hz. The color of the respective power spectra corresponds to color of the respective the SPM-filter used. Note the different vertical axes. The positions of surface corrected Model~S frequencies have been indicated by vertical lines, showing a clear correspondence to the observed peaks in the power spectra. The legend gives the colors and line styles used for the different mode degrees.}
\label{fig:diff_filters}
\end{figure*}

In Figure~\ref{fig:diff_filters}, we show a zoom-in around the frequency of maximum power, $\nu_{\rm max}$, of the power spectra after a 1.8 $\mu$Hz binning has been applied. 

In this figure, we have indicated frequencies computed from the solar structure model \emph{Model~S} \citep{1996Sci...272.1286C}. These frequencies have been corrected for near-surface effects as described by \citet[][Equation (14)]{2012AN....333..914C}.

\begin{figure*}
\centering
\subfigure{
   \includegraphics[scale=0.4] {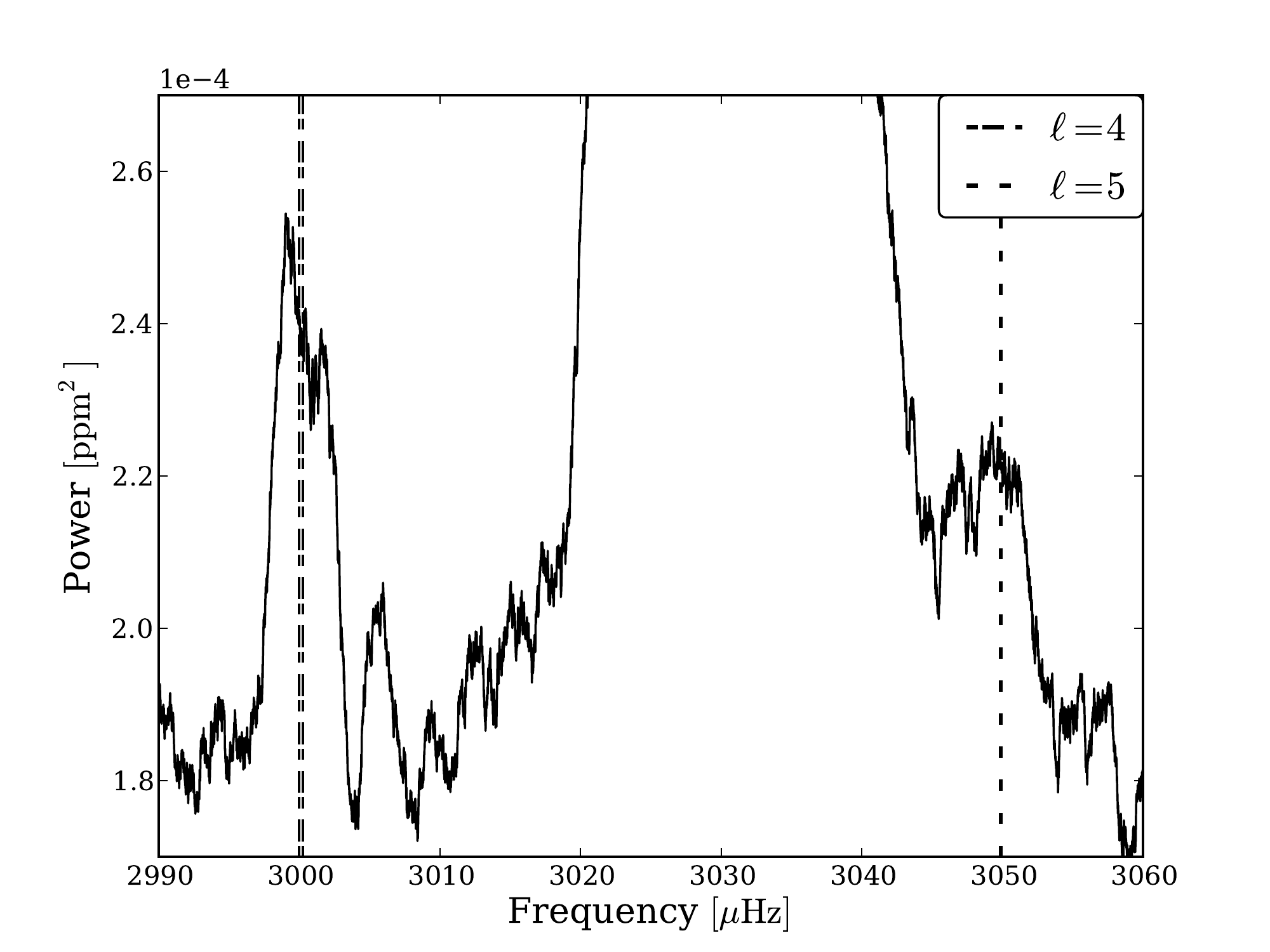}
 }
 \quad
 \subfigure{
   \includegraphics[scale=0.4] {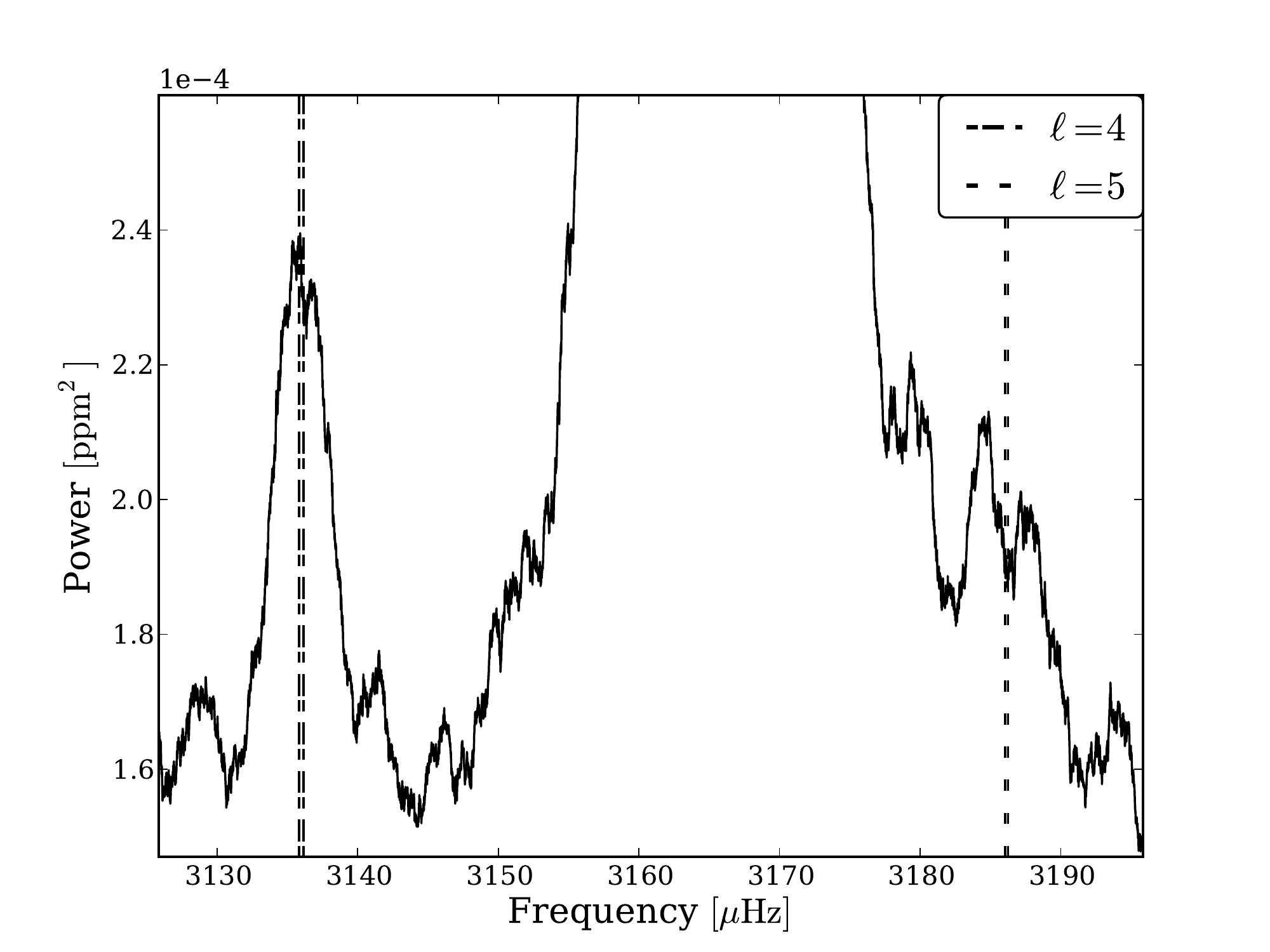}
 }
\caption{\footnotesize Zoom-ins on two central orders ($n=19,\,20$) of the power spectrum from VIRGO data of the red SPM-filter. A 1.8 $\mu$Hz boxcar smoothing has been applied. The vertical lines (two drawn for each degree) show the range in the mode frequency estimates (including errorbars) from LOI \citep{1998ESASP.418...99A}, BiSON \citep{2009MNRAS.396L.100B} (only $\ell=4$), MDI \citep[][]{2008JPhCS.118a2083L}, BiSON+LOWL \citep{1997MNRAS.292..243B}, and GONG (only $\ell=5$) \citep[][]{2000ApJ...543..472K}. As seen, the correspondence between these estimates and the VIRGO data is unequivocal.} 
\label{fig:instruments_comp}     
\end{figure*}

From a mere visual inspection of this figure, it is quite clear that the prominent peaks in the power spectrum agree with the frequencies from Model~S. This is of course to be expected considering that Model~S is calibrated to match the Sun. However, the match also includes many of the Model~ S $\ell=4$ (\textit{hexadecapole}) and $\ell=5$ \citep[\textit{dotriacontapole};][]{1988IAUS..123..111E} modes, confirming that it is indeed signal from $\ell=4$ and $\ell=5$ modes that is seen. The strong peak ${\sim}2777 \, \rm \mu Hz$, which coincides with the frequency of an $\ell=5$ mode, is not of stellar nature but is an artefact, having half the frequency of the signal from the DAS. The observed frequencies also match up with values reported from \eg\ LOI\footnote{The Luminosity Oscillations Imager (on \textit{SoHO}).}  \citep{1995A&A...294L..13A,1998ESASP.418...99A}, BiSON \citep{2009MNRAS.396L.100B} (up to $\ell=4$), MDI\footnote{Michelson Doppler Imager (on \textit{SoHO}).} \citep[see, \eg,][]{2008JPhCS.118a2083L}, BiSON+LOWL \citep{1997MNRAS.292..243B}, and GONG\footnote{Global Oscillation Network Group.} (for $\ell=5$) \citep[see, \eg,][]{2000ApJ...543..472K}. See Figure~\ref{fig:instruments_comp} for an example of the correspondence between the red SPM-VIRGO data and the frequency estimates from the above references for two of the central orders ($n=19,\, 20$).
These $\ell=4$ and $\ell=5$ modes have not previously been reported on the basis of full-disk photometry data, and all of the comparison frequency estimates stem from either observations done in velocity or from photometry where the solar surface is semi-resolved.

\begin{figure}
\centering
\includegraphics[scale=0.4]{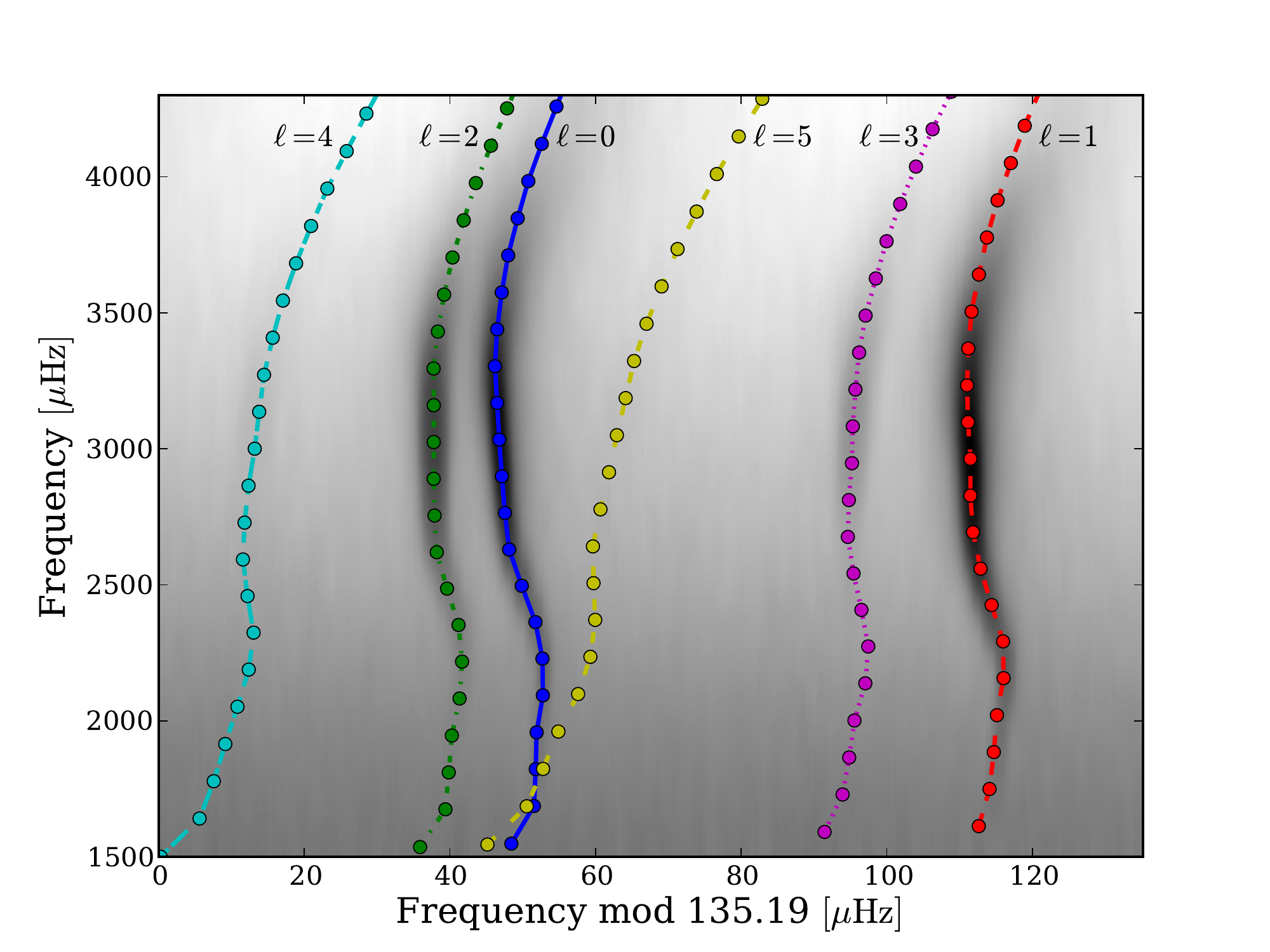}
\caption{\footnotesize \'Echelle diagram of the 1.8 $\mu$Hz smoothed power spectrum, red SPM band. The Model~S frequencies indicated are of degrees; $\ell=0$ (blue), $\ell=1$ (red), $\ell=2$ (green), $\ell=3$ (magenta), $\ell=4$ (cyan), and $\ell=5$ (yellow). The gray scale goes from white (low power) to black (high power), and is given on a logarithmic scale.}
\label{fig:echelle_sun1}
\end{figure}

\begin{figure}
\centering
\includegraphics[scale=0.4]{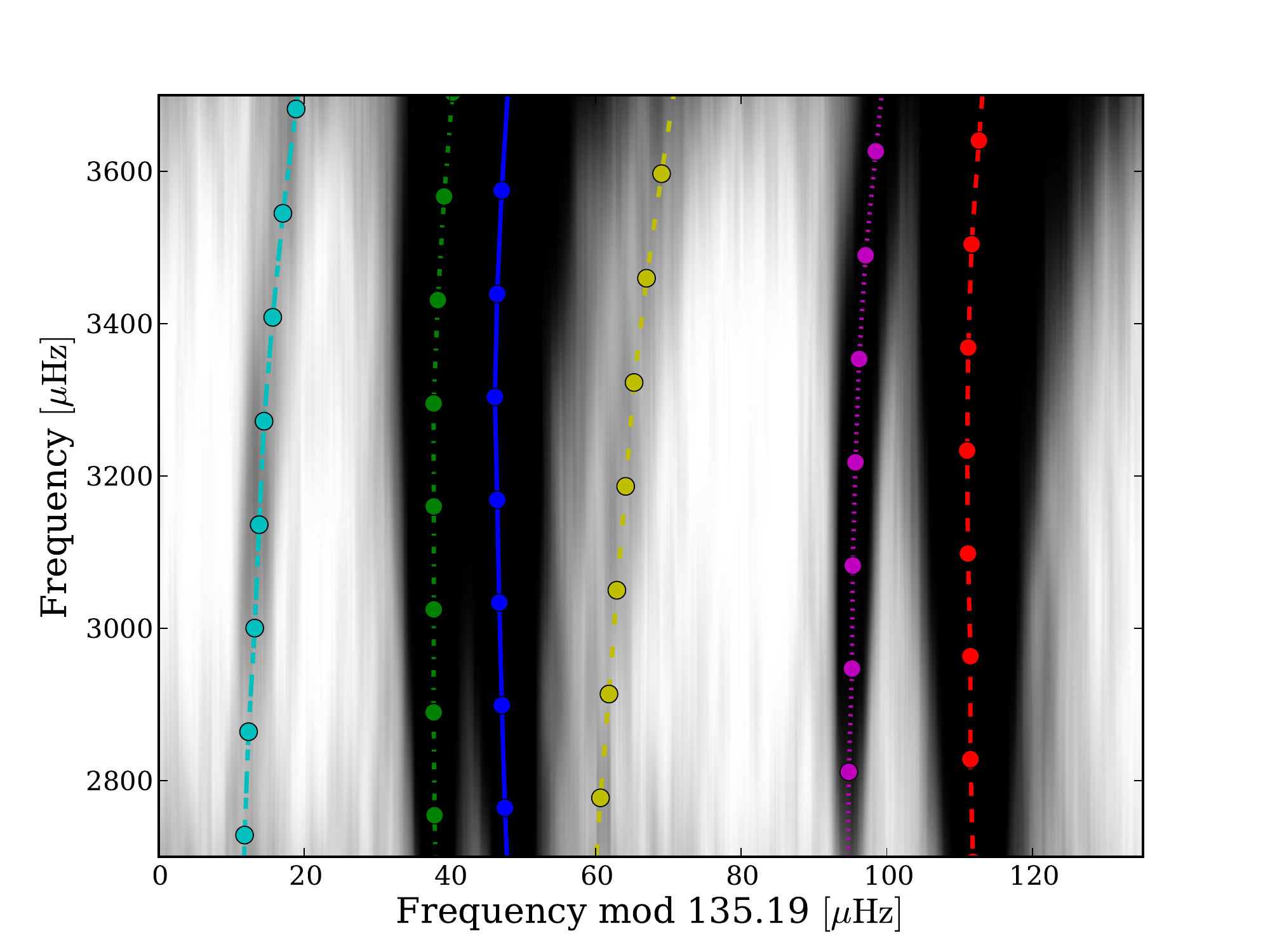}
\caption{\footnotesize Zoom-in on the \'echelle diagram in Figure~\ref{fig:echelle_sun1}. To enhance the visibility of the $\ell=4$, and $\ell=5$ modes the power spectrum has first been divided by a linear function approximating the stellar background, furthermore the colors have been truncated such that the maximum power level (black) corresponds roughly to the highest $\ell=3$ modes. The power excess from the sought for modes clearly lines up with the Model~S frequencies - the convention relating color to degree is the same as adopted in Figure~\ref{fig:echelle_sun1}, as is the order of the degree of the ridges. }
\label{fig:echelle_sun2}
\end{figure}

We can further visualize the power excess from the $\ell=4$ and $\ell=5$ modes in the well-known \'echelle diagram \citep{1983SoPh...82...55G}, where the power spectrum is first divided in segments corresponding to the so-called large separation given by the frequency difference between modes of same degree and consecutive radial order. Subsequently these segments are stacked on top of each other. In practice, the frequency is plotted against its modulo to the large separation - we have on the ordinate plotted the mid frequency of the respective segments. In order to obtain a nice representation of the ridges, it is customary to add an arbitrary value to the frequencies before taking the modulo, thereby allowing a shift of the ridges on the abscissa \citep[see, \eg,][]{2011arXiv1107.1723B}. 
In Figure~\ref{fig:echelle_sun1}, we show the \'echelle diagram of the power spectrum from the red band, here also with the Model~S frequencies over-plotted. 
In Figure~\ref{fig:echelle_sun2}, we give the same \'echelle diagram, though in a narrower frequency range. To make clearer the excess from the $\ell=4$ and $\ell=5$ modes, the power spectrum has been divided by a linear representation of the stellar background, and the colors have been truncated such that the maximum corresponds roughly to the highest $\ell=3$ mode.



\section{Enhancing the detectability}
\label{sec:col}

To make easier the detection of the $\ell=4$ and $\ell=5$ modes in stars other than the Sun, we propose here a simple three step method that should accomplish this.
Firstly, the power spectrum is stretched, this is described in Sections~\ref{sec:para}-\ref{sec:modify}. Secondly, the power spectrum is smoothed in \S~\ref{sec:smo}. Thirdly, the stretched, smoothed power spectrum is collapsed in \S~\ref{sec:col2}. The final spectrum we tentatively name the \textit{SC}-spectrum (\textit{Straightened-Collapsed}). 


\subsection{Parameterization of the Power Spectrum}
\label{sec:para}

In the first step, we parameterize the power spectrum in order to account for the possible departures from the asymptotic description of the constituent modes. The motivation for this step is, that in the end we need to correct for such departures in order to facilitate the co-addition (collapsing) of power from many radial orders of a specific degree .  
This is inspired by a step in the Octave code \citep{2010MNRAS.402.2049H}, where the frequency scale of the power spectrum is modified to account for departures in the asymptotic description and thereby obtain a higher signal in the power-spectrum-of-power-spectrum ($\rm PS\otimes PS$). 

We start from the following version of the asymptotic frequency relation \citep{1980ApJS...43..469T}, which to a good approximation is applicable to acoustic modes of high radial order $n$, and low angular degree $\ell$:
\begin{equation}
\nu_{n\ell} =  \Delta\nu (n + \ell/2 + \epsilon) - \ell(\ell+1) D_0\, .
\label{eq:asym}
\end{equation}
In this equation, $\Delta\nu$ is the large separation, $\epsilon$ is a dimensionless offset sensitive to the surface layers, and $D_0$ is a quantity sensitive to the sound-speed gradient near the core of the star \citep[see, \eg,][]{1983SoPh...82...75S, 1993ASPC...42..347C, 2010CoAst.161....3B}.

It is well-known from both theory and observations \citep[see, \eg,][]{2011A&A...525L...9M, 2012A&A...541A..51K, 2013A&A...550A.126M} that Equation~\ref{eq:asym} is in fact only an approximate description as $\Delta\nu$, $D_0$, and $\epsilon$ all have small dependencies on both frequency and degree. There exist departures from the asymptotic description on both small and large scales. Small-scale oscillations in the large separation originate, \eg, from sharp changes in the sound-speed profile \citep[one source being the He \textsc{ii} ionization zone;][]{2007MNRAS.375..861H} - we will not account for this in the following. The larger scale departures manifest themselves as overall curvatures or tilts of the ridges in the \'echelle diagram. With the larger scale departures in mind, \cite{2010MNRAS.402.2049H} described one way to find the variation in the large separation as a function of the radial order, $d\Delta\nu/dn$ \citep[see also][]{2009A&A...508..877M, 2009A&A...506..435R}. This variation in the large separation was used in \cite{2010MNRAS.408..542C}, where it was introduced in the asymptotic relation as a term quadratic in $n$ - we will follow the same procedure for this term. The small separation ${\delta\nu_{02} \equiv \nu_{n\ell=0}-\nu_{n-1\ell=2} = 6D_0}$, and in turn $D_0$, is for the Sun found to decrease almost linearly with frequency \citep[][]{1990Natur.347..536E}. With this in mind, we introduce to our modified asymptotic relation a term $dD/dn$ which is linear in $n$. 
We end up with the following modified version of the asymptotic relation \citep[see also][]{2011A&A...525L...9M, 2013A&A...550A.126M}: 
\begin{align}\label{eq:asym2}
\rm \tilde{\nu}_{n\ell} = & \Delta\nu_{\rm pivot} (n + \ell/2 + \epsilon) - \ell(\ell+1) D_0\\
 &-\ell(\ell+1) \frac{d D_0}{d n}(n-n_{\rm pivot})\notag \\
&+ (n-n_{\rm pivot})^2 \frac{d \Delta\nu / d n }{2}\, . \notag
\end{align}

We have denoted this model frequency by $\tilde{\nu}_{n\ell}$ to indicate that this is a predicted value only.
$\Delta\nu_{\rm pivot}$ denotes the value of the large separation at $n=n_{\rm pivot}$ \citep[various formulations exist for finding $\Delta\nu_0$, see, \eg,][]{2012A&A...541A..51K}. 
Here $n_{\rm pivot}$ (which is not constrained to be of integer value) can be seen as the pivot point for the variations in both the large separation and $D_0$ - the frequency at which $n \sim n_{\rm pivot}$ is generally coinciding with the frequency of maximum power of the modes, $\nu_{\rm max}$. In the description of Equation~\ref{eq:asym2}, we have assumed that the various parameters are frequency-dependent only and neglected any potential dependencies on angular degree. In the following, we will not touch upon the physical meaning behind the frequency dependencies in $\Delta\nu$ and $D_0$ but merely use them in our parameterization of the frequencies in the power spectrum.

The interested reader is referred for instance to \citet[][]{1980ApJS...43..469T}, \citet[][]{2007MNRAS.375..861H}, and most recently \citet[][]{2013A&A...550A.126M} and references therein for more on the physics behind the departures.


\subsection{Modifying the Frequency Scale}
\label{sec:modify}

\begin{figure}[th]
\centering
\includegraphics[scale=0.45]{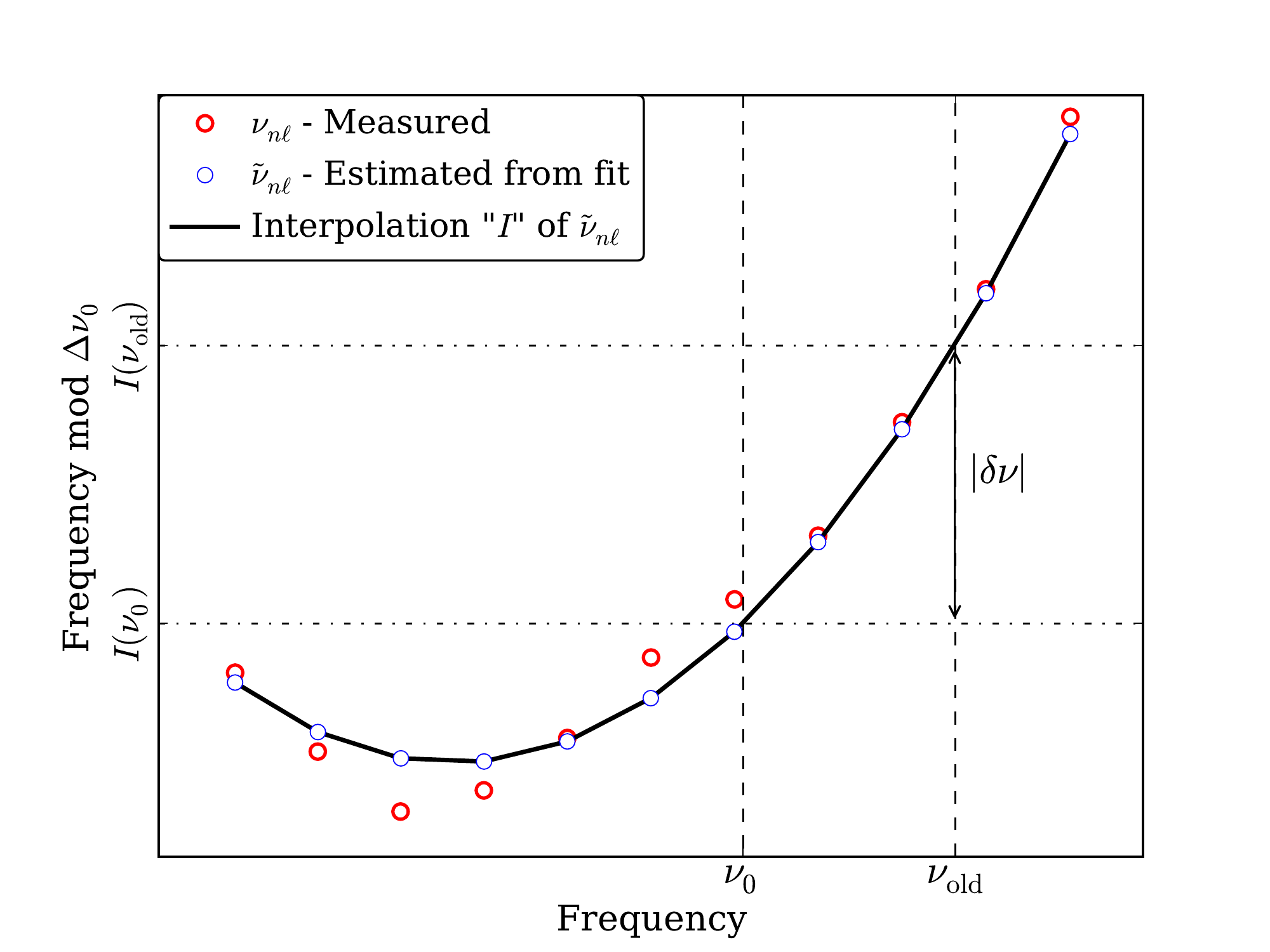}
\caption{\footnotesize Illustration of the concept at the base of the straightening procedure. The figure can be seen as a flipped \'echelle diagram, with the frequency as the independent variable and the modulo of the frequency with the large separation as the dependent variable. The open thick-edge circles give "measured" frequencies, meaning frequencies obtained from either peak-bagging or from a stellar model. Equation~\ref{eq:asym2} is fitted to these. Open thin-edge circles give the "estimated" frequencies, \ie, the ones found with the best fit of Equation~\ref{eq:asym2} for a given degree. The solid black line gives the interpolation between these estimated frequencies. The reference frequency $\nu_0$ defines the correction, $\delta\nu$, that should be applied to a value $\nu_{\rm old}$ on the frequency scale in order to obtain a power spectrum that for a specific targeted degree follows Equation~\ref{eq:asym} more closely.}
\label{fig:stra_expla}
\end{figure}

The modification or straightening of the frequency scale in the power spectrum comes down to three steps:

\begin{enumerate}
\item Fitting of the parameterization given by Equation~\ref{eq:asym2}.

The fitting of the modes is best illustrated in the \'echelle diagram described in \S~\ref{sec:l4l5}, and we will do so from now onward.

If only observed frequencies are available, \eg\ from peak-bagging \citep[\eg][]{2003Ap&SS.284..109A}, the above parameterization is simply fitted to these frequencies, with the free parameters being ${\boldsymbol\Psi = \{d \Delta\nu / d n, D_0, d D_0 / d n, \epsilon, n_{\rm pivot} \}}$. Note that even if the identification of the modes with respect to the radial order is wrong a good fit can still be obtained as $n_{\rm pivot}$ is a free parameter. The large separation $\Delta\nu$ is not included in the optimization as this can be determined relatively easily, and any small deviations from $\Delta\nu_{\rm pivot}$ at $n_{\rm pivot}$ can be accounted for by $\epsilon$.

If on the other hand a stellar model with computed frequencies is available, and which after applying a surface correction \citep[\eg,][]{2008ApJ...683L.175K} matches the observed data well, one can apply the fitting to these modeled frequencies. Here it is then possible to use also the model calculated frequencies for the $\ell=4$ and $\ell=5$ modes in the fitting.

With this first step, we obtain estimates for the parameters entering Equation~\ref{eq:asym2}. 

\item Estimate frequencies for a targeted degree.

The reason for selecting a specific degree (the "targeted" degree) is that as we have included a variation in $D_0$ in our parameterization, we cannot modify the frequency scale \citep[as in][]{2010MNRAS.402.2049H} such that all degrees fulfil Equation~\ref{eq:asym}. The reason for this is the factor of $\ell(\ell+1)$ on the term describing the variation depending on $D_0$, making this term $\ell$-dependent.

From the estimated parameters ($\boldsymbol\Psi$) found above one can now use Equation~\ref{eq:asym2} to estimate the frequencies $\tilde{\nu}_{n\ell}$ of a specific or targeted degree. This could for instance be the $\ell=4$ or $\ell=5$ modes. 
One should be aware that small errors in $\epsilon$ and $D_0$ can cause a big offset in the estimate of, \eg, the $\ell=4$ or $\ell=5$ mode frequencies. The biggest issue is $D_0$ due to the factor of $\ell(\ell+1)$ on this parameter; furthermore, this value is one of the most difficult ones to constrain as many modes of different degree are needed. In contrast, $\epsilon$ (and $\Delta\nu_0$ for that matter) can be fairly well constrained by modes of the same degree.
This issue is mainly a concern if the fit of the parameterization in step (1) is made to observed frequencies only, \eg, of peak-bagged modes up to $\ell=2$, and the desire is to identify power from modes of a higher degree, \eg\, $\ell=4$ or $\ell=5$. 
The concern becomes irrelevant if a well-matching model exists.

\item Correct for deviation from Equation~\ref{eq:asym} for a specific degree. 

We now wish to take out the dependencies on frequency in the modes spacings. This is to get a power spectrum that for modes of a targeted degree follow Equation~\ref{eq:asym} more strictly, \ie\, with equidistance between mode frequencies - equivalent to a straight ridge in the \'echelle diagram. 
This is accomplished by changing the frequency scale in the power spectrum using frequencies computed in step (2) for the modes of the targeted degree.

An interpolation is made between the computed mode frequencies from step (2) as the independent variable and their modulo with the large separation as the dependent variable - this would correspond to a flipped \'echelle diagram, see Figure~\ref{fig:stra_expla}. We denote the obtained interpolation by "$I$".

This interpolation now follows the ridge in the \'echelle diagram from the computed frequencies of the targeted degree. We can now for every frequency ($\nu_{\rm old}$) in the frequency scale of the power spectrum compute a new modified frequency ($\nu_{\rm new}$) by adding a value $\delta\nu$:
\begin{equation}
\nu_{\rm new} = \nu_{\rm old} + \delta\nu\, .
\label{eq:fremodif}
\end{equation}
This added value is obtained from the interpolation as (see Figure~\ref{fig:stra_expla})
\begin{equation}
\delta\nu = I(\nu_{0}) - I(\nu_{\rm old})\, .
\end{equation}
With this value of $\delta\nu$ the reference frequency sets the position on the abscissa of the targeted ridge in the \'echelle diagram - by choice we set the reference frequency equal to the value of $\nu_{\rm max}$.
As we wish to co-add the segmented power spectrum, we need in the end to map the modified power spectrum onto a frequency scale with a regular step size.
\end{enumerate}

\begin{figure*}
\centering
\subfigure{
   \includegraphics[scale=0.4] {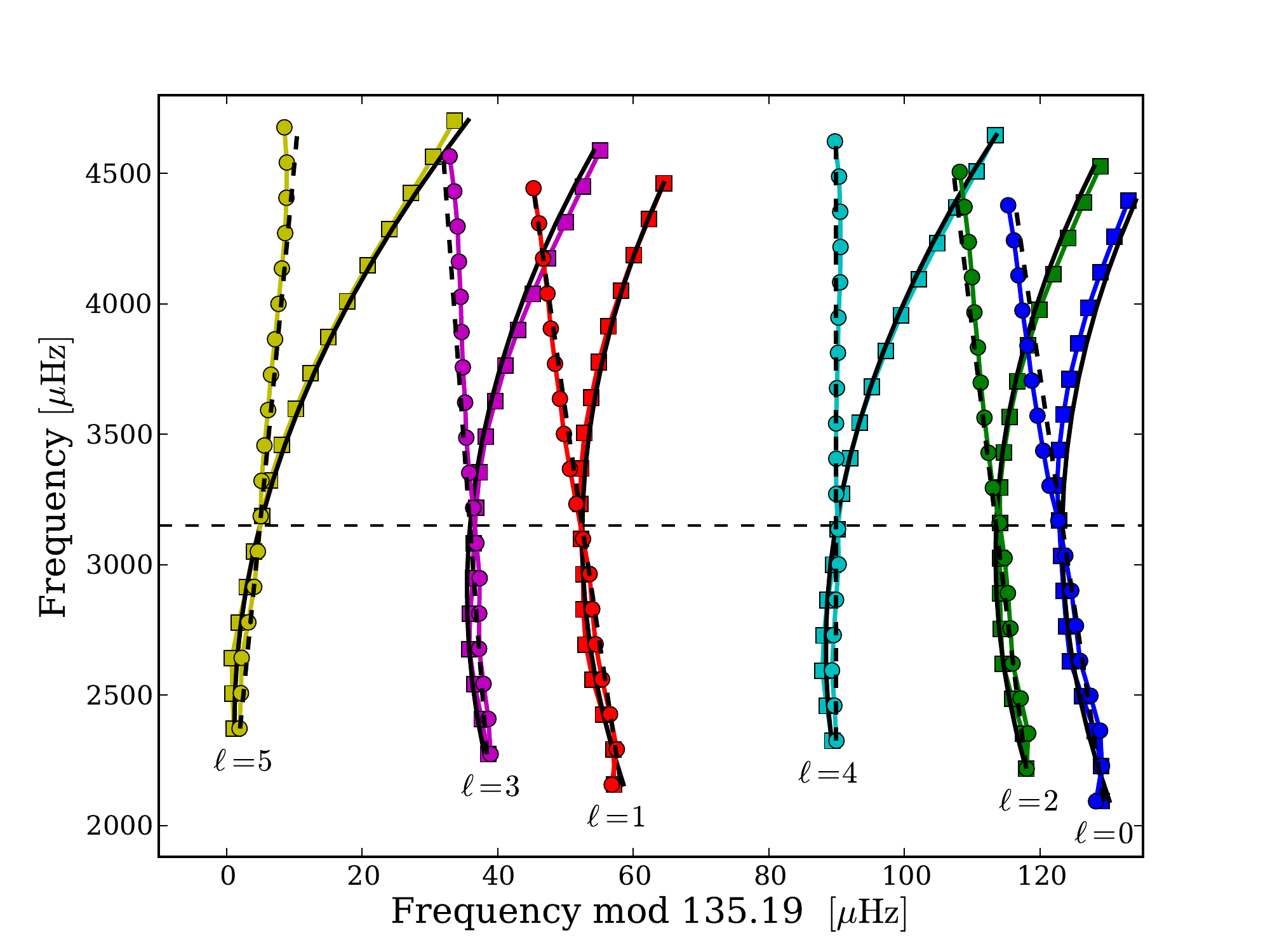}
 }
 \quad
 \subfigure{
   \includegraphics[scale=0.4] {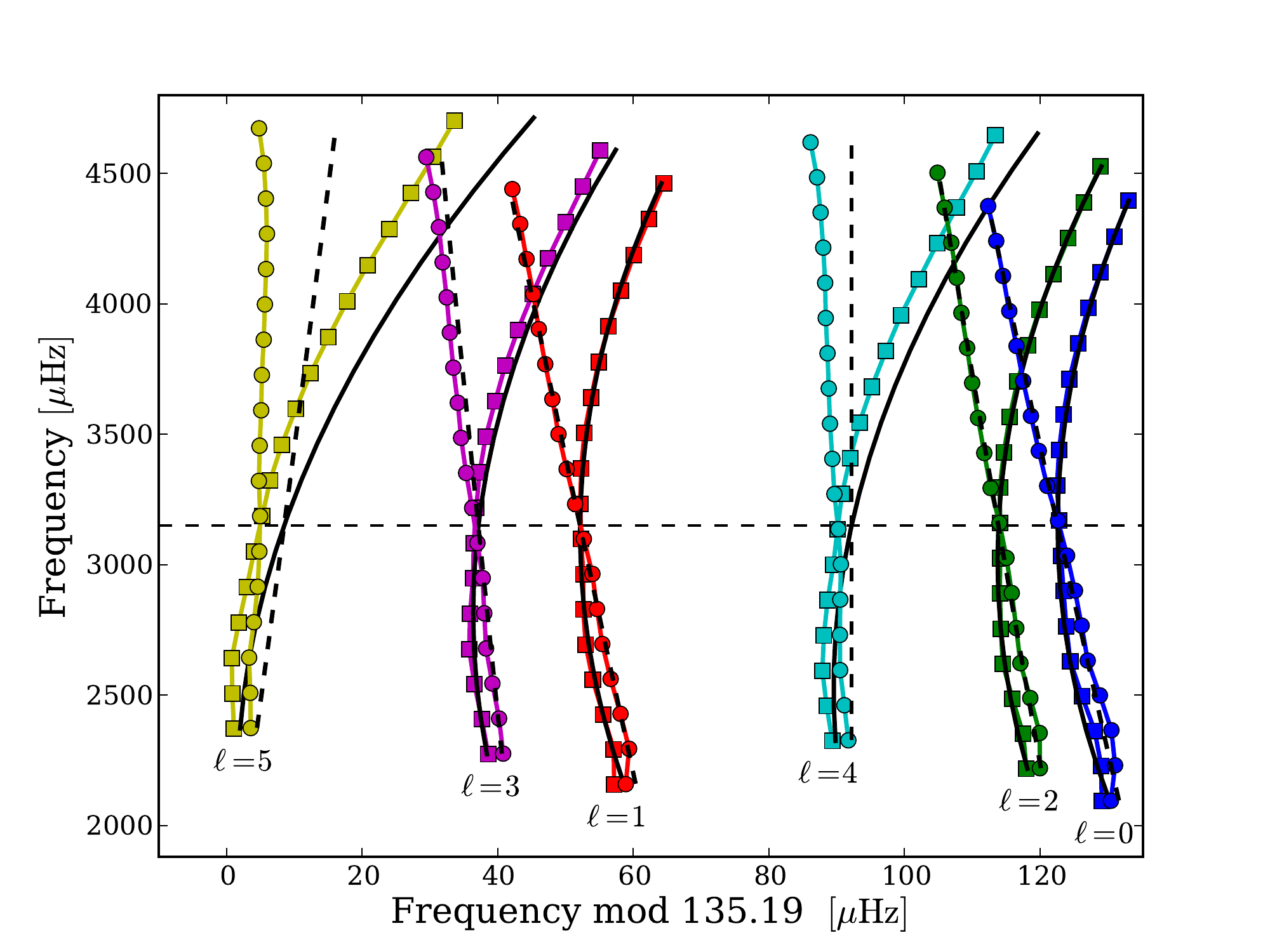}
 }
\caption{\footnotesize Method of straightening applied to Model~S frequencies, targeting modes with $\ell = 4$. In the left panel the fit of Equation~\ref{eq:asym2} is made to Model~S frequencies (squares) with degrees $\ell=0$ (blue), $\ell=1$ (red), $\ell=2$ (green), $\ell=3$ (magenta), $\ell=4$ (cyan), and $\ell=5$ (yellow). The solid black lines following these frequencies illustrate the interpolation of the obtained fit. The dashed black lines give the behavior of the fit after the straightening procedure, with the straightened Model~S frequencies now given as circles. Note that the frequencies of the targeted $\ell = 4$ modes form a vertical line. The value of $\nu_0$ is indicated by the dashed horizontal black line. The right panel shows the same fit, but here only modes of degree $\ell=0-2$ were included in the fit of Equation~\ref{eq:asym2}. As seen, the fit is now noticeably worse and the position of the straightening ridge no longer agrees with the predicted position (dashed line).\\ } 
\label{fig:model_ech}      
\end{figure*}

In Figure~\ref{fig:model_ech}, the method of straightening has been applied to Model~S frequencies. In the left panel the fit of Equation~\ref{eq:asym2} is made to Model~S frequencies (squares) with degrees $\ell=0$ (blue), $\ell=1$ (red), $\ell=2$ (green), $\ell=3$ (magenta), $\ell=4$ (cyan), and $\ell=5$ (yellow). The solid black lines illustrate the interpolation to the frequencies estimated from the obtained fit of Equation~\ref{eq:asym2}. The dashed black lines give the behavior of the fit after the straightening procedure, with the straightened Model~S frequencies now given as circles. The straightening is targeted at the $\ell=4$ modes, and they now form a vertical line. As seen, the position on the abscissa of this vertical line is given by the value of the fit at the chosen reference frequency $\nu_0$. The reference frequency is given by the horizontal dashed black line, and were here set equal to $\nu_{\rm max,\odot}=3150\,\rm \mu Hz$. In the right panel the same procedure is followed, but here only the modes having $\ell=0-2$ were used. The use of less modes leads to a fit (step 1) with a poorly determined value of $D_0$ (mainly), which in turn results in a bad estimation of $\ell=3-5$ modes (step 2). When using the badly estimated values of $\ell=4$ modes in the straightening (step 3), an offset between the actual (straightened Model~S frequencies) and predicted (straightened estimated frequencies) value on the abscissa of the straightened $\ell=4$ modes is seen.


\subsection{Smoothing} 
\label{sec:smo}

For every mode of degree $\ell$, there are $2\ell+1$ degenerate $m$-components ranging in value from $-\ell$ to $+\ell$, with $m$ being the so-called azimuthal order. The degeneracy of these modes is lifted when the star rotates, with the $m$-components being spread out in frequency and thereby taking away power from the central frequency at $m=0$. In the straightening procedure, it is the position of this central $m=0$ component that is found. 

For this reason, we apply a smoothing to the straightened power spectrum in order to boost the detectability, as it is desirable to merge the power contained in the rotationally split $m$-components. The effect from this merging of power to the central frequency will depend on the size of the rotational splitting compared to the mean mode width. If the splitting is very large compared to the mode width the smoothing will not merge much of the power from different $m$-components. However, the smoothing will in any case decrease the point-to-point scatter in the power spectrum from the $\chi^2$ noise, allowing any underlying structures to stand out more clearly. In addition, the smoothing takes out the impact of small wiggles that unavoidably will be present in the ridge for the degree of interest, even after the straightening procedure. The wiggles from for instance the He \textsc{ii} ionization zone will also still be present in the straightened ridge as we did not account for these. 
It is difficult to determine the smoothing level that optimizes a visual detection, as this indeed is a very qualitative measure, and the impact of a given smoothing level will depend on the rotational splitting, mode width and inclination angle of the specific star. Also, a "too high" smoothing can result in a significant merging of power from components of different degrees, which is undesirable.  
To visualize the impact of the rotational splitting we can look at the visibility within a multiplet, which depends on the stellar inclination through the visibility function $\mathcal{E}_{\ell m}(i)$ given by \citep{1977AcA....27..203D, 2003ApJ...589.1009G}
\begin{equation}
\mathcal{E}_{\ell m}(i) = \frac{(\ell - |m|)!}{(\ell + |m|)!}\left[P^{|m|}_{\ell} (\cos i) \right]^2\, ,
\label{eq:epsilon}
\end{equation}
where $P^{|m|}_{\ell}$ is the associated Legendre function, and $i$ is the stellar inclination, giving the angle between the stellar rotation axis and the line of sight, such that a pole-on view corresponds to an inclination of $i=0^{\circ}$.
See, \eg, Appendix~A~of~\cite{2011A&A...527A..56H} for the $\mathcal{E}_{\ell m}(i)$ factors up to $\ell=4$.

\begin{figure}
\centering
\subfigure{
	\includegraphics[scale=0.4] {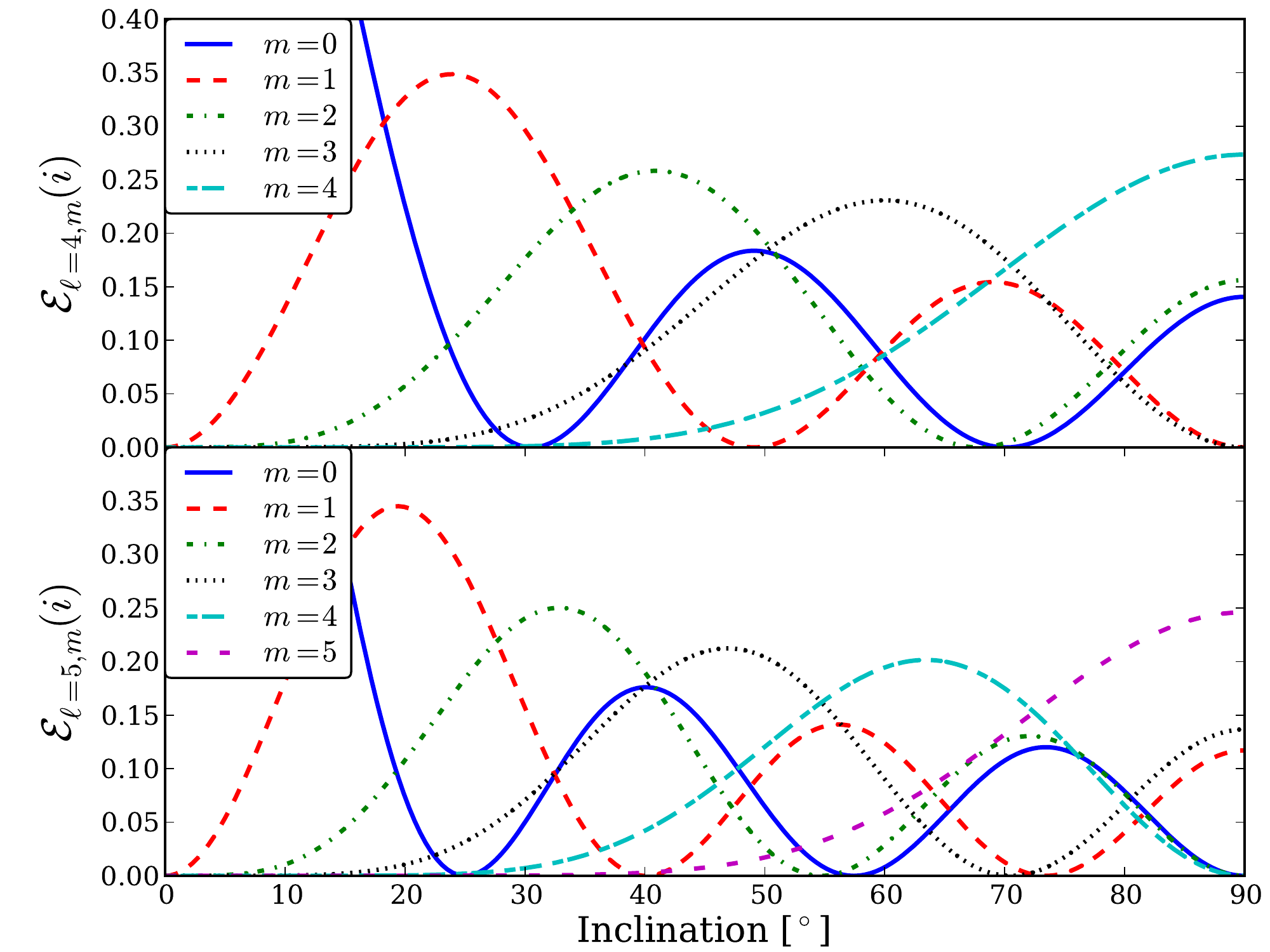}
 }
 \\
 \subfigure{
   \includegraphics[scale=0.4] {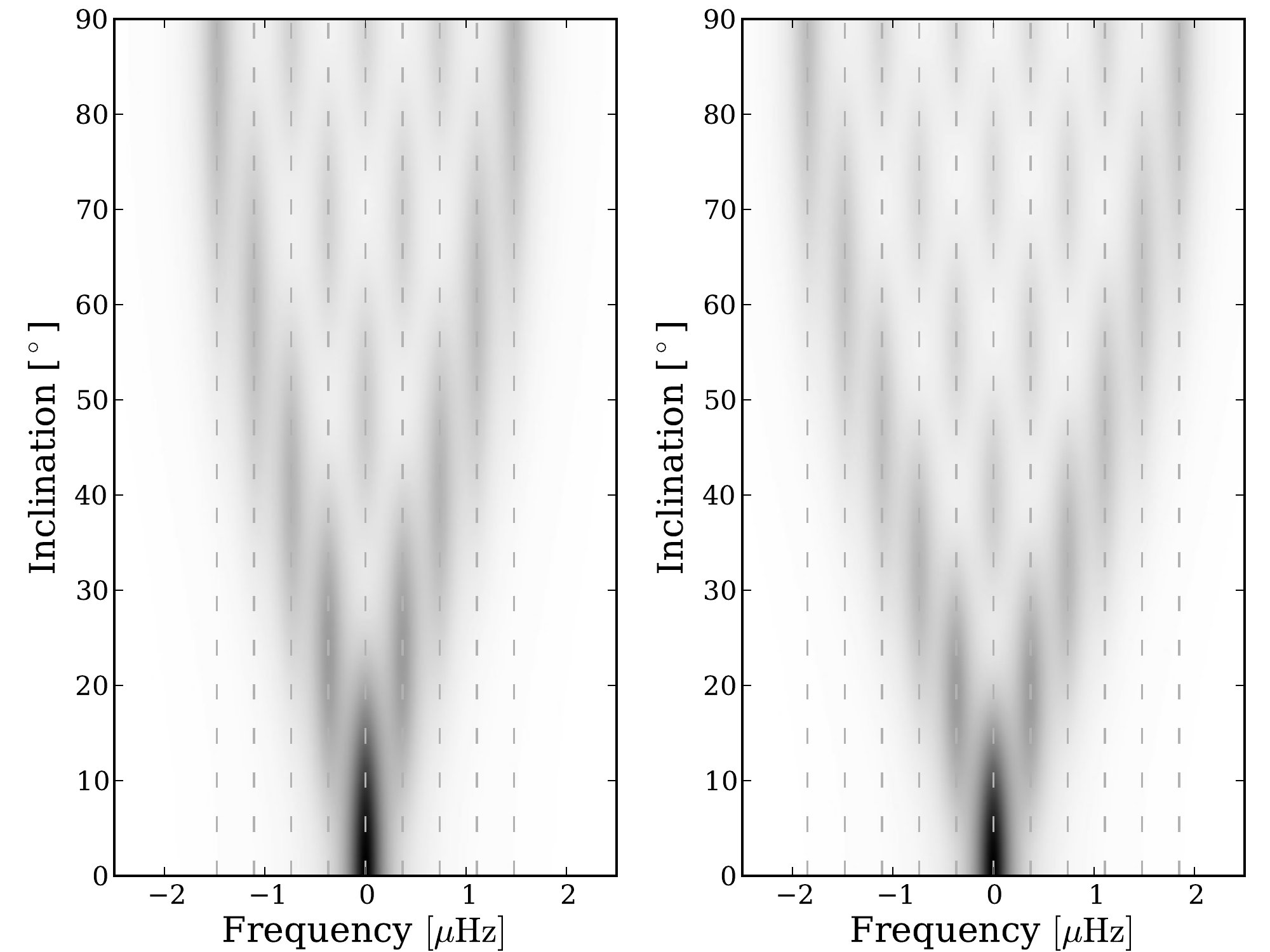}
 }
\caption{\footnotesize Top two panels give the functional form of $\mathcal{E}_{\ell=4, m}(i)$ (first) and $\mathcal{E}_{\ell=5, m}(i)$ (second). The $m=0$ component (blue) goes to a value of 1 at $i=0^{\circ}$. Bottom two panels render the same information, only here a top view of the rotationally split multiplets ($\nu_s = 0.37\, \rm \mu Hz$) is shown, with the visibilities of different azimuthal components given by the gray scale, ranging from black (high visibility) to white (low visibility) \citep{2003ApJ...589.1009G}. The vertical dashed lines show the positions of the split $m$-components.}
\label{fig:geovis_both}      
\end{figure}

The behavior of $\mathcal{E}_{\ell m}(i)$ for $\ell=4$ and $\ell=5$ modes is illustrated in the top panel of Figure~\ref{fig:geovis_both} as a function of the stellar inclination.  
An equivalent representation is given in the lower panel of Figure~\ref{fig:geovis_both}, where the visibilities in rotationally split multiples of $\ell=4$ (left) and $\ell=5$ (right) are plotted in gray scale as a function of inclination \citep[see ][]{2003ApJ...589.1009G}. Here the spread of power in frequency from $m=0$ is quite clear, especially at high inclination angles.

\vspace{0.5cm}
\subsection{Collapsing the Power Spectrum}
\label{sec:col2}

Prior to this final step in the construction of the \emph{SC}-spectrum, we assume that the stellar noise background has been fitted and corrected for.

When collapsing the power spectrum, we will be co-adding the segmented power spectrum. As the modes in the power spectrum have a decreasing signal-to-background ratio (\textsc{sbr}) away from the peak at $\nu_{\rm max}$, we weight the power as a function of the distance from this frequency. This is done in order to not simply add noise to the collapsed spectrum when using frequencies far away from $\nu_{\rm max}$.
 
The envelope of the \emph{p}-modes is most often described by a Gaussian function, and we chose this as our weighting function 
\begin{equation}
G(\nu) = \exp \left(- \frac{(\nu - \nu_{\rm max})^2}{2\sigma_{\rm env}^2} \right)\, .
\label{eq:weight}
\end{equation}
\cite{2012A&A...537A..30M} give the following relationship between $\nu_{\rm max}$ and the full width at half-maximum (\textsc{fwhm}), and thereby the spread of the Gaussian envelope:
\begin{align}
\rm \textsc{fwhm} &\approx 0.66\,\nu_{\rm max}^{0.88} \Rightarrow\\
\sigma_{\rm env} &\approx \frac{0.66\,\nu_{\rm max}^{0.88}}{2\sqrt{2 \ln(2)}}\, .
\end{align}
With this, we can construct the new weighted power spectrum to collapse:
\begin{equation}
P_{\rm new}(\nu) = P_{\rm old}(\nu) \, G(\nu)\, .
\end{equation}
The power spectrum which now has a modified frequency scale and weighted power values is then collapsed to form the \textit{SC}-spectrum, and the steps taken should ensure that the signal from the degree of interest should be better defined in frequency and easier to identify in power. 
Notice, that the application of Equation~\ref{eq:fremodif} on the power spectrum results as wanted in frequency equidistance of the modes for the targeted degree, but for modes of another degree it will generally have the opposite effect; equivalently, when the power spectrum is collapsed in general only the targeted degree will have a more well-defined signal peak while the peaks of other degrees will tend to be smeared out to some extend.  

Even with the above weightening, modes far from $\nu_{\rm max}$ will mainly contribute noise to the \textit{SC}-spectrum. Therefore, one should preferably only include a relatively small number of central orders. The optimum number of overtones to add for a specific target will depend on the \textsc{sbr} as a function of frequency for that specific target, in addition to the large separation, and is therefore not easily generalized.


\subsection{Remarks}
\label{sec:remark}

The choice of modes in the fitting of Equation~\ref{eq:asym2} is important. As seen in the \'echelle diagram of the Sun in Figure~\ref{fig:echelle_sun1} the ridges have in general an "S"-shape from the variation in the large separation. The section of the ridge that is well fitted by a quadratic term in $n$ is only the top part of this "S" (or, equivalently, only the lower part), \ie, from about $\rm 2400 \, \rm \mu Hz$ and up. As seen the curvature of this upper part of the "S" is quite well centred around $\nu_{\rm max}$ at around $\rm 3150 \, \rm \mu Hz$. So, using modes that lie far away from $\nu_{\rm max}$ can result in a bad fit of Equation~\ref{eq:asym2}. 

For more evolved stars, mixed modes should as far as possible be excluded from the fit to the modes - so, \eg, $\ell=1$ modes experiencing the largest effect of the mode bumping should be left out.

Another aspect to be aware of is that the collapsed power from the modes $\ell=0-5$ will to some extend be contaminated with power from even higher degree modes. As also mentioned in, \eg, \citet{1998ESASP.418...99A} $\ell=1$ modes are polluted by $\ell=6$ and $\ell=9$ modes, while $\ell=7$ modes fall in the proximity of the $\ell=4$ modes, $\ell=8$ near the $\ell=5$ modes, and so forth. For this reason, it will furthermore be very unlikely to pick up isolated signals from, \eg, $\ell=6-8$ modes.

If hypothesis testing is desired on the collapsed spectrum, in the form of for instance an H$_0$ test, the last step of weighting could be left out, and a binning of points rather than a smoothing would be more suitable.


\section{Analysis of 16 Cyg A and B}
\label{sec:kep}

We will now apply the method of the \textit{SC}-spectrum to the two solar analogues 16 Cyg A\footnote{KIC 12069424, HR 7503, HD 186408.} (G1.5V) and B\footnote{KIC 12069449, HR 7504, HD 186427.} (G3V). These two very similar stars are in fact part of a hierarchal triple system (16 Cygni), with a faint M dwarf as the third component. The most striking difference between the two stars is that measurements in radial velocity show 16 Cyg B to host a Jupiter-mass planet in a highly elliptical orbit \citep{1997ApJ...483..457C}. The second major difference between the two stars is the Li-abundance, which for some reason seems to be reduced in the A-component by a factor of $\geq4.5$ relative to the B-component \citep[see, \eg,][]{1993A&A...274..825F}. Because of the solar similarity and chemical peculiarities of the two main components this system has been studied by numerous groups, see, \eg, \cite{2011ApJ...737L..32S}, and references therein for examples of the chemical abundance analysis.
With its position in the sky (Cygnus - The Swan), the system is fortuitously observed by the \textit{Kepler} satellite. The \textit{Kepler} magnitudes\footnote{\textit{Kepler} magnitudes are nearly equivalent to \textit{R} band magnitudes \citep{2010ApJ...713L..79K}.} of these stars, 5.864 (16~Cyg A) and 6.095 (16~Cyg B), make them some of the very brightest stars in the \textit{Kepler} field-of-view, and ideal for asteroseismic studies. Such an asteroseismic study has recently been conducted by \citet[][hereafter \M]{2012ApJ...748L..10M}, where both stars were peak-bagged and subsequently modeled - we will use results and data from this study in our analysis.

The reason for choosing these stars as a starting case for the search of $\ell=4$ and $\ell=5$ modes is that both of them have already shown clearly detectable octupole $\ell=3$ modes. The prospects of having detectable modes with $\ell>3$ yields some very strong constraints on the stellar models. This comes in addition to the constraint on the allowed model differences from the binarity of the system.

As of now, it has not been possible to constrain the inclination angles of the stars using asteroseismology, but work is currently being done to resolve this issue (G. Davies 2013, private communication).


\subsection{\textit{Kepler} Data}

For both stars, we used short cadence \citep[SC, $\Delta t = 58.85\, \rm s$;][]{2010PASP..122..131G} data from quarters Q7-Q13, corresponding in the end to ${\sim}643$ days, all downloaded from the KASOC webpage\footnote{Kepler Asteroseismic Science Operations Center: \url{kasoc.phys.au.dk}.}. The stars were not observed in SC in Q0-Q5. As both stars are highly saturated\footnote{The saturation limit for \emph{Kepler} is about $K_p\sim 11.5$ \citep[][]{2010ApJ...713L.160G} large custom aperture masks are needed in order to capture as much flux as possible. However, the SC observations made in parts of Q6 did not make use of custom masks on the CCD which resulted in a rather poor data quality.} These data have therefore not been included in our data sets. 
The data type used is the uncorrected simple aperture photometry (SAP). 
The correction of the time series was done by high-pass filtering individual sub-quarters (${\sim}1$ month) by a one day moving median filter.
Bad data points (or outliers) were then removed, with a bad datum being identified as one having a point-to-point flux difference falling outside $3\sigma$ - with $\sigma$ found as the standard deviation of the point-to-point flux differences of the entire time series \citep{2011MNRAS.414L...6G}. We did not estimate this standard deviation (STD) directly but used instead the more robust median-absolute-deviation (MAD), and from this estimated the STD via the scaling ${\text{STD}=1.4826\times \text{MAD}}$\footnote{The multiplicative factor of $1.4826$ converts (approximately) the MAD for a normal distribution to a consistent measure of the STD, and can be found as $1/\Phi^{-1}(3/4)$, with $\Phi^{-1}$ representing the inverse cumulative distribution function or quantile function.}. The filtering of bad data points was performed iteratively four times. In addition, we used the "Quality"\footnote{The bit values of various known artefacts can be found in \citet{kepman}.} entry in the FITS files for the SAP to remove points with known artefacts. The final time series had duty cycles\footnote{Percentage of final number of points to the total expected number of points given the length and cadence of the time series.} of 85.7 $\%$ (16 Cyg A) and 82.2 $\%$ (16 Cyg B). With these duty cycles in mind the spectral windows for the two stars were checked, and we found no significant leakage of power into side-lobes. In any case, the smoothing of the power spectrum should ensure than any potentially leaked power is still accounted for.

\begin{figure*}
\centering
\subfigure{
   \includegraphics[scale=0.4] {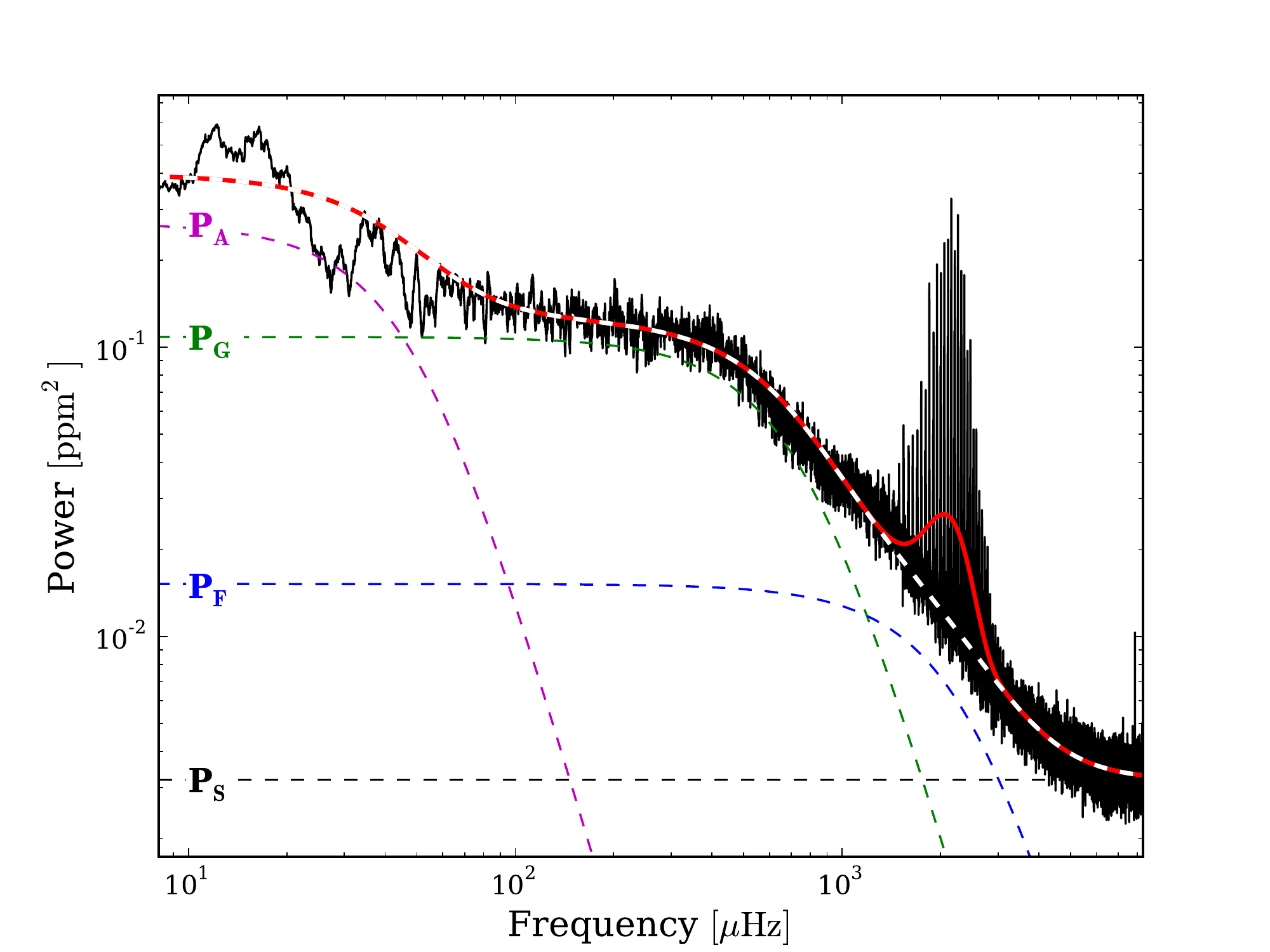}
 }
 \quad
 \subfigure{
   \includegraphics[scale=0.4] {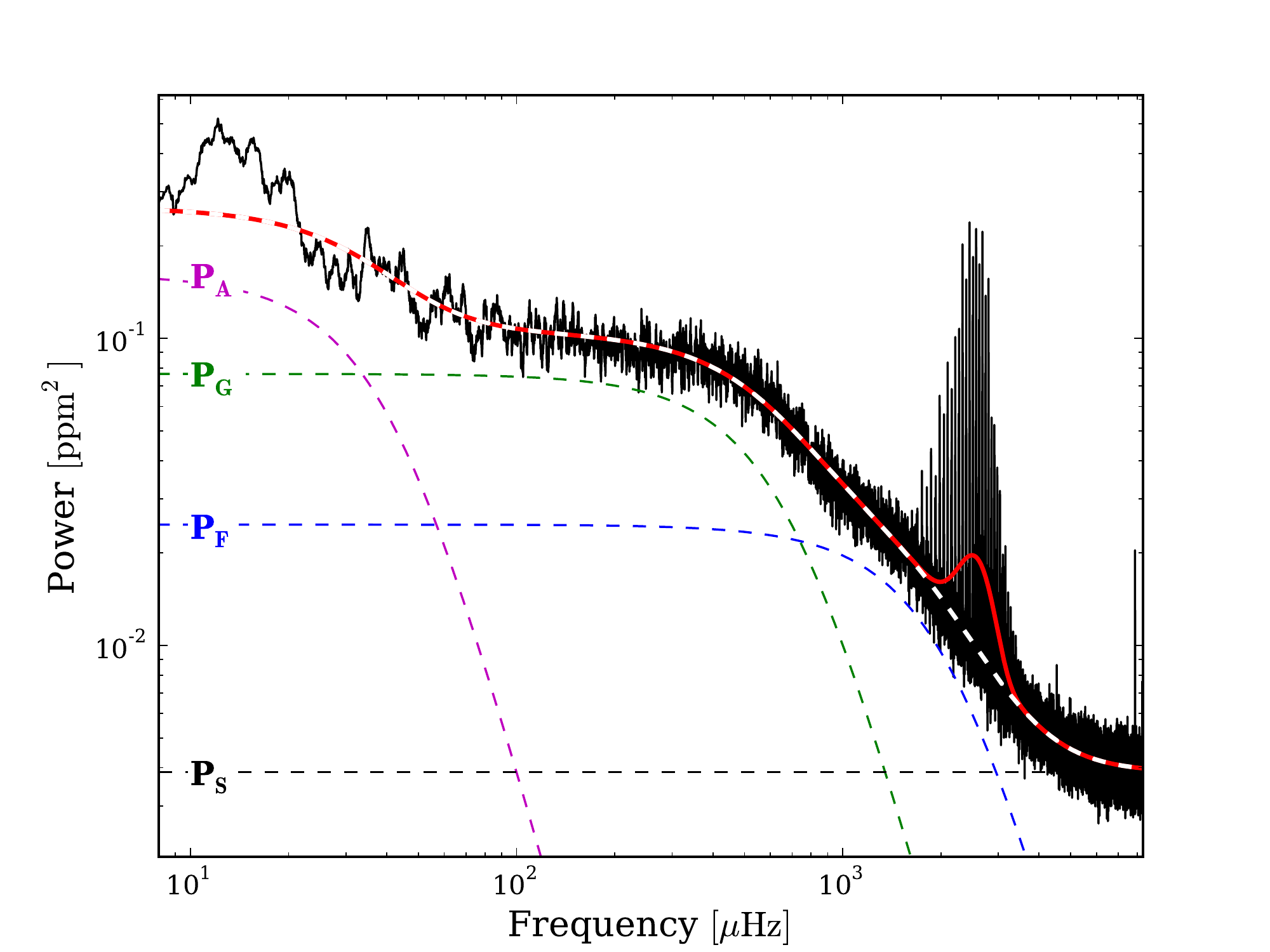}
 }
\caption{\footnotesize Power spectra (black) of 16 Cyg A (left) and B (right) from Q7-Q13 data, smoothed with a 1.8 $\mu$Hz boxcar filter. The optimum fit to the background (red) includes, besides the Gaussian envelope from \emph{p}-modes, the signature from activity (\textcolor[rgb]{0.75, 0, 0.75}{\textbf{P$\rm _A$}}; magenta), granulation (\textcolor[rgb]{0, 0.5, 0}{\textbf{P}$\rm _G$}; green), faculae (\textcolor[rgb]{0, 0, 1}{\textbf{P$\rm _B$}}; blue), and white/shot noise (\textcolor{black}{\textbf{P$\rm _S$}}; black). The dashed white line shows the background fit without the Gaussian envelope.} 
\label{fig:16Cyg}         
\end{figure*}

The power spectra (see Figure~\ref{fig:16Cyg}) were calculated in the same manner as for the solar data, except for the use of statistical weights in the computation.
Weights were computed as $w_i = 1/\sigma_i^2$, with $\sigma_i$ found from a 3 day windowed MAD of the corrected time series.

For the modeling of the stellar background signal, we use a sum of power laws \citep{1985ESASP.235..199H}, here in the version proposed by \cite{KarPhD} \citep[see also][]{2009CoAst.160...74H}, and in addition, we add a Gaussian function to account for the excess \emph{p}-mode power:
\begin{align}\label{eq:backg} 
B(\nu) = \sum_{i=0}^{3} &\frac{4\sigma_i^2\tau_i}{1 + (2\pi\nu\tau_i)^2 + (2\pi\nu\tau_i)^4}\\
 &+ A\, \exp \left(- \frac{(\nu - \nu_{\rm max})^2}{2\sigma_g^2} \right) + B_N, \notag
\end{align}
where $B_N$ is the white shot-noise component, $\sigma_i$ is the rms intensity of the $i$th noise component, $\tau_i$ is the corresponding time scale of the noise component, $A$ and $\sigma_g$ are the amplitude and spread, respectively, of the \emph{p}-mode envelope.
The three background components included, and shown in Figure~\ref{fig:16Cyg}, account for the activity (magenta), granulation (green), and faculae (blue) signals, respectively. 
When correcting the power spectrum for the stellar background, we of course leave out the Gaussian \emph{p}-mode envelope (dashed white) from the full fit (thick red). 

\vspace*{1cm}
\section{Results on detectability from bona fide data}
\label{sec:results1}

\begin{figure*}
\centering
\subfigure{
   \includegraphics[scale=0.4] {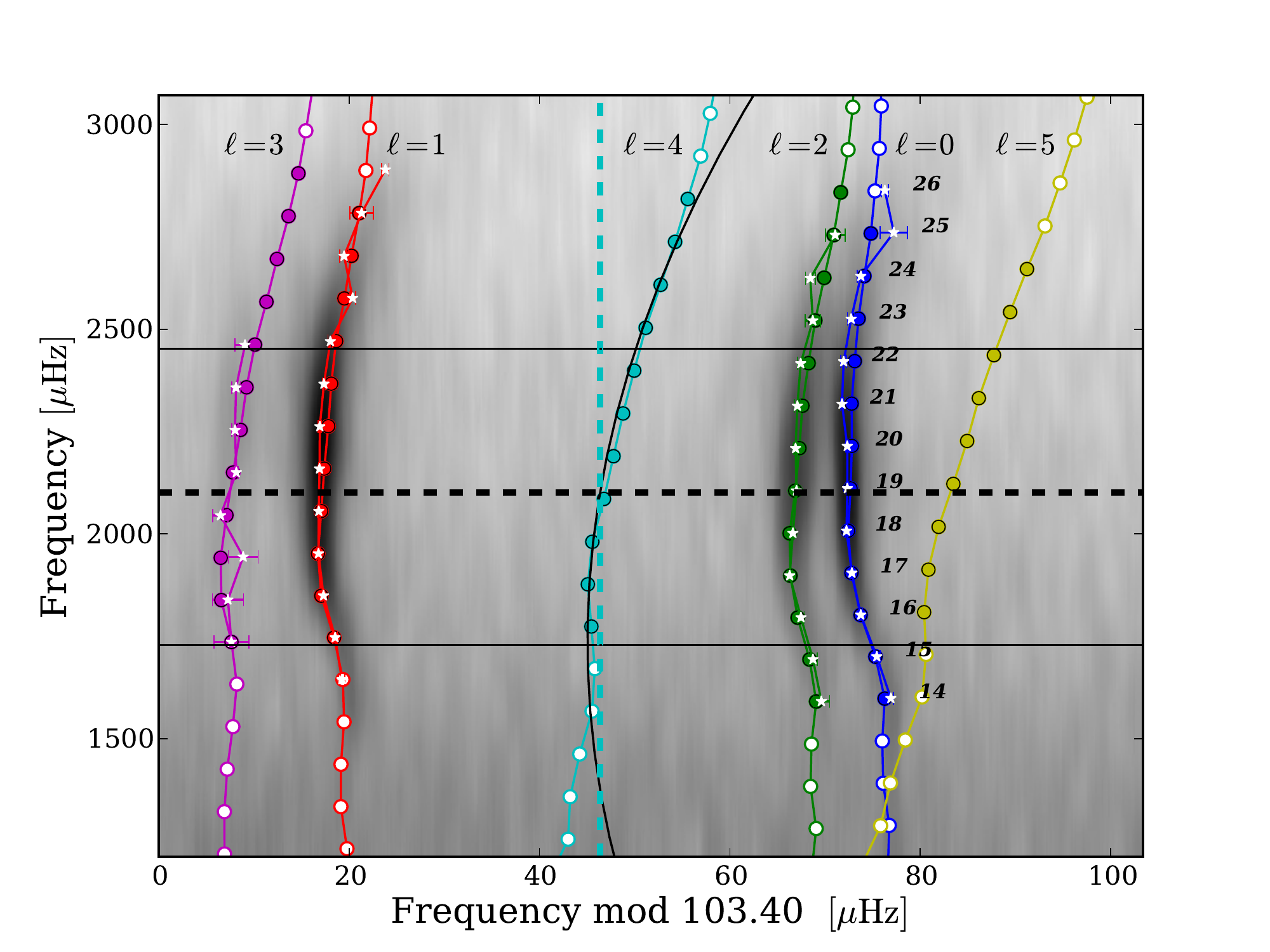}
 }
 \quad
 \subfigure{
   \includegraphics[scale=0.4] {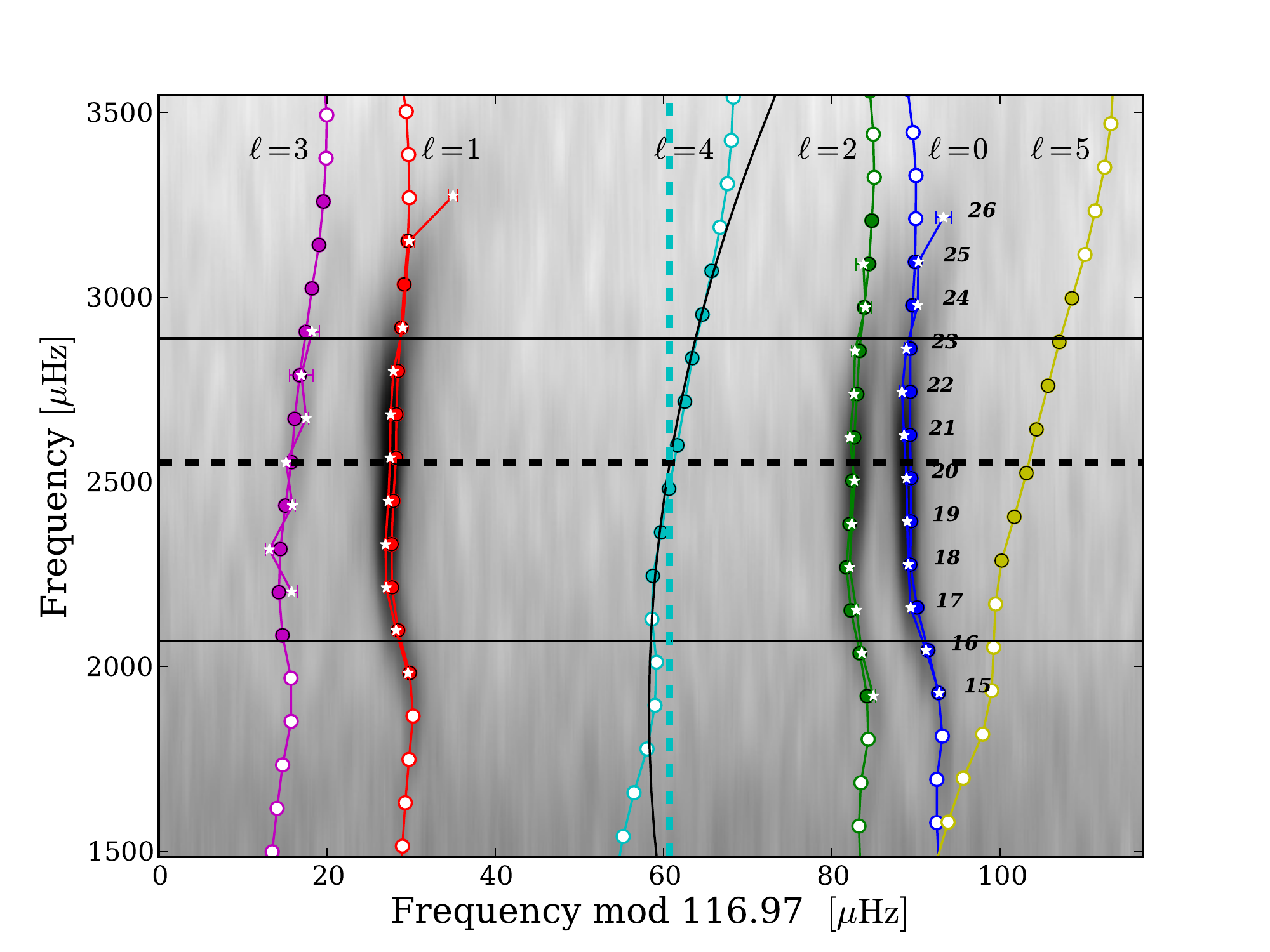}
 }
\caption{\footnotesize Method from \S~\ref{sec:col} applied to \textit{Kepler} data of 16 Cyg A (left) and B (right). The gray scale in these \'echelle diagrams range from white at low power to black at high power. Observed mode frequencies are given as white stars with associated errorbars. Model frequencies (see \S~\ref{sec:modelling}) are given as circles, with filled circles used in the fitting of Equation~\ref{eq:asym2}, while white circles were left out. The black line following the $\ell=4$ ridge illustrate the obtained fit to estimated frequencies. The radial order, $n$, of the $\ell=0$ modes is indicated by the numbers. The two horizontal black lines give the range in frequency that was collapsed in making the \emph{SC}-spectrum, while the black dashed line indicate the frequency of $\nu_{\rm max}$. The vertical dashed cyan line gives the positions of the $\ell=4$ ridge after the straightening.}
\label{fig:res1}           
\end{figure*}

We applied the method presented in \S~\ref{sec:col} to the \textit{Kepler} data of 16 Cyg A and B. 
Figure~\ref{fig:res1} gives the \'echelle diagrams of the two stars, with the same color convention for the different degrees as used in Figure~\ref{fig:model_ech} for the Sun. In the figure, observed mode frequencies are given as white stars with associated errorbars. Model frequencies (see \S~\ref{sec:modelling}) are given as circles, with the distinction that filled circles were used in the fitting of Equation~\ref{eq:asym2}, while white circles were left out - see \S~\ref{sec:remark} for the considerations made in the selection of modes to include in the fit. In the plot, we have indicated the radial order $n$ of the $\ell=0$ modes. The fit to the model frequencies was made with the straightening of the $\ell=4$ ridge in mind, and the black line depicts the obtained fit (similar curves of course exist for the other degrees, see Figure~\ref{fig:model_ech}). The dashed cyan line gives the position of the ridge after the straightening, with the reference frequency $\nu_0$ (dashed black line) set equal to the value of $\nu_{\rm max}$ obtained from fitting Equation~\ref{eq:backg} to the power spectra, viz., $2101 \, \rm \mu Hz$ (16 Cyg A) and $2552 \, \rm \mu Hz$ (16 Cyg B).
The regions between the two thin horizontal lines give the range chosen in the collapsing of the straightened power spectra, \ie, the ${\sim} 7$ orders closest to $\nu_{\rm max}$. This region largely coincides with the region wherein $\ell=3$ modes are readily observed.

\begin{figure*}
\centering
\subfigure{
   \includegraphics[scale=0.4] {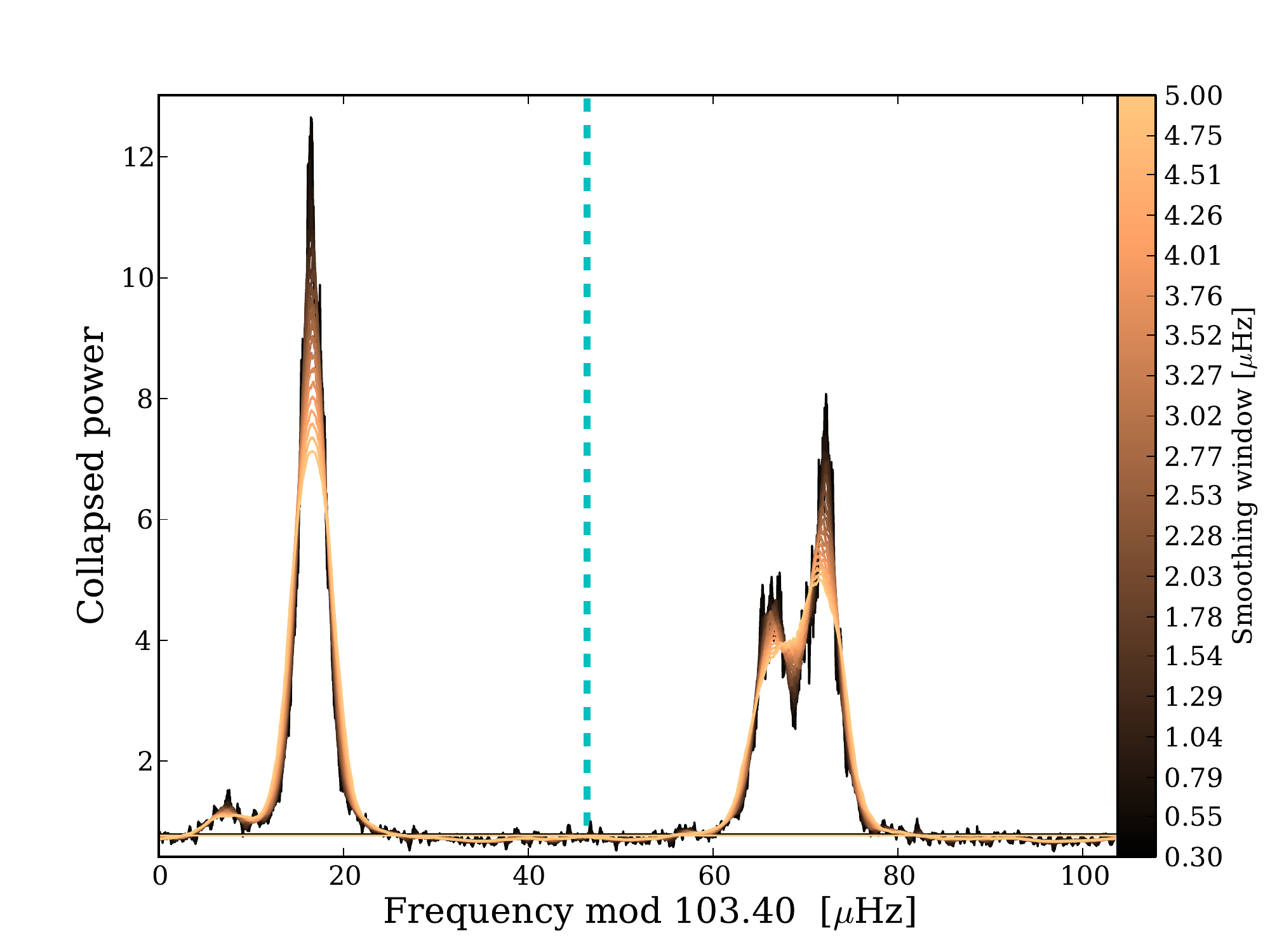}
 }
 \subfigure{
   \includegraphics[scale=0.4] {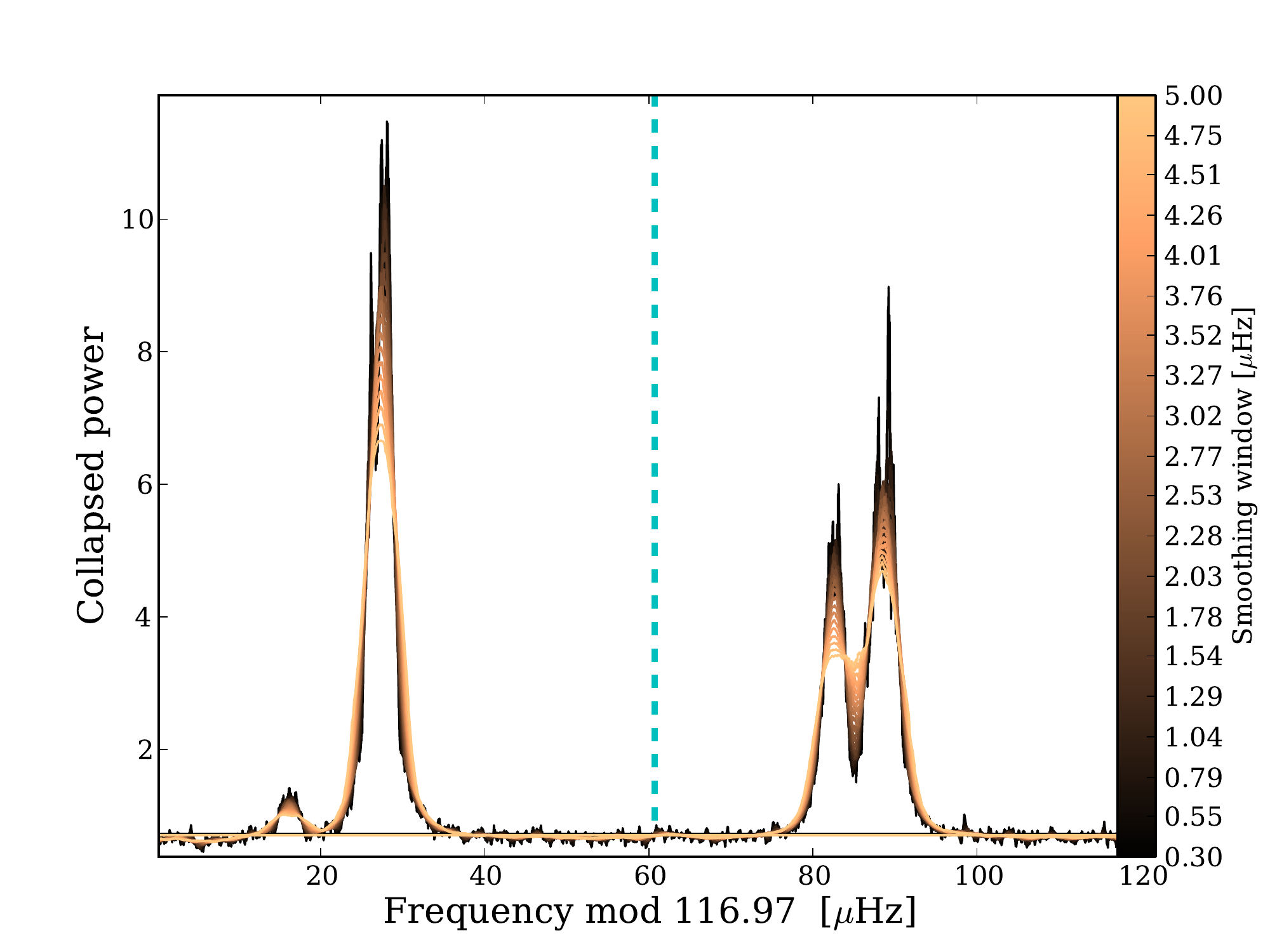}
 }\\
\subfigure{
   \includegraphics[scale=0.4] {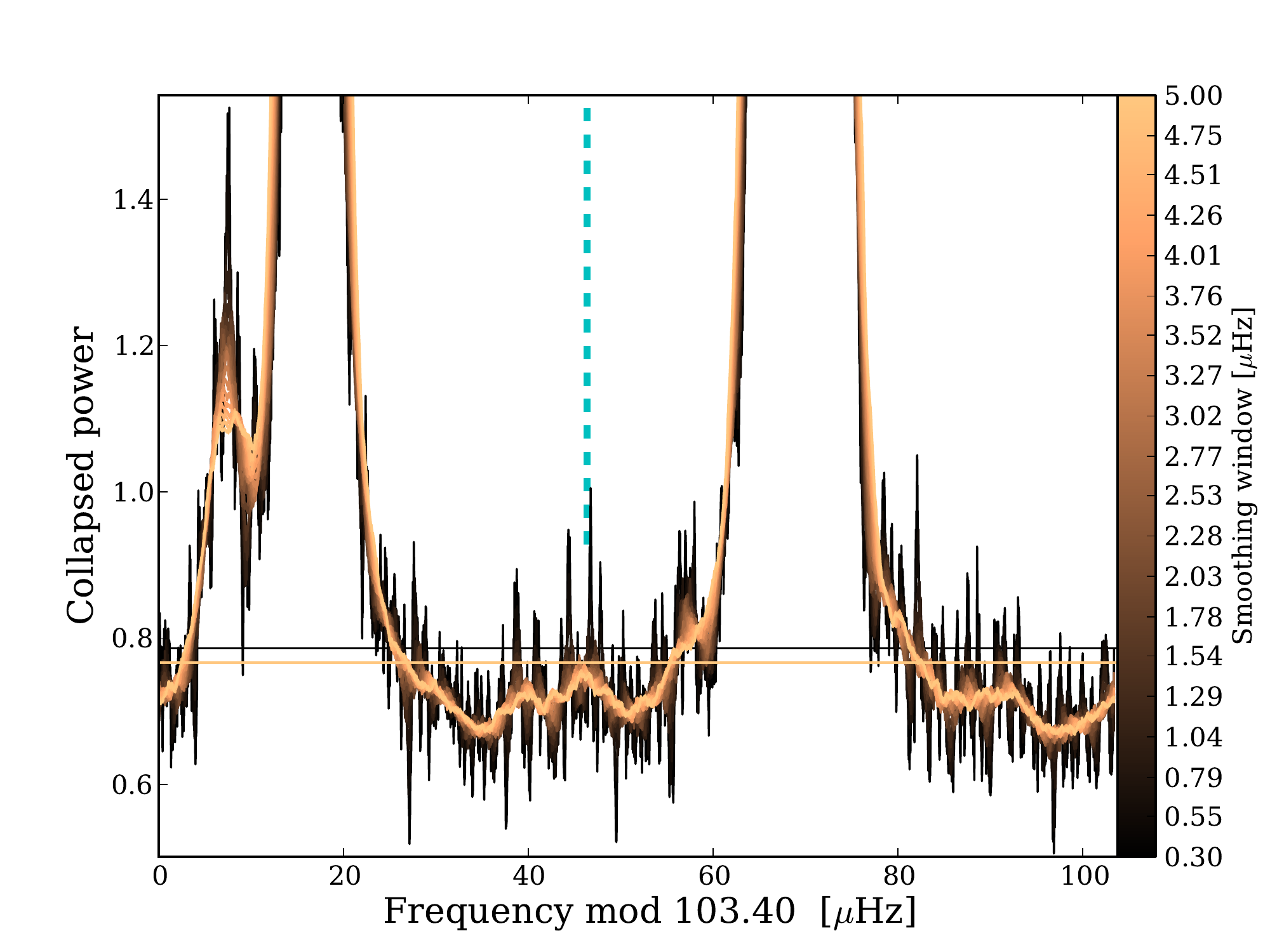}
 }
 \subfigure{
   \includegraphics[scale=0.4] {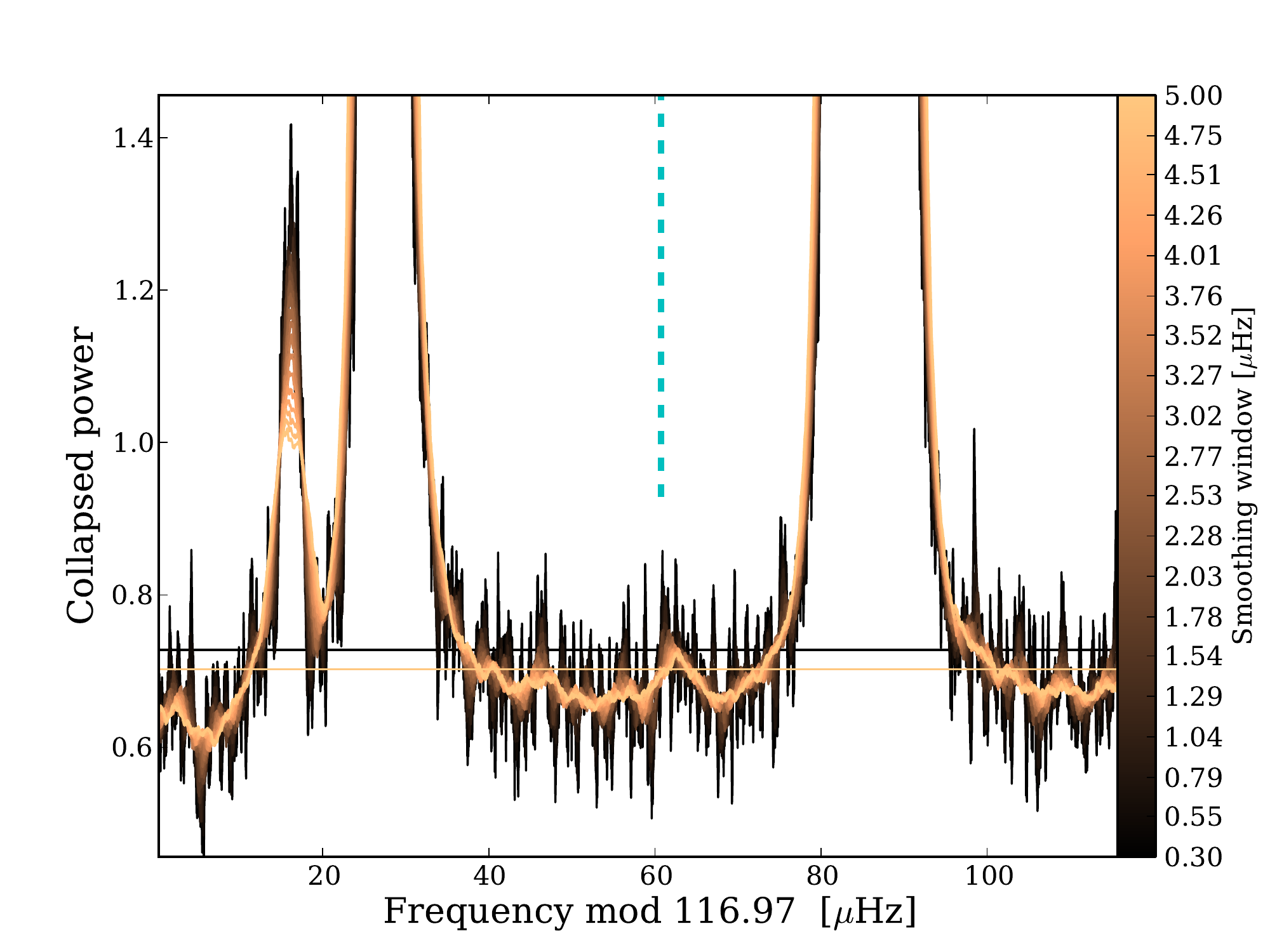}
 } 
\caption{\footnotesize \textit{SC}-spectra for 16 Cyg A (left) and B (right) with different amounts of applied smoothing. The top two panels give the full \emph{SC}-spectra, while the bottom two panels show zoom-in versions. The dashed cyan line indicate as in Figure~\ref{fig:res1} the position of the straightened $\ell=4$ ridge and hence the expected position of a possible excess from $\ell=4$ modes. The horizontal lines give the median values of the \textit{SC}-spectra in the minimum and maximum smoothing cases. }
\label{fig:res2}               
\end{figure*}

In Figure~\ref{fig:res2} the \textit{SC}-spectra are shown, and rendered in a multitude of different amounts of applied smoothing. The top two panels give the full \textit{SC}-spectra, while the bottom two panels show zoom-in versions. The dashed cyan line indicate as in Figure~\ref{fig:res1} the position of the straightened $\ell=4$ ridge, and thereby give the expected position of the possible excess from $\ell=4$ modes. The horizontal lines give the median value of the \textit{SC}-spectrum in the minimum and maximum smoothing cases. These have been added simply to establish a reference level in the spectrum.
For both stars we see a rise in the collapsed power at the expected position of the $\ell=4$ ridge, while no clear excess power is seen directly in the power spectra for the expected frequencies of individual modes. 

We also applied the method targeting the $\ell=5$ ridge, but here no noticeable rise was seen in the collapsed power. The $\ell=5$ ridge is in any case more difficult to capture due to its close proximity to the strong $\ell=0$ modes.     

Model frequencies were obtained from the stellar model described in \S~\ref{sec:modelling} below, and no other model was tested for the straightening. However, we note that the small separation $\delta \nu_{02}$ seems to be correctly reproduced as the model frequencies match well with the observed values for both the $\ell=0$ and $\ell=2$ ridges. From the definition of $\delta \nu_{02}$ as $6 D_0$ (see \S~\ref{sec:para}) this means that $D_0$ is approximately correct and the $\ell=4$ ridge should therefore be located at nearly the same place in the \'echelle diagrams for all the models that correctly reproduce $\delta \nu_{02}$.


\section{Simulated power spectra}
\label{sec:simulation}

In order to check if a potential excess power in the \textit{SC}-spectrum is in line with what can be expected for $\ell=4$ or $\ell=5$ from the current amount of observed data, we simulated the power spectra of 16 Cyg A and B.
This also enables a test on the impact of longer observing times on the detectability of the $\ell=4$ and $\ell=5$ modes, and thereby hint to the needed amount of data for a solid detection.
Furthermore, we can investigate the role of the currently unknown stellar inclination angles, and test the effect of applying different amounts of smoothing.
In the following, the various aspects underlying the simulations will be described, to wit, the stellar modeling, the synthetic power spectrum, visibilities, and noise properties.  


\subsection{Stellar Modeling}
\label{sec:modelling}

For the stellar models needed to calculate the pulsation frequencies we use pre-existing models\footnote{16 Cyg A: \url{amp.phys.au.dk//browse/simulation/191}\\ 16 Cyg B: \url{amp.phys.au.dk//browse/simulation/189}.} from the asteroseismic modeling portal \citep[AMP;][]{2009ApJ...699..373M, 2010arXiv1011.6332W} - see Table~\ref{tab:AMP_models} for the parameters of the adopted models.
The models were in AMP found by optimizing \citep[using a parallel genetic algorithm (GA), see][]{2003JCoPh.185..176M} the fit of the observed oscillation frequencies (published in \M) and observational constraints to modeled oscillation frequencies. The spectroscopic constraints listed in Table~\ref{tab:AMP_models} are from \cite{2009A&A...508L..17R}.
We refer the reader to \M\ and references therein for the input physics and specific details of the modeling.
Notice, that the final AMP parameters quoted in \M\ were found after using a localized Levenberg-Marquardt optimization algorithm (utilizing singular-value-decomposition) to adjust the stellar parameters from the values found by the GA, viz., the values given in Table~\ref{tab:AMP_models}. For this reason the values in Table~\ref{tab:AMP_models} will differ slightly from the ones given in \M.
With the best-fitting AMP models, we use the "Aarhus adiabatic pulsation package" \citep[ADIPLS;][]{2008Ap&SS.316..113C} to compute the oscillation frequencies including $\ell=4$ and $\ell=5$ modes.
Finally, we apply an empirical correction for surface effects in order to have model frequencies that resemble the observed frequencies of 16 Cyg A and B as closely as possible. 
Following the prescription by \citet{2008ApJ...683L.175K}, the frequency correction applied to the model frequencies is given by
\begin{equation}
\delta \nu \equiv \nu_{\rm obs}(n) - \nu_{\rm model}(n) = a \left(\frac{\nu_{\rm obs}(n)}{\nu_0} \right)^b\, ,
\label{eq:surface}
\end{equation} 
where $\nu_0$ is a reference frequency which we set equal to the mean of the observed radial modes; $b$ is set to the solar calibrated value of $4.823$ \citep{2012ApJ...749..152M}. The value of $a$ is found using Equation~10\footnote{Using $r=1$.} in \citet{2008ApJ...683L.175K}. Note that when using Equation~\ref{eq:surface} to correct model frequencies outside the range of observed modes one should be aware that this parameterization of the surface correction is not suitable at frequencies far from the observed values due to a frequency-range dependence in the exponent $b$ as the frequency difference does not exactly follow a simple power law \citep{2008ApJ...683L.175K}.

\begin{table}
\caption{\small AMP Modeling}
\centering
\begin{tabular}{lcc}
\hline \hline\\[-0.3cm]
Parameter  &  Model A &  Model B \\[0.05cm]
\hline \\[-0.3cm]

\multicolumn{3}{c}{Results} \\[0.05cm]
\hline     
$M\, (M_{\odot})$	   	& 1.10  	& 1.07   \\
$R\, (R_{\odot})$		& 1.2361    & 1.1256  \\
$L\, (L_{\odot})$		& 1.5669  	& 1.2616  \\
$\tau \rm \, (Gyr)$     & 6.5425  	& 5.8162  \\
$\alpha_{\rm ML}$  			& 2.06    	& 2.00  \\
$Y_0$         		& 0.2510  	& 0.2430\\
$Z$		    			& 0.02239 	& 0.02032 \\
$T_{\rm eff} \, (\rm K)$	& 5814.01 	& 5771.32  \\

\hline\\[-0.3cm]
\multicolumn{3}{c}{Constraints in Optimization}   \\[0.05cm]
 \hline
$T_{\rm eff} \, (\rm K)$	&	5825 $\pm$ 50   		& 5750 $\pm$ 50  \\
$\rm [Fe/H]$			&	0.096 $\pm$ 0.026   	& 0.052 $\pm$ 0.021  \\
$\log g$			&	4.33 $\pm$ 0.07 		& 4.34 $\pm$ 0.07\\
$L\, (L_{\odot})$\footnote{See \M\ and references therein for the calculation of the luminosity.}		&	1.56 $\pm$ 0.05	    & 1.27 $\pm$ 0.04\\
No. of fitted modes	&   46  				& 41  \\

\hline\\[-0.3cm]
\multicolumn{3}{c}{Surface Correction Parameters}  \\[0.05cm]
 \hline
a	$(\rm \mu Hz)$			&	-3.100  	& -4.927  \\
$\nu_0$	$(\rm \mu Hz)$		&	2215.812   	& 2569.344  \\
\hline
\end{tabular}
\label{tab:AMP_models}
\end{table}


\subsection{The Synthetic Limit Spectrum}
\label{sec:synspec}

Even though it has not yet been possible to extract mean rotation periods from the power spectra of 16 Cyg A and B, we can with gyrochronology make a rough estimate for the rotation period.
The mean rotation period is found using the expression of \cite{2007ApJ...669.1167B}:
\begin{equation}
P_{\rm rot} = t^n \, a \, [(B-V)_0 - c]^b\, ,
\label{eq:gyro}
\end{equation}
with parameters $n = 0.519$, $a=0.773$, $b=0.601$, and $c=0.4$. Here $t$ is the stellar age in Myr, for which we use the model values given in Table~\ref{tab:AMP_models}.
The values for $(B-V)_0$ are $0.64$ (16 Cyg A) and $0.66$ (16 Cyg B), both adopted from \cite{1953ApJ...117..313J}.
Using Equation~\ref{eq:gyro} on the model results, gives for both models a rotation period of $P_{\rm rot} \approx 31$ days, equivalent to a rotation frequency of $\nu_s\approx 0.37\, \rm \mu Hz$.
Using instead the finally adopted common age of $6.8\pm0.4$ Gyr from \M\ results only in minor differences.
In setting up the model power spectrum, the effect of rotation on a mode of radial order $n$, degree $\ell$, and azimuthal order $m$ is included to first order as \citep{1951ApJ...114..373L}
\begin{equation}
\nu_{n\ell m} = \nu_{n\ell} + m(1 - C_{n\ell}) \nu_s\,  ,
\end{equation}
where the effect of the Coriolis force is represented by the dimensionless parameter $C_{n\ell}$.
For high order, low degree acoustic modes, as the ones seen in the solar analogues 16 Cyg A and B, the rotational frequency splitting is dominated by advection and the parameter $C_{n\ell}$ is set equal to 0.

In the simulated power spectra, we describe individual modes as a standard Lorentzian \citep[see, \eg,][]{1990ApJ...364..699A} given by
\begin{equation}
L_{n\ell m}(\nu) = S_{n\ell m} \left[1+\left(\frac{(\nu-\nu_{n\ell m})}{\Gamma_{n\ell}/2}\right)^2 \right]^{-1}\, .
\label{eq:lorentz}
\end{equation}
This shape is appropriate for describing a damped-driven oscillation such as the stochastically excited \emph{p}-modes. In Equation~\ref{eq:lorentz} $\Gamma_{n\ell}$ is the damping rate for the mode and gives the \textsc{fwhm} value of $L_{n\ell m}(\nu)$. Furthermore, the mode lifetime is given by $\tau_{n\ell} = (\pi \Gamma_{n\ell})^{-1}$. 
No mode asymmetries have been included in our description.

For the width of the individual modes, there is a general consensus in the field of a high temperature dependence, commonly given as a power law $\Gamma_{\rm max} \propto T_{\rm eff}^n$ with $\Gamma_{\rm max}$ as the mode linewidth at $\nu_{\rm max}$. Also, there is a frequency dependence to $\Gamma$, with a local minimum at $\nu_{\rm max}$ and with decreasing widths toward lower frequency and increasing width toward higher frequencies \citep[see, \eg,][]{1986ssds.proc..223I,1988ApJ...334..510L,1997MNRAS.288..623C}. Typically, no dependence is assumed for the degree $\ell$ of the mode \citep[][]{1988ApJ...334..510L,1999A&A...351..582H}. There is, however, not an unequivocal value for the exact size of the dependence (given by $n$), and this is indeed still a matter of great debate in the literature. For MS stars \citet{2009A&A...500L..21C} found a temperature exponent of $n \approx 4$, \citet{2011A&A...535C...1B, 2011A&A...529A..84B} found using \emph{CoRoT}\footnote{COnvection ROtation and planetary Transits.} observations $n \approx 16 \pm 2$, while \citet{2012A&A...537A.134A} from \emph{Kepler} observations found a value of $n \approx 15.5 \pm 1.6$ if the measurement was made at maximum mode amplitude and $n \approx 13.0 \pm 1.4$ if it was made at maximum mode height. \citet{2012A&A...540L...7B} found from a theoretical approach an exponent of $n \approx 10.8$, and with the expression for $\Gamma$ including a small dependence on surface gravity \citep[see also][]{2013EPJWC..4303009B}. See also \citet{2012ApJ...757..190C}, who adopt an exponential scaling as a function of $T_{\rm eff}$. 

In this work, we have chosen to scale the width of the individual modes from solar values\footnote{From BiSON observations.} and use a temperature exponent of $n=7.5$ (G. Houdek 2012, private communication), whereby $\Gamma(\nu)$ is found according to
\begin{equation}
\Gamma(\nu_{n\ell,*}) = \left(\frac{T_{\rm eff}}{5777\, \rm K} \right)^{7.5} \times \tilde{\Gamma}_{\odot}(\nu_{n\ell,*}/\nu_{\rm max,*})\, .
\end{equation}
Here $\tilde{\Gamma}_{\odot}$ gives the solar values for the linewidth on a frequency scale of $\nu_{n\ell,{\odot}}/\nu_{\rm max,{\odot}}$. The asterisk subscript denotes the stellar values in the equation. Note that this equation is not the same as the linewidth relation given in, \eg, \citet{2012A&A...537A.134A}, which instead gives the mode linewidth at maximum mode height/amplitude as a function of effective temperature.

The exponent on the temperature dependence is in general a very important input parameter as it impacts the height and by extension the detectability of a given mode in the power spectrum (see, Equation~\ref{eq:height}). However, as we are here dealing with solar analogues having effective temperature comparable to the Sun, the difference in mode linewidths from different exponents of the temperature dependence is negligible. As an example, with the temperature of $T_{\rm eff}=5825\, \rm K$ for 16 Cyg B the difference in mode linewidth at $\nu_{n\ell,*}=\nu_{\rm max,*}$ between using an exponent of $n=7.5$ and one of $n=15.5$ \citep{2012A&A...537A.134A} amounts to a difference in linewidth of $\rm {\sim} 0.073 \, \mu Hz$. Furthermore, the smoothing applied in the making of the \textit{SC}-spectrum ensures that any small differences in mode linewidth will be rendered unimportant.

In assuming equipartition of energy between the components of a $(n\ell m)$-multiplet, the height $S_{n\ell m}$ in Equation~\ref{eq:lorentz} can be written as
\begin{equation}
S_{n\ell m} = \mathcal{E}_{\ell m}(i)S_{n\ell} = \mathcal{E}_{\ell m}(i) \tilde{V}_{\ell}^2 \alpha_{n\ell}\, ,
\label{eq:vis}
\end{equation}
where $\mathcal{E}_{\ell m}(i)$ is the geometrical function given in Equation~\ref{eq:epsilon}.

$\tilde{V}^2_{\ell}$ is the square of the so-called relative mode visibility, \ie, the squared amplitude (power) ratio between different $\ell$-components normalized to the radial modes - we calculate our own values for the visibilities in \S~\ref{sec:vis}, as tabulated values in the literature seldom include calculations for $\ell=4$ and $\ell=5$ \citep[see ][for calculations of visibilities for velocity measurements]{1982MNRAS.198..141C}. 
The last factor in Equation~\ref{eq:vis} represents a frequency dependent amplitude modulation. We describe this modulation by a Gaussian function, $G(\nu)$, such as the one given in Equation~\ref{eq:weight}:
\begin{equation}
\alpha_{n\ell}(\nu) =H_{\rm max}\, G(\nu)\, .
\label{eq:alpha_mod}
\end{equation}
In this formula, $H_{\rm max}$ is the maximum height of the radial ($\ell=0$) modes found at the corresponding frequency $\nu_{\rm max}$.

The height (in power density) is found as \citep{2006MNRAS.371..935F, 2008A&A...485..813C}
\begin{equation}
H(T) = \frac{2A^2/\pi\Gamma}{[1 + (2/\pi\Gamma T)]}\, ,
\label{eq:height}
\end{equation}
where $A$ is the mode amplitude and $T$ is the observing length.

The maximum amplitude $A_{\rm max}$ is for both models estimated from the power spectra of 16 Cyg A and B. 
This is done following the prescription given in \citet{2008ApJ...682.1370K}, which is a revised treatment of the \citet{1995A&A...293...87K} procedure \citep[see also][]{2009A&A...495..979M,2012A&A...537A..30M}. In brief, the power spectrum is first heavily smoothed to produce a single power bump. For the smoothing, we used a boxcar filter with a width of $4\Delta\nu$, with estimated values for $\Delta\nu$ of $103.4 \, \rm \mu Hz$ (16 Cyg A) and $116.97 \, \rm \mu Hz$ (16 Cyg B). The same smoothing is done on the previously fitted background function in Equation~\ref{eq:backg} (without Gaussian envelope). Power (ppm$^2$) is then converted to power spectral density (ppm$^2 \, \rm \mu Hz ^{-1}$) by multiplying the power spectrum with the effective observation length (equivalent to the area under the spectral window). Now the smoothed background function is subtracted from the smoothed power spectrum, leaving ideally only the oscillation bump left. The maximum value of the power bump is now multiplied by $\Delta\nu$, thereby giving the total power contained in all modes within one large separation. As we would like to determine the amplitude per oscillation mode at the peak in power ($\nu_{\rm max}$), we divide by the sum $\tilde V_{\rm tot}^2$ of squared relative visibilities \citep[$c$ factor in][]{2008ApJ...682.1370K} as this gives the effective number of modes per order.
This can be written as
\begin{equation}
\langle A_{\ell=0,\rm max} \rangle = \sqrt{\frac{P_{\rm max}\, \Delta\nu}{\tilde{V}_{\rm tot}^2}}\, ,
\end{equation}
where $P_{\rm max}$ give the peak value of the smoothed power bump, in units of power density.

In the fit of Equation~\ref{eq:backg} to the power spectra we obtain direct values for $P_{\rm max}$ from the value of $A$, and we tested that this value indeed corresponds to the value obtained from the above procedure. The fit also directly gives us the ingredients for Gaussian function $G(\nu)$, specifically $\sigma_g$ and $\nu_{\rm max}$.

The final limit spectrum is now constructed as the sum of Lorentzian functions from the individual modes: 
\begin{equation}
\mathcal{P}(\nu_j) = \sum_{n=n_{a}}^{n_{b}}\sum_{\ell=0}^{5}\sum_{m=-\ell}^{\ell} \frac{\mathcal{E}_{\ell m}(i)S_{n\ell}}{1+\frac{4}{\Gamma_{n\ell}^2}(\nu_j-\nu_{n\ell m})^2 } + N(\nu_j)\, . 
\label{eq:limitspec} 
\end{equation}
In this equation, $N(\nu_j)$ gives the added noise term (see \S~\ref{sec:noise} below) comprising both the instrumental and stellar noise contributions.


\subsection{Visibilities}
\label{sec:vis}

The visibilities used in the simulations of the power spectra are naturally of great importance and deserve special attention.

We calculate theoretical visibilities following the method described in \citet{2011A&A...531A.124B} [hereinafter \B].
To test this method, we first calculate the solar visibilities and then compare to values measured directly from the power spectra of the Sun.

The visibility of a mode of degree $\ell$ can be written as \citep{1977AcA....27..203D, 2003ApJ...589.1009G}
\begin{equation}
V_{\ell} = \sqrt{\pi (2\ell+1)} \int_0^{1} P_{\ell}^0(\mu) W(\mu)  \mu d\mu\, .
\label{eq:visibil}
\end{equation}
In this equation $P_{\ell}^0$ is the $\ell$th order Legendre polynomial, and $\mu$ is measure of the projected distance of a surface element to the stellar limb given by $\mu = \cos \phi$, with $\phi$ being the angle between the line of sight and the normal to the surface at the position of the element. Thus, $\mu$ varies from 0 at the limb to 1 at the center. $W(\mu)$ is linked to the limb-darkening (LD) function for the star, given by the relative intensity at a specific wavelength $\lambda$ to the center, \ie, $g_{\lambda}(\mu) = I_{\lambda}(\mu)/I_{\lambda}(1)$.

\begin{figure}
\centering
\includegraphics[scale=0.4]{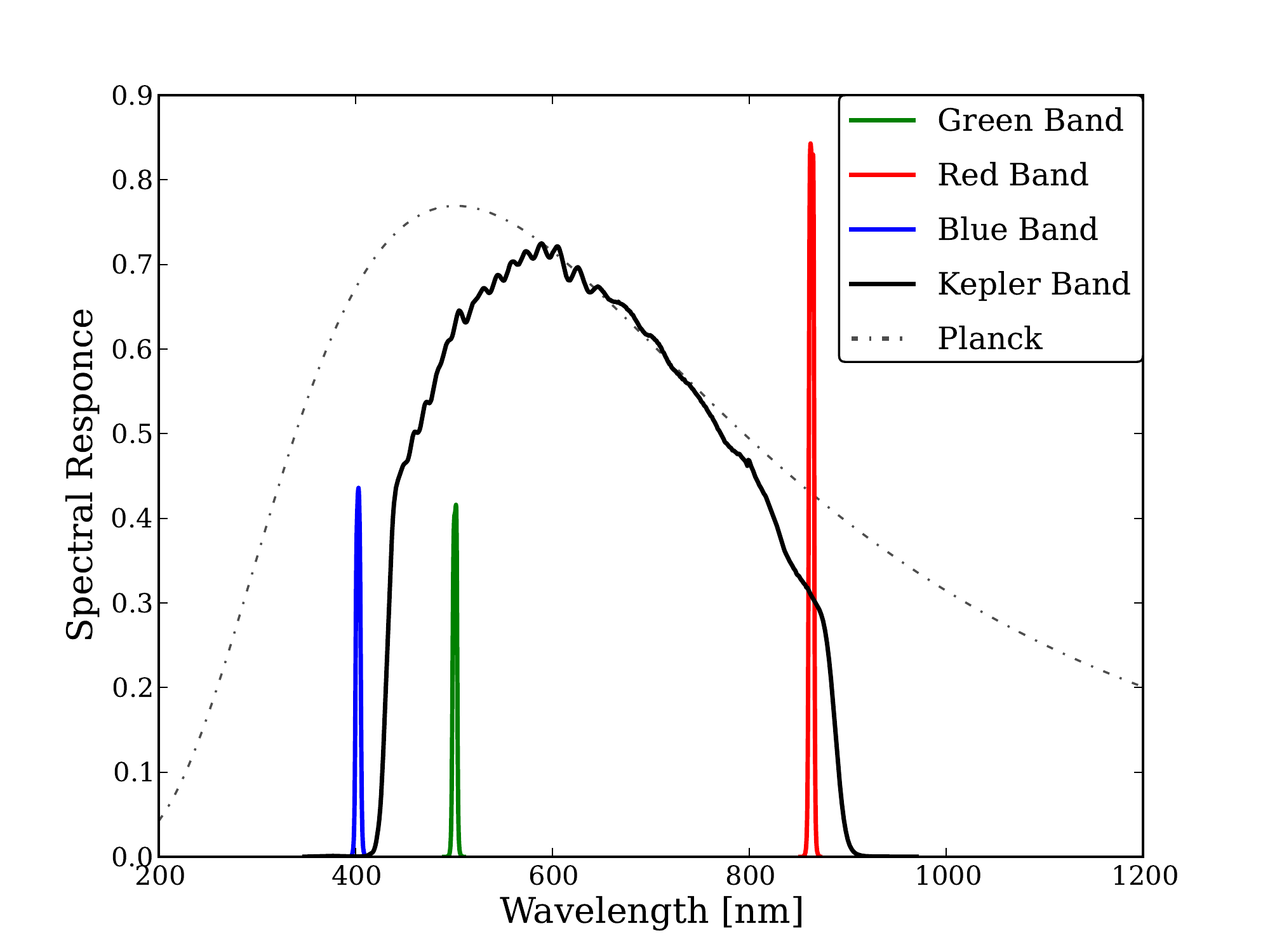}
\caption{\footnotesize Overview of the spectral response functions of the \textit{Kepler} photometer (black) and of the three VIRGO-SPMs, with colors corresponding to the wavelengths of the specific filters. The dashed curve shows (in arbitrary units) the black-body (Planck spectrum) curve of the Sun.} 
\label{fig:filters}
\end{figure}

When the observation is performed over a wavelength band, $W(\mu)$ can be approximated by the factor $W_K(\mu)$ given by \citep[see also,][]{1990A&A...227..563B,2009A&A...495..979M}
\begin{equation}
W_K(\mu) = \frac{\int \mathcal{T}_K(\lambda) T_{\rm eff} \frac{\partial B}{\partial T_{\rm eff}}(\lambda, T_{\rm eff}) H_{\lambda}g_{\lambda}(\mu) d\lambda} {\int \mathcal{T}_K(\lambda) B(\lambda, T_{\rm eff}) H_{\lambda}G_{\lambda} d\lambda}
\label{eq:wk}
\end{equation}
where
\begin{equation}
G_{\lambda} = \int_0^{1} \mu g_{\lambda}(\mu)  d\mu \quad \text{and} \quad H_{\lambda} = \left( \int_0^{1} g_{\lambda}(\mu)  d\mu \right)^{-1}\, .
\label{eq:G}
\end{equation}
In Equation~\ref{eq:wk} $B$ is the Planck function. $\mathcal{T}_K(\lambda)$ is the transfer function and is given by
\begin{equation}
\mathcal{T}_K(\lambda) = \mathcal{E}_K(\lambda)/E_{\nu} = \mathcal{E}_K(\lambda)\lambda/ hc\, ,
\end{equation}
where $\mathcal{E}_K(\lambda)$ is the spectral response of the detector with which the observations are obtained as a function of wavelength.
In Figure~\ref{fig:filters}, we show the spectral response function for the three VIRGO-SPM detectors and for the \textit{Kepler} detector. The dashed line give (in arbitrary units) the black-body spectrum corresponding to the solar effective temperature.  

\begin{figure}
\centering
\includegraphics[scale=0.45]{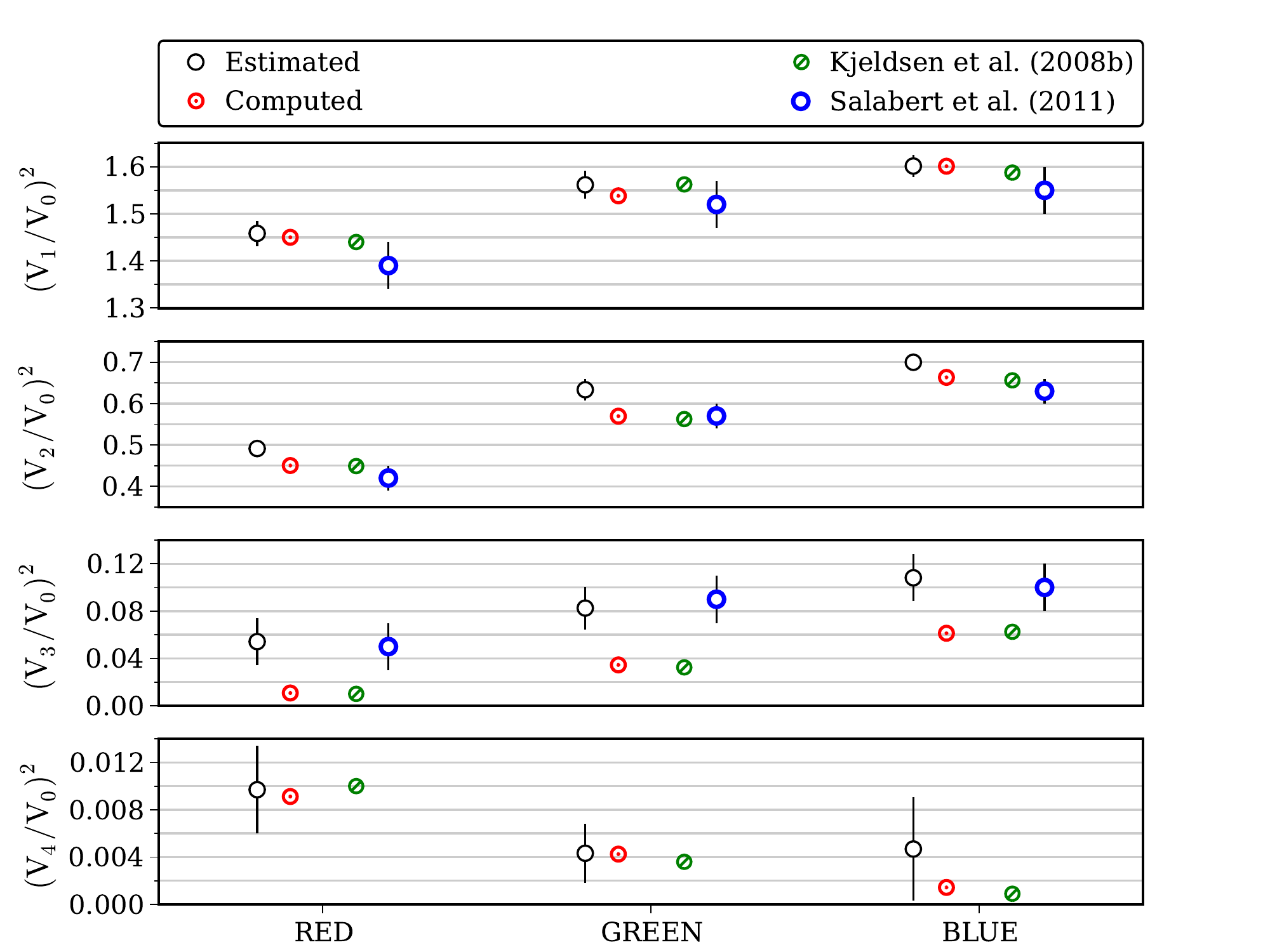}
\caption{\footnotesize Visibilities up to $\ell=4$ for the VIRGO-SPM filters. Illustrated are: theoretical (computed) values (obtained via the \B\ approach), values estimated from the power spectrum with their corresponding error bars, values obtained in \citet{2011A&A...528A..25S} from fitting to the power spectrum (up to $\ell=3$), and theoretical values calculated in \citet{2008ApJ...682.1370K}.} 
\label{fig:visibility}
\end{figure}

The form of the visibility that is used in Equation~\ref{eq:vis} is as mentioned the relative visibility normalized to $V_0$; we give this as
\begin{equation}
\tilde{V_{\ell}} = V_{\ell}/V_0.
\end{equation}
For the solar calculations, we use the following six-parameter LD function:
\begin{align}\label{eq:NL}
g_{\lambda}(\mu) &= A_0 + A_1\mu + A_2\mu^2 + A_2\mu^3 + A_4\mu^4 + A_5\mu^5\, , \\
\sum_0^5{A_n} &= 1\, , \notag
\end{align}
with wavelength-dependent parameters, $A_n$, given in \cite{1994SoPh..153...91N}.
In the integrations over wavelength we interpolate linearly between the values tabulated in \cite{1994SoPh..153...91N}.
The computed visibilities for the different SPM filters are shown in Figure~\ref{fig:visibility} and Table~\ref{tab:visi3}.
Upon comparison with solar visibilities calculated in \citet{2008ApJ...682.1370K} (up to $\ell=4$), we see that our values are in accord with these, see Figure~\ref{fig:visibility}.

To test if these values match what is actually observed for the Sun, we make a crude estimation of the visibilities from the data in the following simple manner:
The frequency of the $m=0$ component is located for modes in the vicinity of $\nu_{\rm max}$. We now sum all power in a range of $\pm [\ell\nu_s + 3.5\, \Gamma_{n\ell}]$ around the $m=0$ frequency, with the rotational frequency splitting set to $\nu_s=0.4\, \rm \mu Hz$. The factor of 3.5 was chosen to ensure that all power from a split multiplet was collected even if using a somewhat wrong estimate for the rotational splitting. The value of $\nu_s=0.4\, \rm \mu Hz$ corresponds to the rotation period of the Sun at about $45^{\circ}$ latitude.

To estimate the power contained in the noise background in the vicinity of $\nu_{\rm max}$ we first sample the power spectrum at frequency ranges midway between the included $\ell=3$ and $\ell=1$ modes.
A linear function is fitted to these background estimates (a fair approximation close to $\nu_{\rm max}$), and we then subtract the power contained under this fitted function in the frequency ranges used for the respective modes.

To get the squared visibilities $\tilde{V}^2_{\ell}$, we now interpolate the $\ell=0$ estimated power values in frequency. The values obtained for other modes are now simply divided by the value of the $\ell=0$ modes at the interpolated frequency of the mode. Finally, the median of the estimated single-mode visibilities for a given degree is adopted as the final visibility, with an error bar estimated by the standard deviation of these values around the median value. 

The visibilities obtained from this approach are also given in Figure~\ref{fig:visibility} and Table~\ref{tab:visi3}. As seen, the observations do not match the calculated values very convincingly, with the highest deviation seen for the $\ell=3$ modes. \citet{2011A&A...528A..25S} perform the same test for the Sun but estimated the visibilities by extracting the heights \citep{2004A&A...413.1135S} of all $2\ell+1$ components of multiplets up to $\ell=3$. We also give their results in Figure~\ref{fig:visibility} and note that our estimates are generally in line with these within the quoted errors. Clearly, the method of fitting the modes directly should give more accurate estimates of the visibilities, but at the cost of a much higher computational effort, especially if methods such as MCMC are used, we find that the simple method described above serves well the purpose of this analysis. 

\begin{figure}
\centering
\includegraphics[scale=0.45]{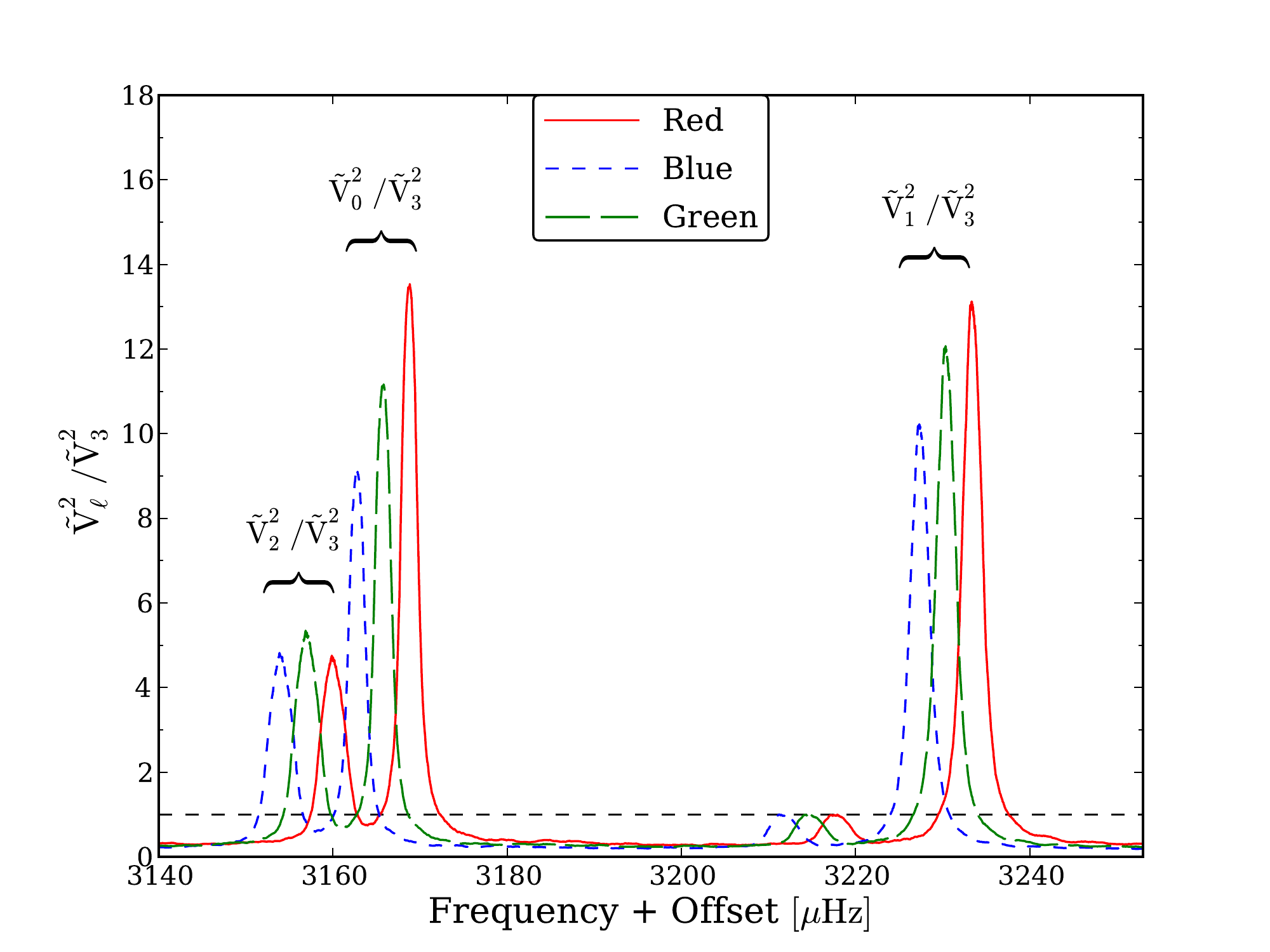}
\caption{\footnotesize Comparison between the relative visibilities for the different SPM filters. The relative squared visibilities have for each band been normalized to the value for the $\ell=3$ modes by dividing the $1.8\, \rm \mu Hz$ boxcar smoothed power spectrum for one of the central orders with the peak height at the position of the $\ell=3$ modes, giving the $\ell=3$ modes a peak height of 1. A frequency shift of $-3\, \rm \mu Hz$, and $-6\, \rm \mu Hz$ has been applied to the green and blue bands respectively to separate the peaks from different filters.} 
\label{fig:filters_vis_comp}
\end{figure}

From the theoretical values plotted in Figure~\ref{fig:visibility}, it is evident that there are differences between the different SPM detectors - especially for the relative difference between $\tilde{V}^2_2$ and $\tilde{V}^2_3$ in the red band as compared to the blue and green bands. The fact that there are differences between the filters is confirmed in the data, see Figure~\ref{fig:filters_vis_comp}. However, we see much smaller relative differences between for instance $\tilde{V}^2_{2}$ and $\tilde{V}^2_{3}$, and the largest relative difference is in fact seen between $\tilde{V}^2_{1}$ and $\tilde{V}^2_{3}$. In Figure~\ref{fig:filters_vis_comp}, we compare the three filters in a very simple manner by dividing one of the central orders in the $1.8\, \rm \mu Hz$ boxcar smoothed power spectrum by the peak in power of the $\ell=3$ mode. For the $\ell=0$ mode, we simply get the inverse of the relative squared $\ell=3$ visibility, \ie, $\tilde{V}^2_0 / \tilde{V}^2_3 = 1 / \tilde{V}^2_3$. The comparison can, however, be made for the $\ell=1$ and 2 modes. As seen, the greatest difference is evident for the $\ell=1$ modes, while the $\ell=2$ modes show little difference - unlike what is expected from theory. 

\begin{figure*}
\centering
\subfigure{
   \includegraphics[scale=0.4] {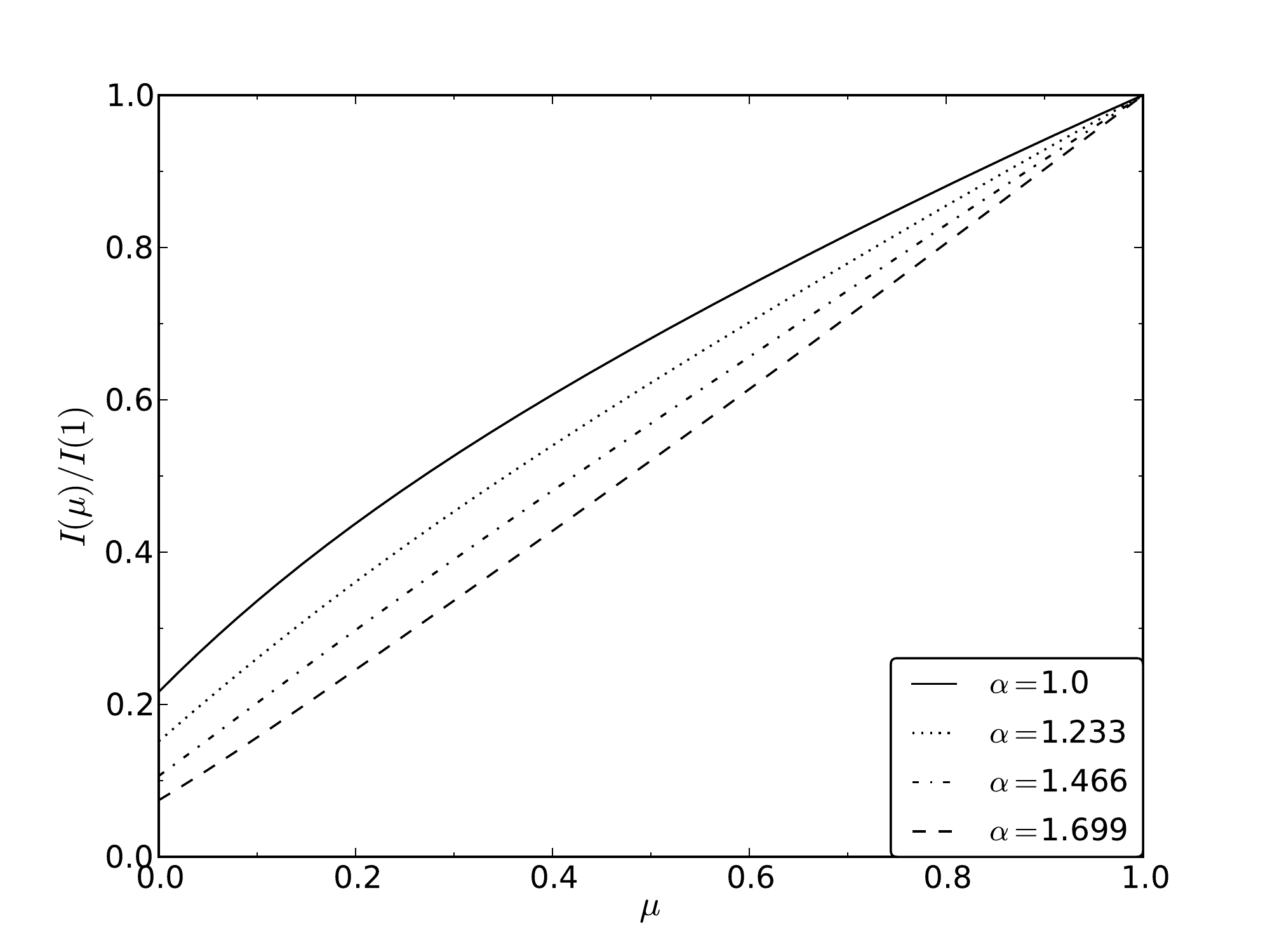}
 }
 \quad
 \subfigure{
   \includegraphics[scale=0.4] {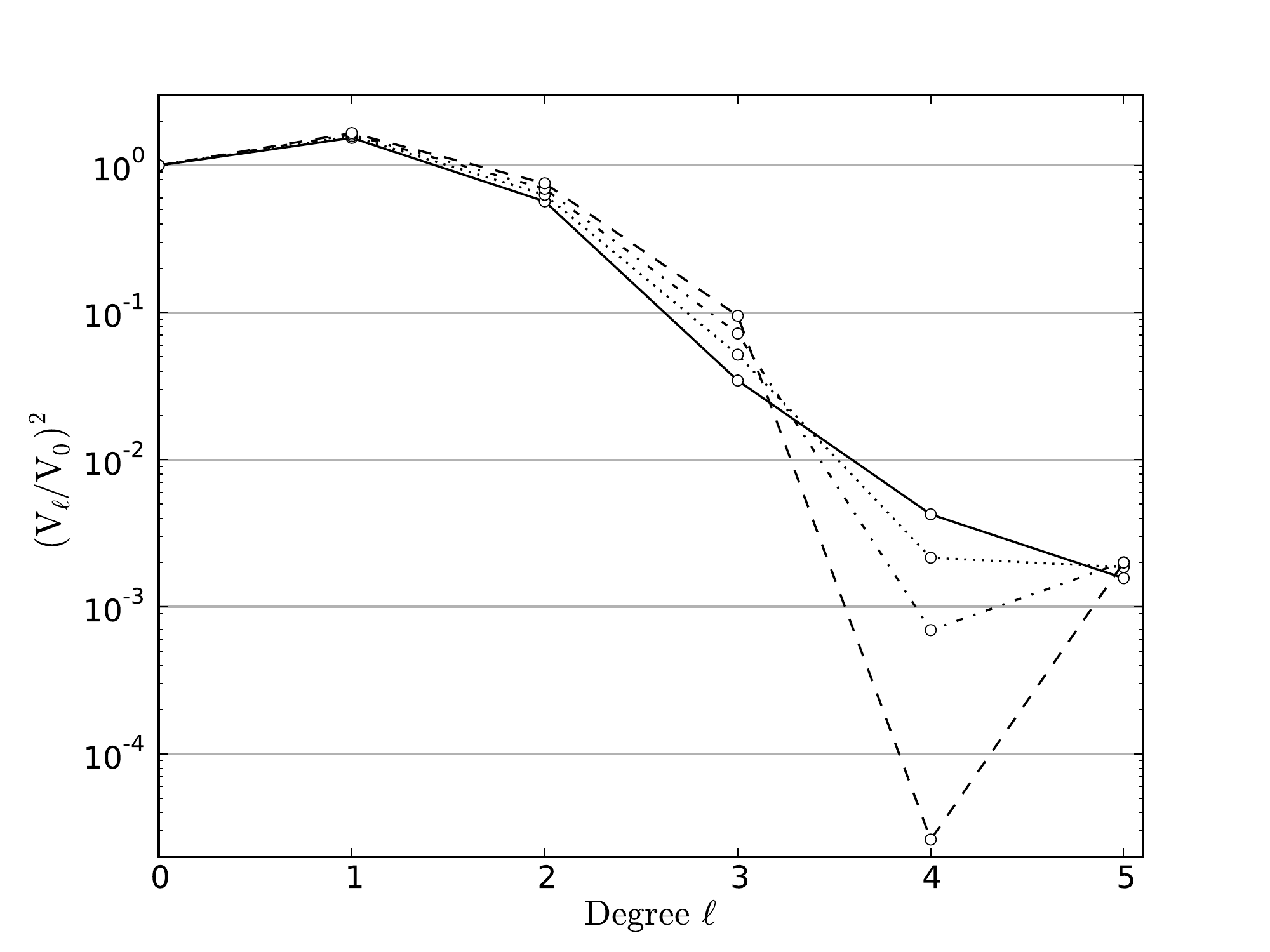}
 }
\caption{\footnotesize Change in the LD profile (left) for the green SPM band as a function of $\mu$ from adding an exponent $\alpha$ to the solar LD law in Equation~\ref{eq:NL}. For each value of $\alpha$, the corresponding relative squared visibilities are given (right) for $\ell$-values up to $\ell=5$. }
\label{fig:LDcomp_alpha}            
\end{figure*}

This mismatch in visibilities is likely caused, in part at least, by an erroneous treatment of the LD close to the limb of the star. 
The $\ell=3$ modes are relatively more sensitive to the LD, as compared to, \eg, $\ell=1$ modes, due to the fact that the symmetry of the spherical harmonic function of $\ell=3$ modes results in total cancellation in the absence of LD. To test if a small change in the solar LD function can enable a sufficiently high value of the $\ell=3$ modes, we add an exponent $\alpha$ on the solar LD function in Equation~\ref{eq:NL}, \ie, we replace $g_\lambda(\mu)$ by $g_{\lambda}(\mu)^{\alpha}$ in Equations~\ref{eq:wk} and \ref{eq:G}, and increase $\alpha$ until a value of $\tilde{V}^2_{3}=0.1$ is reached. To obtain this increase, values of $\alpha=3.1$ (red band), $\alpha=1.7$ (green band), and $\alpha=1.35$ (blue band) were needed. The result of this procedure is shown in Figure~\ref{fig:LDcomp_alpha} for the case of the green band. In all cases, the shape of the LD function change greatly in appearance, ending in a nearly straight line for high values of $\alpha$. Furthermore, the change results in much too high values of the squared relative visibilities for $\ell=1$ and $\ell=2$ modes, while the visibility for $\ell=4$ modes is greatly diminished.

The mismatched theoretical values are therefore likely not due solely to the LD function. Probably the effects of non-adiabaticity are an important contributing factor, especially for modes with frequencies near $\nu_{\rm max}$ and by extension near the acoustic cut-off frequency where the stellar atmosphere is affected to a greater extent by the oscillations. 

\begin{figure}
\centering
\includegraphics[scale=0.45]{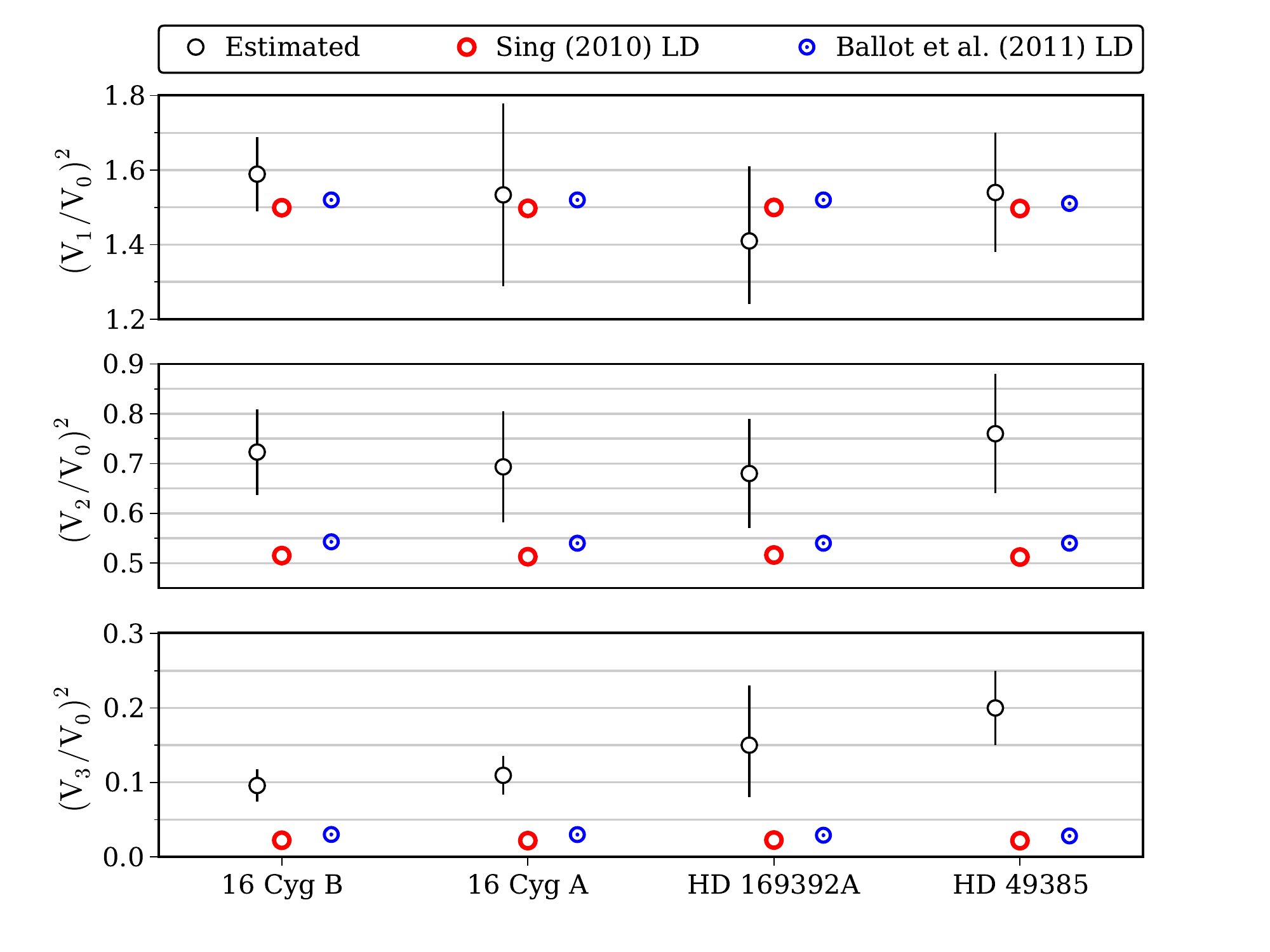}
\caption{\footnotesize Visibilities up to $\ell=3$ for 16 Cyg A, 16 Cyg B, HD 169392A \citep{2013A&A...549A..12M}, and HD 49385 \citep{2010A&A...515A..87D}. Illustrated are: values estimated from the power spectrum with their corresponding error bars, theoretical values from \B, and theoretical values obtained from using the four-parameter LD law, with parameter values from \cite{2010A&A...510A..21S}.} 
\label{fig:visibility2}
\end{figure}

The same procedure as above was then followed for 16 Cyg A and B. When it comes to the LD functions used in the theoretical calculation, we take on a simpler approach than the one in \B\ and use tabulated values instead of a grid of atmosphere models. For Model A and B we use a \textit{four-parameter} non linear LD law and adopt LD parameters calculated for the \textit{Kepler} band by \cite{2010A&A...510A..21S}. These parameters are tabulated for different values of effective temperature, $\log g$, and metallicity. For the two stars we interpolated the values in temperature, while $\log g$ and $\rm [Fe/H]$ were set to the nearest tabulated value. With this approach $g_{\lambda}(\mu)$ becomes a function of the distance to the limb only, and the wavelength integrated LD law, $g_K(\mu)$, is used in Equation~\ref{eq:visibil} as $W(\mu)=g_K(\mu)$.
\B\ makes a comparison in their paper with the values that would be obtained from the approach taken here, namely, using the \cite{2010A&A...510A..21S} LD values, and find that values obtained in this manner are generally lower than the ones found using atmosphere models. This is also the trend we observe (see Figure~\ref{fig:visibility2}).

Before going to the estimates obtained from the power spectra of 16 Cyg A and B, we note that while the calculations in \B\ predict a decrease in the visibilities with increasing temperature, studies of red giants in \cite{2012A&A...537A..30M} find first of all a large scatter in observed visibilities and in addition temperature dependencies differing from the theoretical predictions in \B . Most notably, an overall increase with temperature is found for the visibilities of $\ell=3$ octupole modes suggesting already that visibilities for these might be significantly underestimated when following the approach above. 

\begin{figure*}
\centering
\subfigure{
   \includegraphics[scale=0.4] {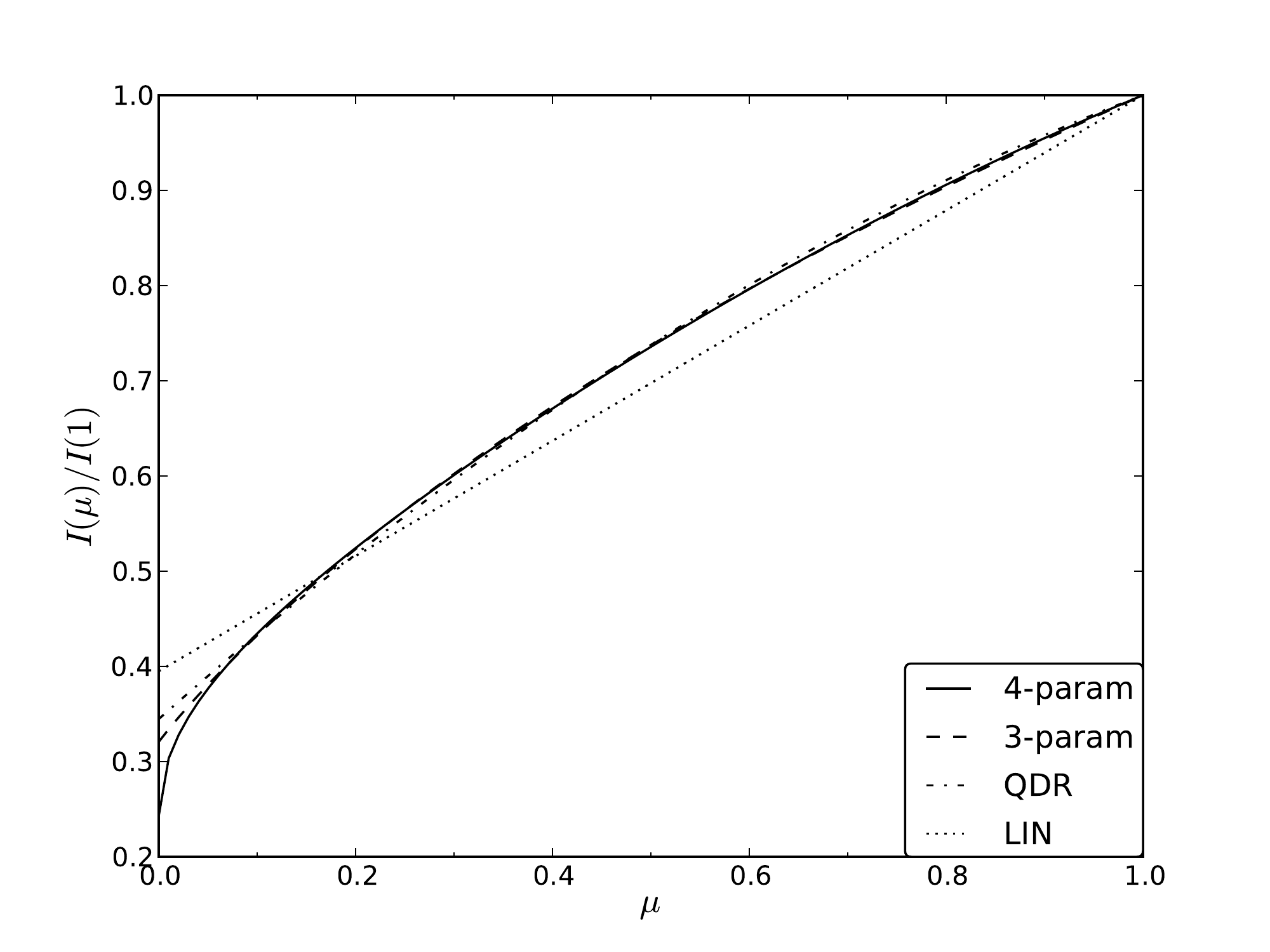}
 }
 \quad
 \subfigure{
   \includegraphics[scale=0.4] {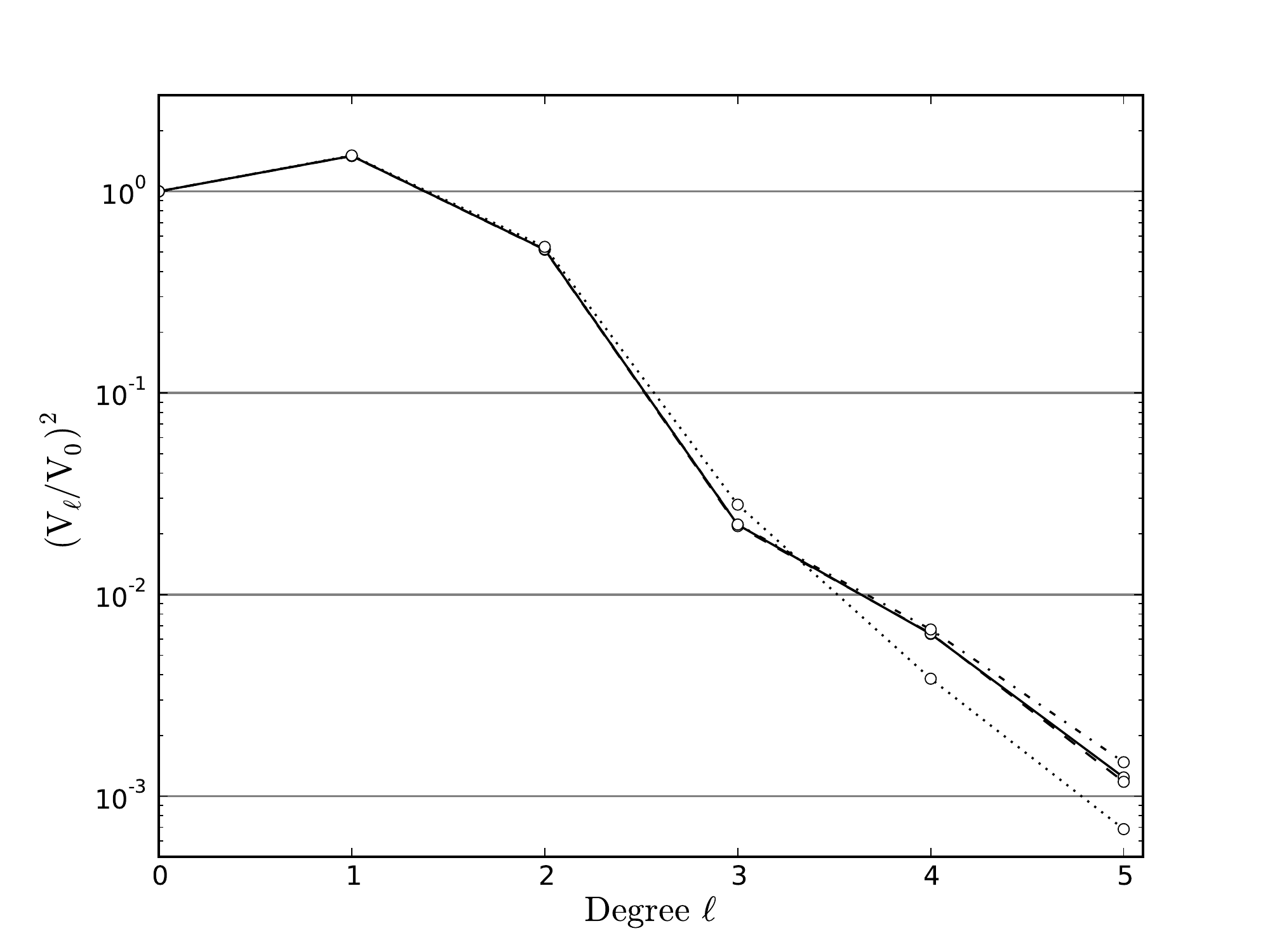}
 }
\caption{\footnotesize Appearance of different LD laws for 16 Cyg B (left). As seen, the greatest difference between the different laws is in general found near the stellar limb ($\mu=0$). For each LD law, the computed squared relative visibilities are given (right) for $\ell$-values up to $\ell=5$. }
\label{fig:LDcomp_law}               
\end{figure*}

As for the Sun, we tested the impact of the LD law, here by comparing the visibilities obtained by using different laws, namely, a \textit{linear} law, a \textit{quadratic} law, and a \textit{three-parameter} law. See Figure~\ref{fig:LDcomp_law} for this procedure applied to 16 Cyg B. For all of these we used the LD parameters from \cite{2010A&A...510A..21S} and found that only the linear LD law deviated significantly from the others. So, even though the quadratic and non linear LD laws differ near the limb, they still give very similar results for the visibilities.

In Figure~\ref{fig:visibility2}, we show both the theoretically computed values using the four-parameter LD law and the values obtained from the power spectra of 16 Cyg A and B. As we have no good knowledge of the mode linewidths, we here simply summed power in a range of $\pm [\ell\nu_s + 2.5\rm\, \mu Hz]$ and using $\nu_s = 0.37 \, \rm \mu Hz$. From Figure~\ref{fig:visibility2}, it is clear that, as for the Sun, the theoretical predictions do not agree well with the observations. Again the largest relative deviation is seen for $\ell=3$ modes, but also for $\ell=2$ modes a non-negligible deviation is seen. We have also illustrated the values that would be obtained for the two stars from the tabulated values in \B.   

\cite{2010A&A...515A..87D} also found deviations between measured and theoretically predicted values in the analysis of the solar-like \textit{CoRoT} target HD 49385, where visibilities were estimated in the same manner as for the Sun in \cite{2011A&A...528A..25S}. We have included the estimated values for this star in Figure~\ref{fig:visibility2}, and also show the theoretical values obtained when using the four-parameter LD law together with the \textit{CoRoT} calibrated LD parameters in \cite{2010A&A...510A..21S}. 
\cite{2013A&A...549A..12M} recently analyzed the binary system HD 169392, also observed with \textit{CoRoT}, and were able to estimate the mode visibilities for the A-component of the system. These values have also been included in Figure~\ref{fig:visibility2}, and again accompanied by theoretical predictions.
Both of these stars are very similar to 16 Cyg A and B in terms of effective temperature and metallicity but are likely a bit more evolved having slightly lower surface gravities. The similarities can also be seen in the visibilities predicted from theory.
For both these stars, the same trend is observed in the deviations as for 16 Cyg A and B.

Unfortunately, the trend in the deviation is not smooth, and we are therefore not in a position to estimate the expected deviation for the $\ell=4$ modes from simple extrapolation of the deviation. Even though an estimate of the $\ell=4$ visibilities was obtained for the Sun (see Figure~\ref{fig:visibility}), the error estimates on these values make them unfit for any inference on the trend in the deviations. We are furthermore not convinced of the validity of these estimates as the $\ell=4$ modes are very close to the background noise level. Because of this lack of predictive power we err on the side of caution, and choose for the $\ell=4$ and $\ell=5$ visibilities the theoretically predicted values, keeping in mind that these are likely underestimated. For the lower-degree modes, we adopt the values obtained from the power spectrum.

\begin{table*}[bt]
\caption{\small Theoretical and Estimated Visibilities Calculated Using the Methods Outlined in \S~\ref{sec:vis}.}
\centering
\begin{tabular}{lcccccc}
\hline \hline \\ [-0.25cm] 
Star  & $\tilde{V}^2_{\rm tot}$ & $\tilde{V}^2_{1}$ & $\tilde{V}^2_{2}$ & $\tilde{V}^2_{3}$ & $\tilde{V}^2_{4}$ & $\tilde{V}^2_{5}$ \\[0.1cm] 
\hline\\[-0.25cm] 
\multicolumn{7}{c}{Theoretical Values}\\[0.1cm] 
\hline
16 Cyg A 	& 3.04 & 1.50 & 5.13E-1 & 2.17E-2 & 6.48E-3 & 1.23E-3 \\
16 Cyg B 	& 3.04 & 1.50 & 5.15E-1 & 2.22E-2 & 6.36E-3 & 1.24E-3 \\
Sun (red) 	& 2.92 & 1.45 & 4.50E-1 & 1.08E-2 & 9.12E-3 & 6.71E-4 \\
Sun (green)	& 3.15 & 1.54 & 5.70E-1 & 3.45E-2 & 4.25E-3 & 1.57E-3 \\
Sun (blue) 	& 3.33 & 1.60 & 6.63E-1 & 6.13E-2 & 1.43E-3 & 2.08E-3 \\[0.05cm] 
\hline \\[-0.25cm] 
\multicolumn{7}{c}{Estimated Values}\\ [0.1cm] 
\hline
16 Cyg A 	& 3.34 $\pm$ 0.38 & 1.53 $\pm$ 0.25 & 0.69 $\pm$ 0.11 & 0.11 $\pm$ 0.03  & - 					& - \\ 
16 Cyg B 	& 3.42 $\pm$ 0.21 & 1.59 $\pm$ 0.10 & 0.72 $\pm$ 0.09 & 0.10 $\pm$ 0.02 	& - 					& - \\ 
Sun (red) 	& 2.92 $\pm$ 0.07 & 1.46 $\pm$ 0.03 & 0.49 $\pm$ 0.02 & 0.05 $\pm$ 0.02  & 0.01  $\pm$ 0.003		& - \\
Sun (green)	& 3.15 $\pm$ 0.08 & 1.56 $\pm$ 0.03 & 0.63 $\pm$ 0.03 & 0.08 $\pm$ 0.02  & 0.004 $\pm$ 0.008	& - \\
Sun (blue) 	& 3.33 $\pm$ 0.07 & 1.60 $\pm$ 0.02 & 0.70 $\pm$ 0.02 & 0.11 $\pm$ 0.02  & 0.005 $\pm$ 0.007	& - \\[0.05cm] 
\hline \\ \\
\end{tabular}
\label{tab:visi3}
\end{table*}


\subsection{Noise Properties}
\label{sec:noise}

The noise level in our simulations is of course of great importance as it, given a certain limit spectrum, sets the signal-to-noise level in the power spectrum, which ultimately determines if the signal of a given mode will stand out from the noise.  

For the noise in the synthetic spectrum, we first tested the prescription by \cite[][see also \cite{2011ApJ...732...54C, 2011ApJS..197....6G}]{2010ApJ...713L.160G} for the instrumental "shot" noise in \textit{Kepler} SC observations as a function of \emph{Kepler} magnitude $K_p$, with the noise level in the amplitude spectrum given as
\begin{align}\label{eq:shot}
\sigma_{\rm amp} &= \frac{10^6}{c  N^{1/2}}\left(c+9.5\times 10^5 (14/K_p)^5 \right)^{1/2} \, \rm ppm  \\
c &= 1.28 \times 10^{0.4(12-K_p)+7}\, , \notag
\end{align}
with $ N$ being the number of data points in the time series.
This will results in a noise level in the power spectrum of \citep{1995A&A...293...87K}:
\begin{equation}
N_{\rm instr} = \frac{4\sigma_{\rm amp}^2}{\pi} \rm \, \rm  ppm^2\, . 
\end{equation}
When comparing the noise obtained from this description with the value estimate from Equation~\ref{eq:backg}, we found that Equation~\ref{eq:shot} underestimated the shot noise by a factor ${\sim}8$. It should be noted that Equation~\ref{eq:shot} is only intended to give a minimal noise term. Furthermore, as 16 Cyg A and B are both highly saturated targets with flux collected from a large custom aperture, it is not guaranteed than Equation~\ref{eq:shot} is at all applicable. For this reason, we have chosen to use the noise estimated from Equation~\ref{eq:backg}, which we scale to the considered observing length $T_{\rm obs}$ by dividing by the factor $\sqrt{T_{\rm obs}/643\, d}$.
For the background component, we add the signal extracted for the two stars in fitting Equation~\ref{eq:backg} to the power spectra.

In calculating the realization noise in the power spectrum, we follow the method given in \citet{1990ApJ...364..699A} and \citet{2003ApJ...589.1009G} where the Box-Muller transform is used, and a realization of the power spectrum given as
\begin{equation}
P(\nu_j) = -\ln(U_j) \, \mathcal{P}(\nu_j)\, .
\end{equation} 
Here $U_j$ is a uniform distribution on $[0, 1]$ and $\mathcal{P}(\nu_j ; \boldsymbol \Theta)$ is the limit spectrum given in Equation~\ref{eq:limitspec}.
This approach ensures a power spectrum obeying the generally assumed $\chi^2$ 2 dof. (degrees-of-freedom) statistic \citep{1984PhDT........34W}.



\section{Results on detectability from simulated data}
\label{sec:results2}
\begin{figure}
\centering
\includegraphics[scale=0.45]{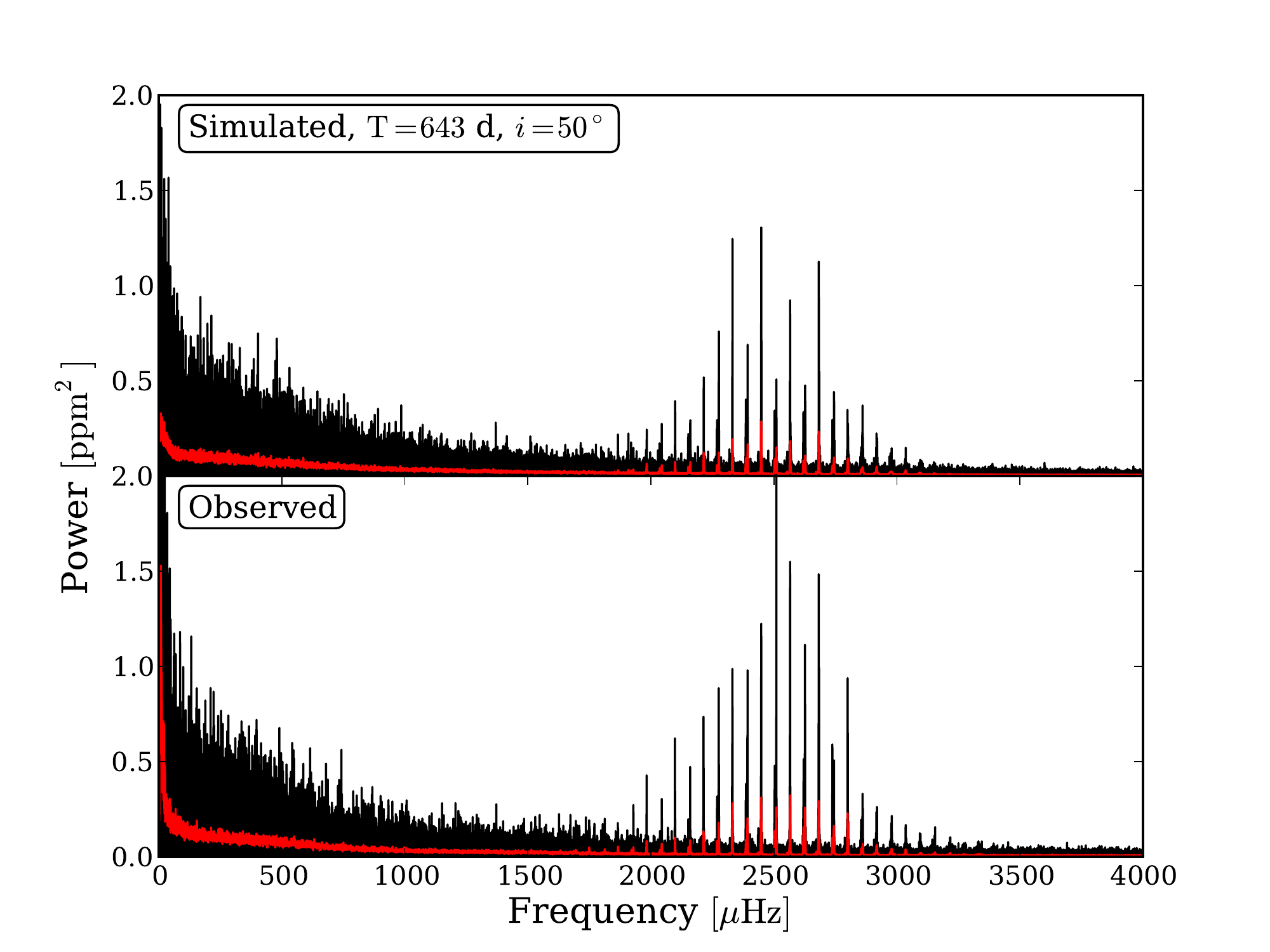}
\caption{\footnotesize Example of a simulated power spectrum of 16 Cyg B (black), corresponding the an observing length of 643 days, \ie, the same length as our data sets for 16 Cyg A and B. In red the $1\,\rm \mu Hz$ boxcar smoothed version is shown. As seen, an inclination angle of $i=50^{\circ}$ was used in this particular simulation. The appearance of the non-smoothed power spectra naturally deviate somewhat due to the $\chi^2$-noise.} 
\label{fig:sim_power1}
\end{figure}

\begin{figure*}
\centering
\subfigure{
   \includegraphics[scale=0.4] {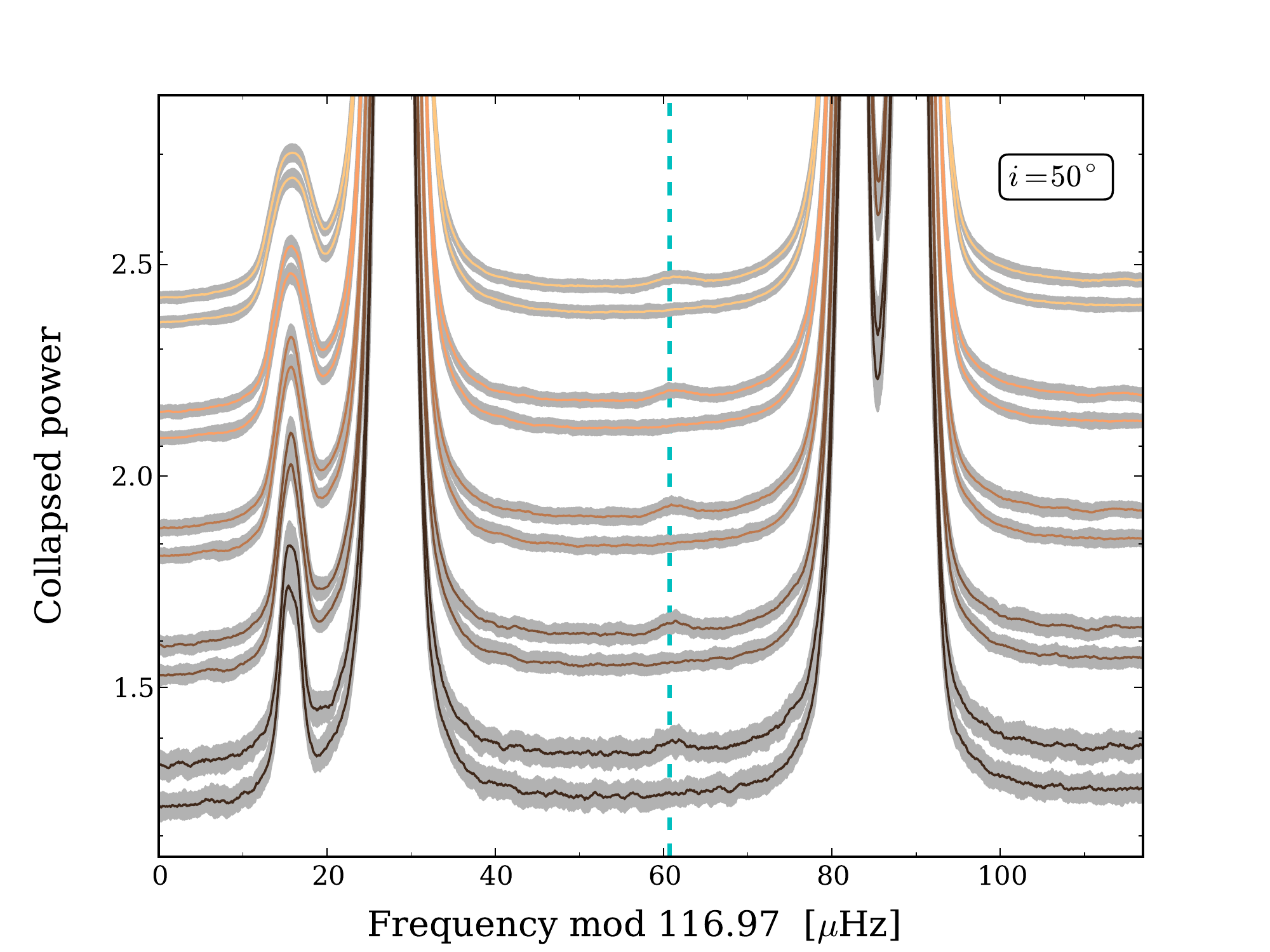}
 }
 \quad
 \subfigure{
   \includegraphics[scale=0.4] {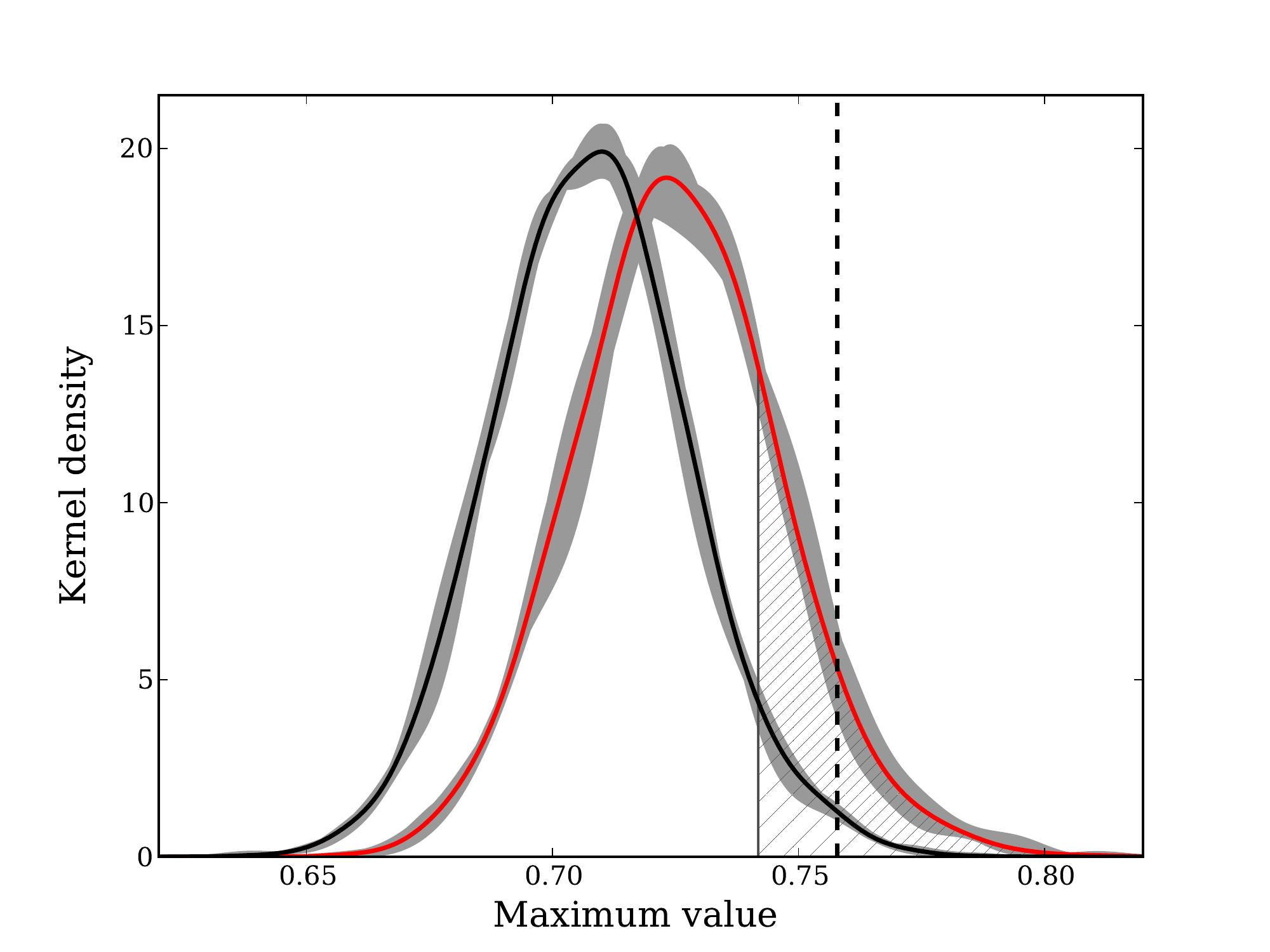}
 }
\caption{\footnotesize Left: \textit{SC}-spectra for a MC set of simulated power spectra, here with an adopted inclination of $i=50^{\circ}$. Two hundred power spectra were simulated and half of these only included modes of degree $\ell=0-3$, while half included modes of degree $\ell=0-5$. The five sets of curves correspond to five smoothing levels applied, going from $1\rm \mu Hz$ at the bottom to $5\, \rm \mu Hz$ at the top.  The dashed cyan line shows the position of the straightened $\ell=4$ ridge, as in Figure~\ref{fig:res1}, and hence the expected position of a possible excess from the $\ell=4$ modes. From the simulated power spectra the mean \textit{SC}-spectrum was found for each smoothing level, given by the solid curves, in addition to the mean absolute deviation from this curve, given by the gray regions around the respective mean curves. For each set, the bottom curve is obtained from the simulated data only included modes of degree $\ell=0-3$, while the top curve is for data including $\ell=0-5$ modes. For clarity offsets have been added to separate the curves. Right: distribution (kernel density) for the maximum value in a $\pm2\,\rm \mu Hz$ window around the expected position of $\ell=4$ modes for $2\,\rm \mu Hz$ smoothed collapsed simulated power spectra. The black curve gives the mean distribution from 5000 spectra with inclinations of $i=10^{\circ},30^{\circ},50^{\circ},70^{\circ},90^{\circ}$ (1000 spectra for each) when including modes of degree $\ell=0-3$, while modes of degree $\ell=0-5$ were included for the red curve. The gray regions around these mean curves give the range in the individual distributions from the different values of the inclination. The sparsely hatched region gives the $>95\%$ area under the black curve, and thus the region where the \emph{null hypothesis} ($H_0$) can be rejected. The densely hatched region gives the corresponding region under the red curve and indicates how often one will be in a position to correctly reject $H_0$. The dashed black line gives the value measured for 16 Cyg B.} 
\label{fig:MCbatch}                             
\end{figure*}
Because of the high similarity of the two stars and their power spectra, we chose to only simulate the power spectrum of 16 Cyg B.
In Figure~\ref{fig:sim_power1}, an example of a simulated power spectrum can be seen, computed following the description of \S~\ref{sec:synspec}. The top panel shows the simulation when using an inclination of $i=50^ {\circ}$ and a frequency resolution corresponding to the length of the observed data, \ie, $643$ days. Also shown is the $1\,\rm \mu Hz$ smoothed version, and as seen the two spectra are very similar. The non-smoothed spectra naturally look somewhat different due to the $\chi^2$-noise. A likely contribution to the difference between the simulation and the observation comes from the inclination angle, which might be different from $i=50^ {\circ}$.

From the simulated power spectra, we are in a position to test if the $\ell=4$ modes are likely to be found in \emph{Kepler} data. To address this question, we made a Monte Carlo (MC) set of simulated power spectra with a frequency resolution corresponding to an observing length of 643 days. 

For an inclination angle of $i=50^{\circ}$, we simulated 100 power spectra which included both $\ell=4$ and $\ell=5$ modes in addition to 100 power spectra including only degrees up to $\ell=3$. For all power spectra, we computed the \emph{SC}-spectrum with smoothing levels from $1-5\, \rm \mu Hz$ and found for each smoothing level the mean \emph{SC}-spectrum from the 100 simulations. The result of this can be seen in the left panel of Figure~\ref{fig:MCbatch}. From this MC set, it is found that the power spectra including $\ell=4$ and $\ell=5$ modes in mean has a noticeable excess power in the \emph{SC}-spectrum at the predicted position from the straightening when compared to power spectra not including these modes. The choice of optimum smoothing level is again difficult to estimate as the smoothing in general smears out the signal over a larger frequency range while at the same time reducing the noise. We note than an excess is seen for all included inclination angles and all smoothing levels.

To quantify the detectability of the $\ell=4$ modes further, we simulated 2000 power spectra for each of the inclinations $i=10^{\circ},30^{\circ},50^{\circ},70^{\circ},90^{\circ}$. Half of these included modes of degree $\ell=0-5$, while half only included modes of degree $\ell=0-3$. For all of these simulated power spectra, we computed the collapsed spectrum and chose a single smoothing level of $2\,\rm \mu Hz$. This smoothing level will encompass all $m$-components of $\ell=4$ assuming the modes are split by less than $0.5\,\rm \mu Hz$. The maximum value was then found in each $SC$-spectrum in a $\pm2\,\rm \mu Hz$ window around the expected position of the collapsed $\ell=4$ power. Two distributions for this maximum value were obtained for each of the inclinations used - one for the $SC$-spectra including only $\ell=0-3$ modes and one for the $SC$-spectra that includes $\ell=0-5$. Before the maximum values were found, we divided the individual $SC$-spectra with the ratio of their median value to the median of the $SC$-spectrum of 16 Cyg B (Figure~\ref{fig:res2}). This is done such that the maximum value found in the $SC$-spectrum of 16 Cyg B can be more readily compared to the distributions of maximum values from the simulations.
The mean kernel density estimations (KDEs)\footnote{Using \emph{Silverman's rule of thumb} \citep[][]{MR848134} for determining the kernel bandwidth.} from the different inclination angles are given by the black ($\ell=0-3$) and red ($\ell=0-5$) curves in the right panel of Figure~\ref{fig:MCbatch}. The gray region around each mean curve gives the range for the KDEs of the different inclinations used. With the obtained KDE we can now better test the significance of the signal seen for 16 Cyg B, and estimate how often one would be in a position to see such a signal. We take as our \emph{null hypothesis} ($H_0$) that only noise is present and thus that the black KDE holds for our maximum value. The $H_0$ hypothesis will only be rejected in favor of the alternative $H_1$ hypothesis (not only noise is present) if the \emph{p}-value\footnote{The probability of obtaining a result equal to or more extreme than what was actually observed under the assumption of the \emph{null hypothesis}.} of a given observed maximum value falls below a given significance level $\alpha$. The lower the value of $\alpha$ the less likely one is to make so-called \emph{Type I} errors where $H_0$ is erroneously rejected in favor of $H_1$. The significance level $\alpha$ is often set to $0.05$ ($5\%$ chance of \emph{Type I} errors) giving an observation ``significant at the $5\%$ level'' if $H_0$ can be rejected. The sparsely hatched region below the black curve in Figure~\ref{fig:MCbatch} indicates the region where $H_0$ can be rejected in favour of $H_1$ at the $5\%$ level. Assuming now that $H_1$ is true and that there is additional power present, we can find the probability of correctly rejecting $H_0$ in favour of $H_1$ at the $5\%$ level. This probability is given by the combined area of the densely and sparsely hatched regions under the red curve. From this, we find that in the case where additional power (here from $\ell=4$ modes) is present, there is a ${\sim}21\%$ chance of correctly rejecting $H_0$ in favor of $H_1$ at the $5\%$ level. Correspondingly, there is a ${\sim}79\%$ chance that $H_0$ will not be rejected even though $H_1$ is true (a \emph{Type II} error), and one will not be able to claim a significant detection. The black vertical dashed line gives the value obtained from 16 Cyg B when following the same procedure as for the simulated $SC$-spectra. Under the assumption that our simulations indeed give an accurate description of the observations for 16 Cyg B, we can from the maximum value of 16 Cyg B reject $H_0$ at the $5\%$ level. In fact, the maximum value comes, with a \emph{p}-value of $0.01041$, very close to the $1\%$ significance level. However, we note that such a direct comparison between observations and simulations should be done very cautiously, as we indeed have a rather poor handle on some of the important input parameters such as relative visibilities and mode linewidths. Also, we note that the maximum value for 16 Cyg B is in the high end of what would be expected from the simulations with a \emph{p}-value of $0.14$ with respect to the red curve. This could possibly indicate that the value used for the relative visibility of $\ell=4$ is underestimated.

We can complement the above estimate of the significance of the detection (or rather the rejection of $H_0$) in a Bayesian manner using Bayes’ theorem to calculate the \emph{posterior probability}. A proper Bayesian analysis, using, \eg, MCMC or MultiNest \citep[][]{2009MNRAS.398.1601F} to approximate the posterior, would require assumed priors for the visibilities and the parameters entering Equation~\ref{eq:limitspec}.
Assuming these parameters are known and correctly set in the models the posterior probability for the hypothesis $H_i$ given the measured peak value of the \emph{SC}-spectrum, $x$, can be readily estimated from the MC simulations. For $H_0$, and equivalently for $H_1$, the posterior is given as \citep[see, \eg,][]{Berger:1987:TPNa,cowan1998statistical,2009A&A...506....1A,2010MNRAS.406..767B} 
\begin{equation}
P(H_0 | x) = \frac{\pi_0 P(x| H_0)}{\pi_0 P(x| H_0)+\pi_1 P(x| H_1)} \, ,
\end{equation}
where $\pi_i$ is the \emph{prior probability} that a given hypothesis $H_i$ is true, while $P(x| H_i)$ (the likelihood) is the probability of observing the data obtained under the assumption of the hypothesis $H_i$. Assuming no prior preference for $H_0$ over $H_1$ or vice versa (\ie, $\pi_0/\pi_1=1$), the posterior probabilities, \ie, the probabilities in favor of a given hypothesis after actually making an observation, then gives $P(H_0 | x)\approx 14\%$ and $P(H_1 | x) \approx 86\%$.
We can quantify the meaning of these values further by the \emph{posterior odds ratio} in favor of $H_0$ against $H_1$ given the measured value $x$ as
\begin{equation}
\mathcal{O}_{0,1} \equiv \frac{P(H_0 | x)}{P(H_1 | x)} = \frac{\pi_0}{\pi_1} \frac{P(x| H_0)}{P(x| H_1)}= \frac{\pi_0}{\pi_1} \mathcal{B}_{0,1} \, .
\end{equation}
Here $\mathcal{B}_{0,1}$ is the so-called \emph{Bayes factor}. Assuming again no prior preference for $H_0$ over $H_1$ the posterior odds ratio is simply given by the Bayes factor. From the simulated distributions and the observed peak value of the \emph{SC}-spectrum ($x$), we obtain a posterior odds ratio in favor of $H_1$ against $H_0$ of $\mathcal{O}_{1,0} = 6.2$ (note that $\mathcal{O}_{1,0}=1/\mathcal{O}_{0,1}$). The evidence for $H_1$ over $H_0$ can be judged from the obtained odds ratio using the Jeffreys' scale \citep[][]{Jeffreys:1961}. According to this scale a value for $\mathcal{O}_{1,0}$ between 3 and 10 can be interpreted as ``substantial" evidence in favor of $H_1$ over $H_0$. An odds ratio between 10 and 30 is needed for a ``strong" evidence in favor of $H_1$ over $H_0$, while a ratio ${>}100$ is needed for ``decisive" evidence.

\paragraph*{}

\begin{figure*}
\centering
\subfigure{
   \includegraphics[scale=0.45] {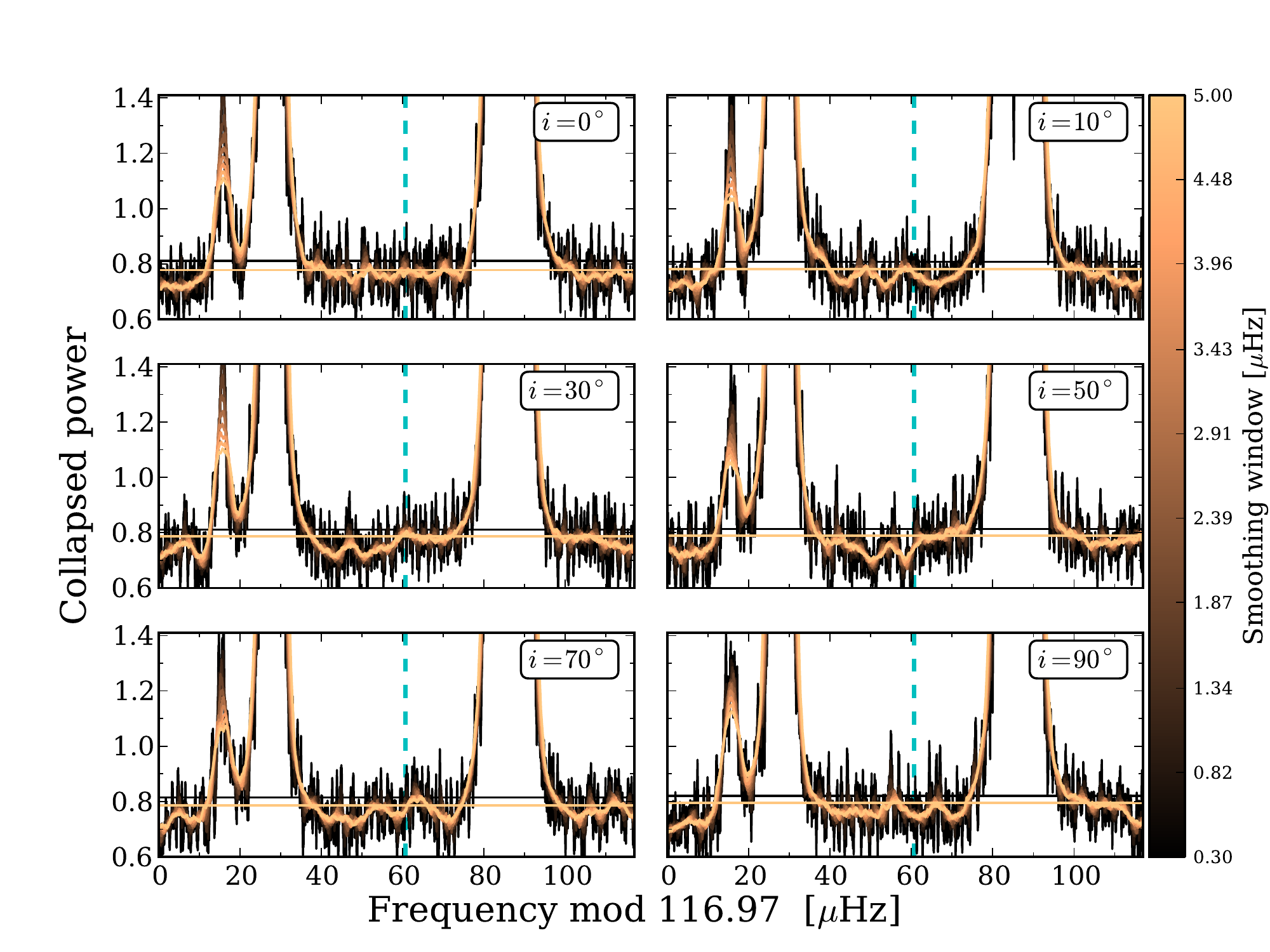}
 }
 \\
 \subfigure{
   \includegraphics[scale=0.45] {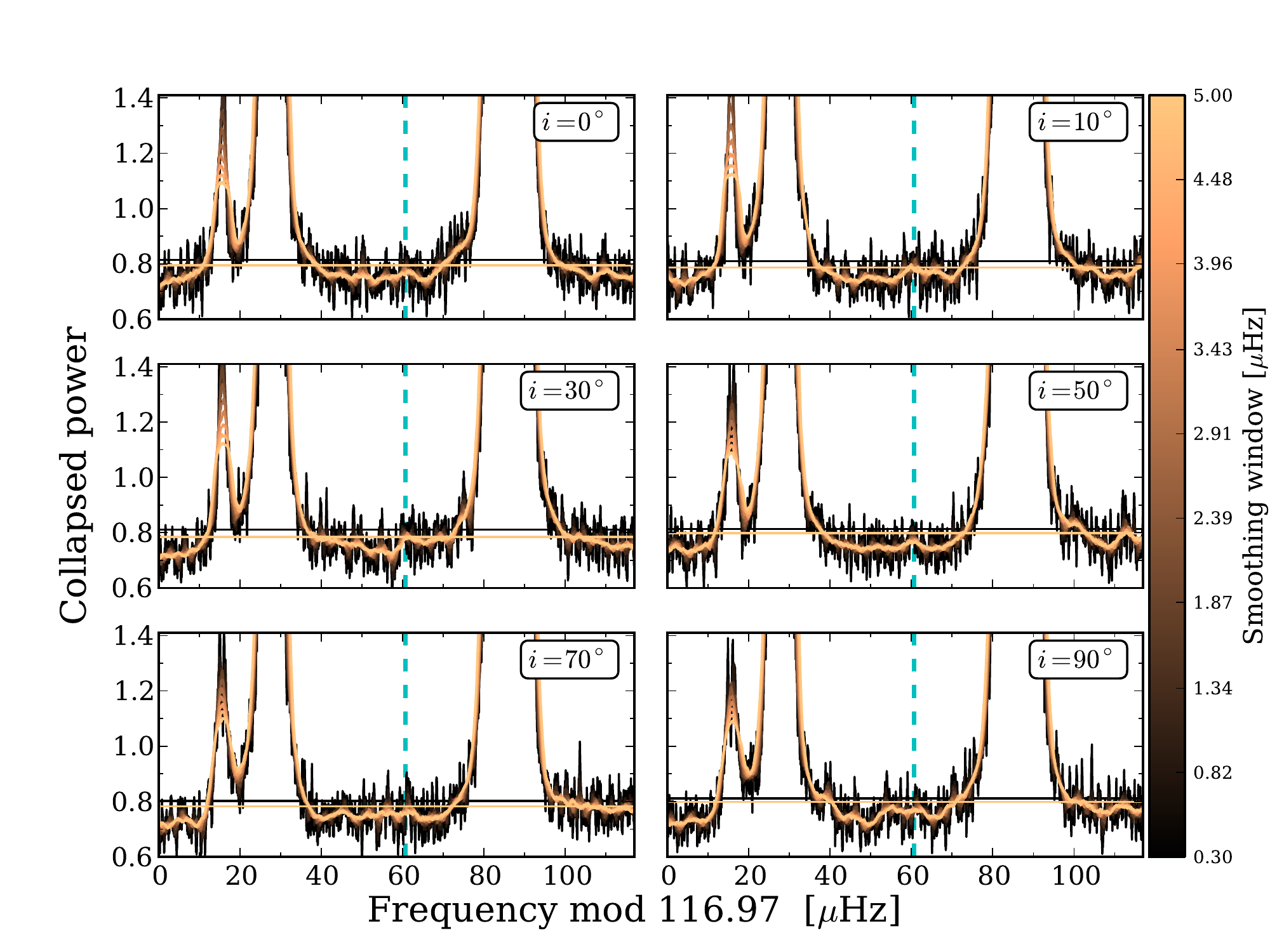}
 }
\caption{\footnotesize Top: \textit{SC}-spectra for simulated 16 Cyg B power spectra having a frequency resolution corresponding to an observing length of 643 days and for different amounts of applied smoothing. For the different panels a specific value for the inclination has been used. The dashed cyan line shows the position of the straightened $\ell=4$ ridge, as in Figure~\ref{fig:res1}, and hence the expected position of a possible excess from the $\ell=4$ modes. The horizontal lines give the median values of the \textit{SC}-spectra in the minimum and maximum smoothing cases. Bottom: same as top panel but here with a frequency resolution corresponding to an observing length of 1286 days, \ie, twice the length of the current data set.}
\label{fig:simu1}            
\end{figure*}

\begin{figure*}
\centering
\subfigure{
   \includegraphics[scale=0.45] {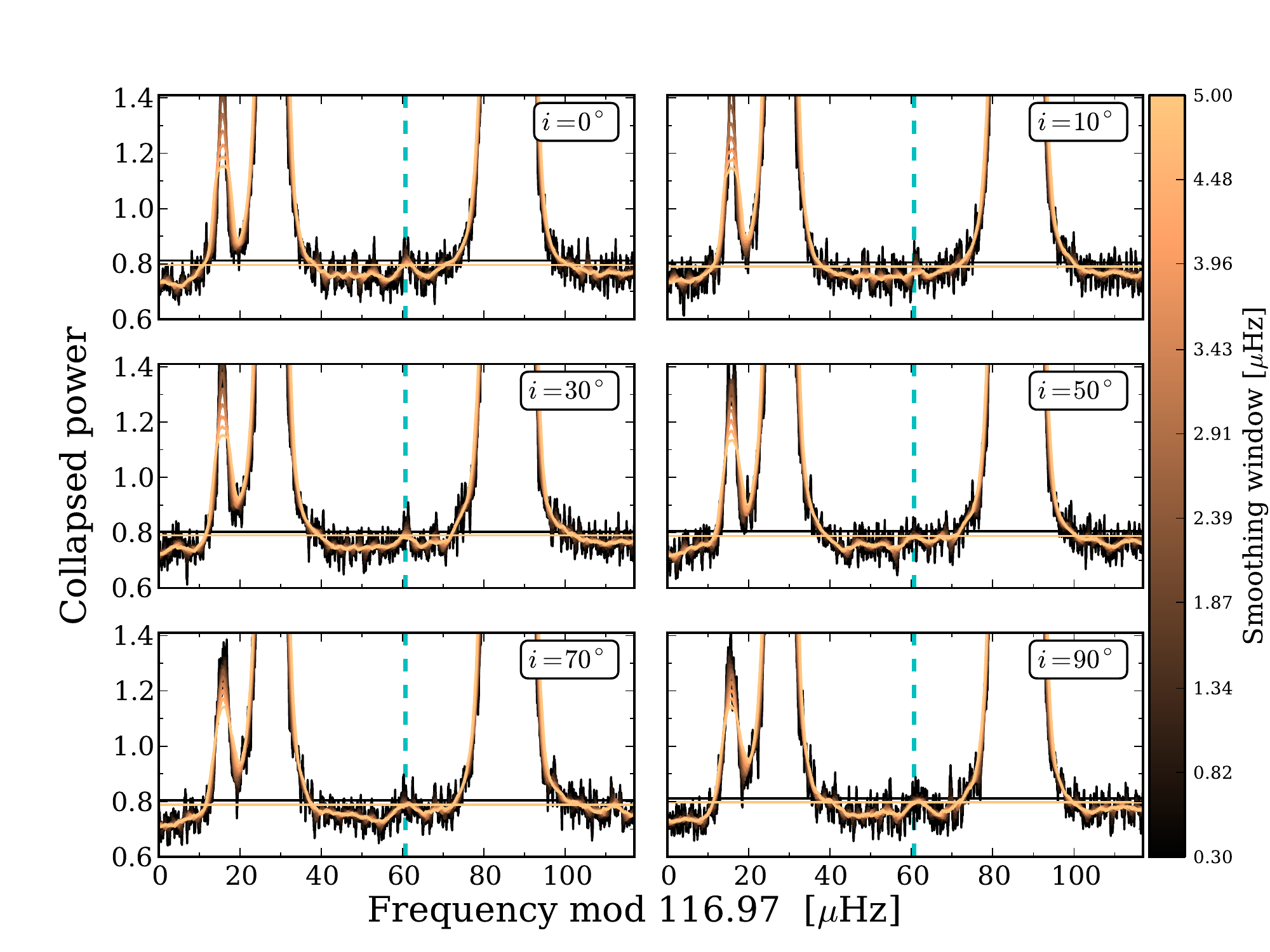}
 }
 \\
 \subfigure{
   \includegraphics[scale=0.45] {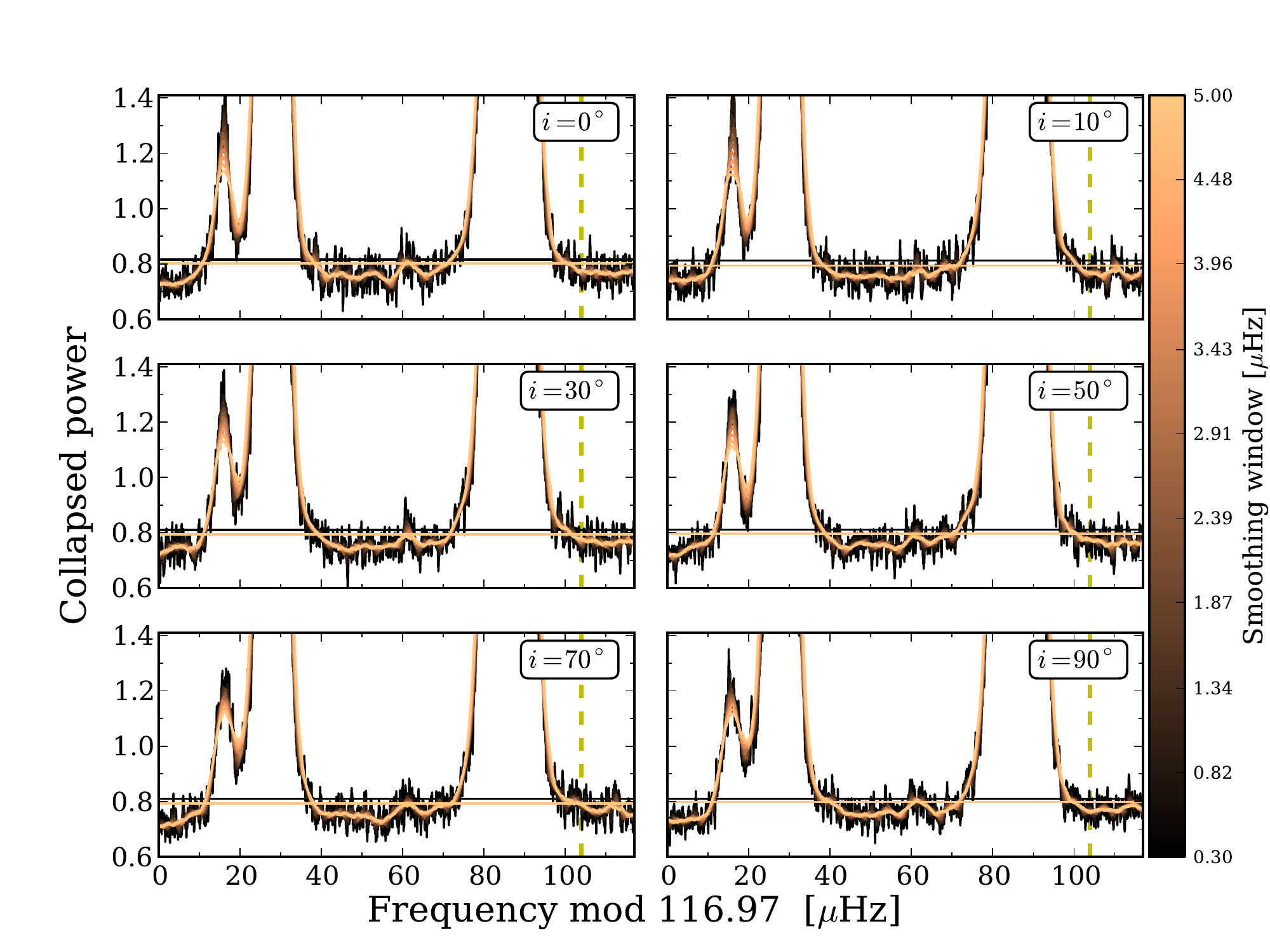}
 }
\caption{\footnotesize Top: \textit{SC}-spectra for simulated 16 Cyg B power spectra having a frequency resolution corresponding to an observing length of 2572 days, \ie\ four times the length of the current data set, and for different amounts of applied smoothing. For the different panels a specific value for the inclination has been used. The dashed cyan line shows the position of the straightened $\ell=4$ ridge, as in Figure~\ref{fig:res1}, and hence the expected position of a possible excess from the $\ell=4$ modes. The horizontal lines give the median values of the \textit{SC}-spectra in the minimum and maximum smoothing cases. Bottom: same as top panel but here the straightening is applied to the $\ell=5$ modes. The dashed yellow line shows the position of the straightened $\ell=5$ ridge, and hence the expected position of a possible excess from the $\ell=5$ modes.}
\label{fig:simu2}                         
\end{figure*}
  
In the top panel of Figure~\ref{fig:simu1} we give a subset of the \textit{SC}-spectra from the simulated 16 Cyg B data, with a frequency resolution corresponding to an observing length of 643 days and using the same range in frequency in the collapsing of the power spectra as for the real data. The different panels of this figure show the impact of a change in the adopted stellar inclination angle, where we have chosen a minimal set of six inclination angles: $i=0^{\circ},10^{\circ},30^{\circ},50^{\circ},70^{\circ},90^{\circ}$. Again the dashed cyan lines give the position of the straightened $\ell=4$ ridge. Comparing these \textit{SC}-spectra with the mean profiles from the MC set it is clear that the noise realization has a rather large impact.

In the bottom panel of Figure~\ref{fig:simu1}, we give the \textit{SC}-spectra as in the top panel, but here using a frequency resolution corresponding to an observing length twice the current length, \ie, ${\sim} 1286$ days. As the noise has only been reduced by a factor $\sqrt{2}$ in doubling the observing length there is still much variation in the \textit{SC}-spectra as compared to Figure~\ref{fig:MCbatch}. 
In Figure~\ref{fig:simu2}, we used a simulated observing length four times the current observing length, \ie, ${\sim} 2572$ days, whereby the shot noise is reduced by a factor two. Note that this corresponds roughly to the amount of data that would have been available had the \textit{Kepler} mission continued uninterrupted until its eighth year of operation. However, in light of the loss of a second reaction wheel, needed for the hitherto fine-pointing stability of the spacecraft \citep[][]{1996JGR...101.9297K}, and the planned continuation of the mission (dubbed ``K2"), this will not be possible. In the top panel, we show the \textit{SC}-spectrum targeted at the $\ell=4$ modes. The signal from these $\ell=4$ modes here stands out very clearly and should be readily observable in the power spectrum. In the bottom panel, we targeted the \textit{SC}-spectrum to the $\ell=5$ modes. Here small indications of excess is seen from the $\ell=5$ modes and a detection could be possible, but still the $\ell=0$ modes overshadow the signal.

\begin{figure*}
        \centering
        \includegraphics[scale=0.5]{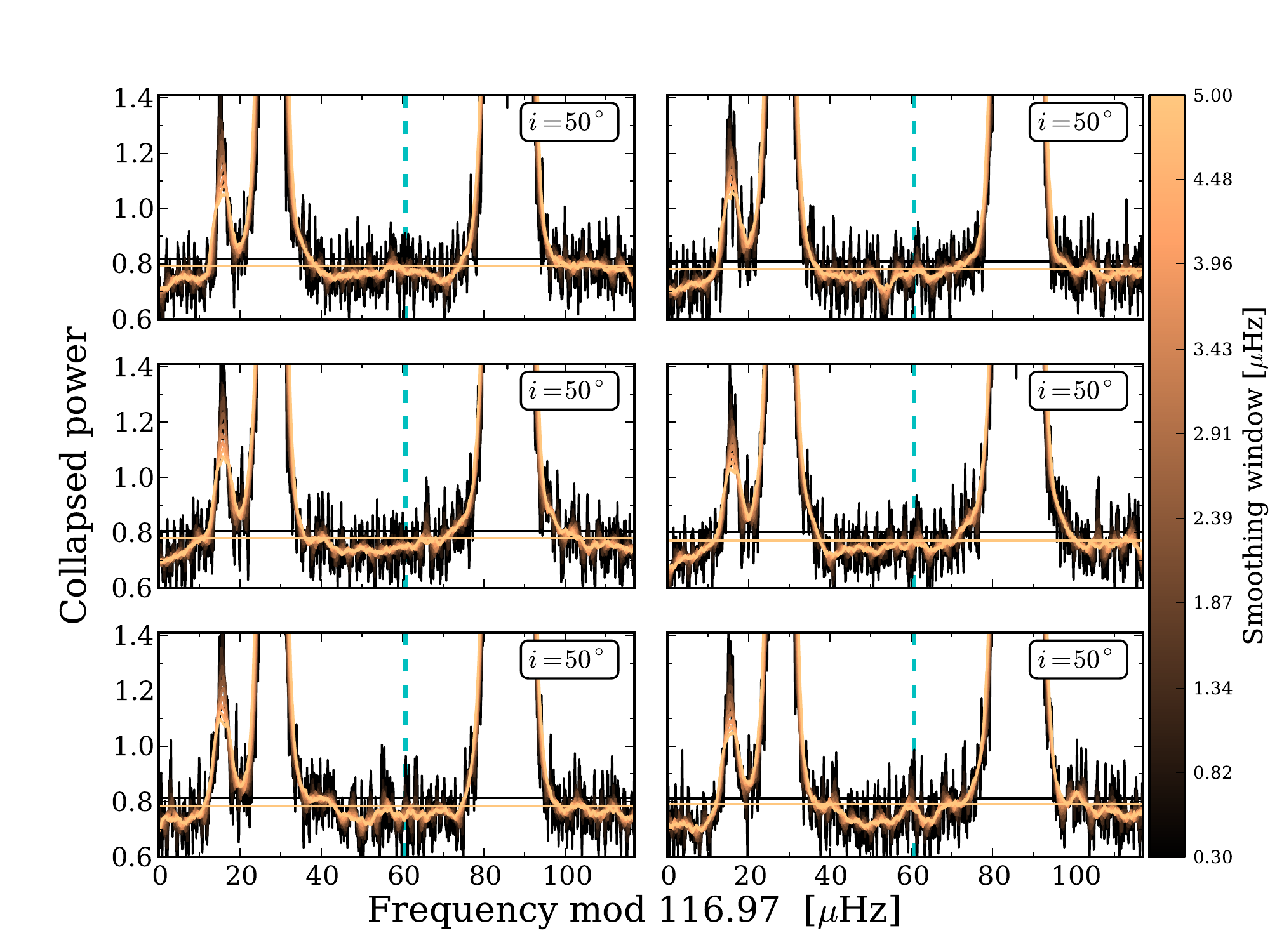}

\caption{\footnotesize Effect of the noise realization for simulated spectra having an observation length of 643 days and an inclination of $i=50^{\circ}$. As seen the realization of the noise has an considerable impact on the appearance of the \textit{SC}-spectrum.}
\label{fig:simu_noise}              
\end{figure*}

The impact of the noise realization hinted above is further illustrated in Figure~\ref{fig:simu_noise} where the \textit{SC}-spectra from six realizations of simulated data having an observation length of 643 days and in inclination of $i=50^{\circ}$ are shown. The noise realization has a non-negligible impact on the \textit{SC}-spectrum, where some cases (\eg, the bottom-right panel) shows an excess comparable to what is seen in 16 Cyg A and B, and then again some show no signs of an excess (\eg\ the middle-left panel). It can also be seen that the noise in some cases aids in the making of rather spiked features, see, \eg, the middle-left panel at around $65\,\rm \mu Hz$. However, the fact that the peak is very narrow speaks against the origin being that of $\ell=4$ modes. The reason for this is first of all that the straightening of modes is not perfect, there will still be small deviation and consequently a small broadening in the \textit{SC}-spectrum. Secondly, the $\ell=4$ modes will in most cases have some of the power placed in $m$-components that lie away from the central $m=0$ component; this too will result in a broadening of the $\ell=4$ excess in the \textit{SC}-spectrum.

\section{Inference from segmented solar data}
\label{sec:solardat}

\begin{figure*}
\centering
\subfigure{
   \includegraphics[scale=0.4] {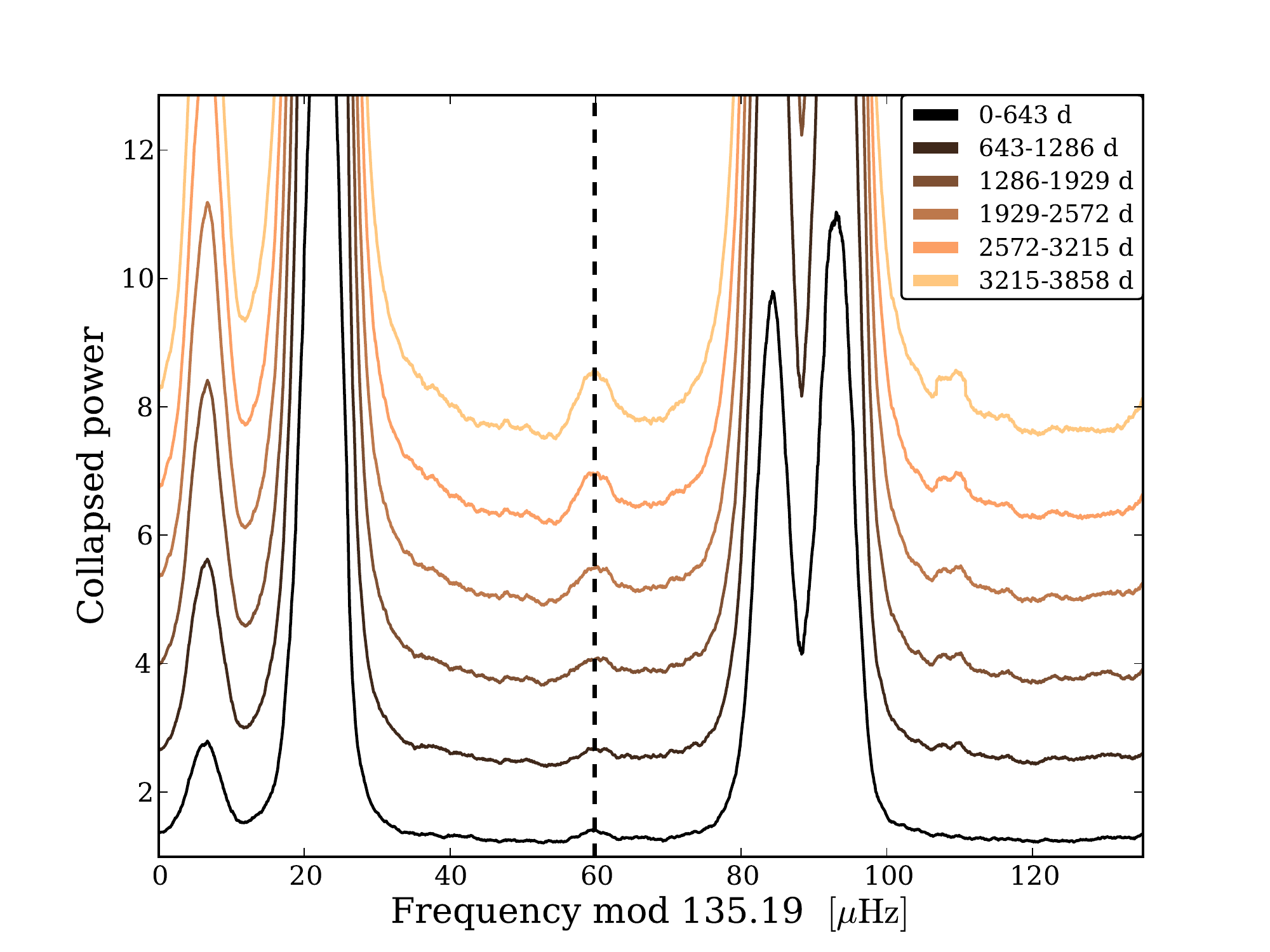}
 }
 \quad
 \subfigure{
   \includegraphics[scale=0.4] {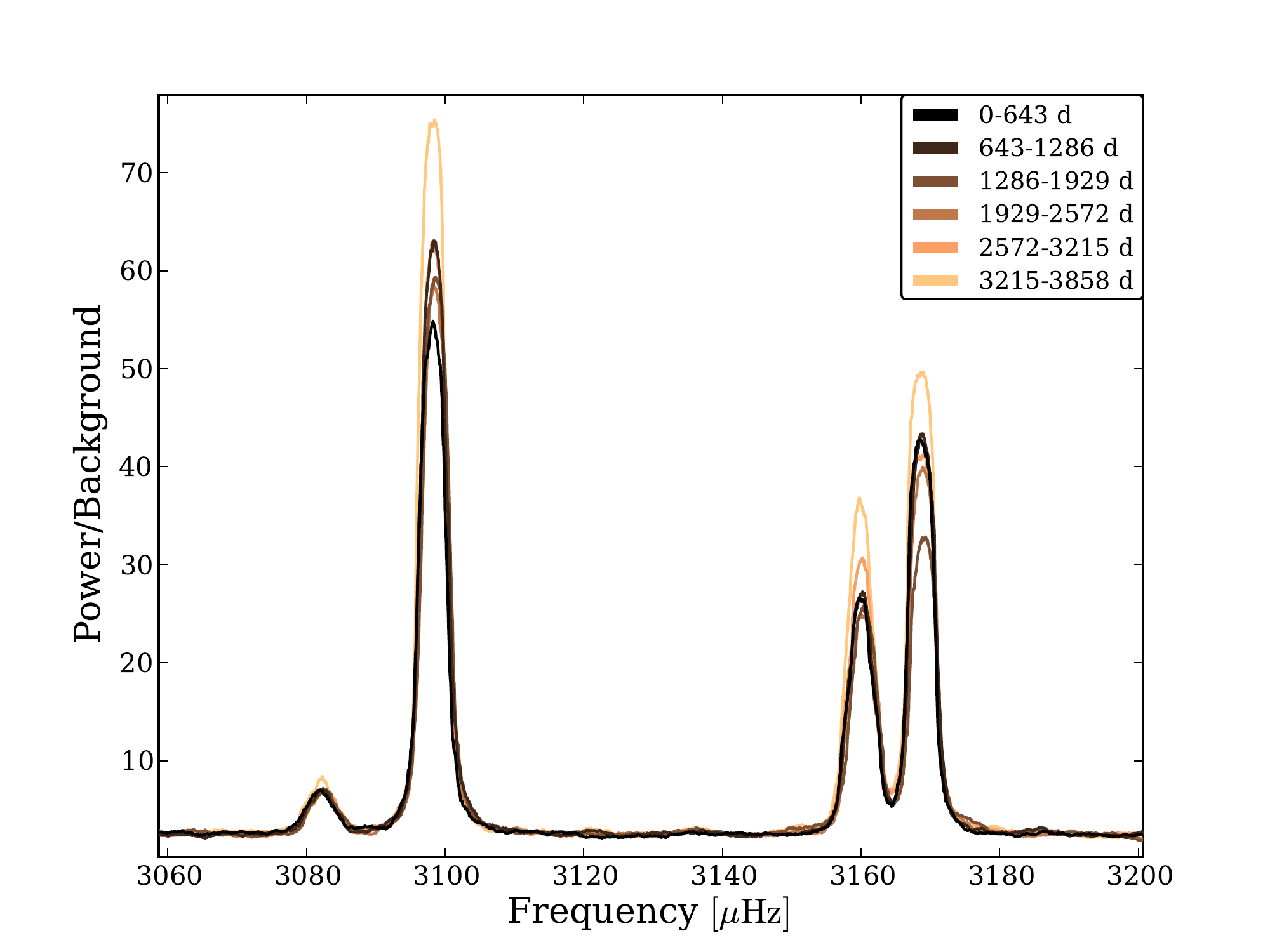}
 }
\caption{\footnotesize Left: \textit{SC}-spectra for the six segments of the blue SPM band solar time series. A smoothing of $4\,\rm \mu Hz$ has been applied, and in the collapsing the central 12 orders was used, from $2400\,\rm \mu Hz$ to $4022\,\rm \mu Hz$. The vertical dashed line shows the position of the expected peak from $\ell=4$. The power at the position of $\ell=5$ modes at around $105\,\rm \mu Hz$ in the \textit{SC}-spectrum is mainly caused by the instrumental peak from the Data Acquisition System. Right: change in mode power for one of the central orders of the power spectrum. Here it is clearly seen how the peak value in the power spectrum changes with time.}
\label{fig:simu_noise2}                      
\end{figure*}

To further test the validity of the observed signal for 16 Cyg A and B, we investigate how the signal from $\ell=4$ modes is seen in the solar data. The solar time series from the blue SPM filter was first divided into segments of 643 days length, resulting in separate small time series. To each of these, we added normally distributed noise by an amount that results in a shot noise level in the power spectrum equal to the level observed in 16 Cyg B. For the six power spectra computed, we calculate the corresponding \textit{SC}-spectrum, here using the fit to the Model~S mode shown in the left panel of Figure~\ref{fig:model_ech}. In the collapsing of the spectrum, we used the central 12 modes from about $2400\,\rm \mu Hz$ to $4022\,\rm \mu Hz$. In the left panel of Figure~\ref{fig:simu_noise2} the six \textit{SC}-spectra are shown after applying a boxcar smoothing of $4\,\rm \mu Hz$, with the vertical dashed line giving the expected position of the $\ell=4$ ridge. As seen, the mean level of the \textit{SC}-spectrum increases with time. The reason for this increase in the mean level can be seen in the right panel of Figure~\ref{fig:simu_noise}, which shows one of the central orders in the power spectrum. Here the relatively large variation in the power of the modes along with clear frequency shifts are seen, owing to both the stochastic nature of the excitation and the solar activity cycle \cite[see, \eg,][]{2002A&A...394..285G}. In the \textit{SC}-spectra we see in all cases an excess from the $\ell=4$ modes comparable in width and general appearance to the possible signal seen in 16 Cyg A and B. Furthermore, the amount of excess is seen to follow the increase in the mean level of the spectra. The relatively strong signal seen at the position of $\ell=5$ is greatly dominated by the instrumental peak described in \S~\ref{sec:sol}.


\section{Discussion}
\label{sec:discuss}

We have found clear evidence for $\ell=4$ and $\ell=5$ modes in the Sun from all VIRGO-SPM filters using a time series of 12 yr. Furthermore, we find indications, albeit no conclusive proof, for the $\ell=4$ modes in the \textit{Kepler} data of the solar analogues 16 Cyg A and B.
The credibility of our findings is supported by our simulations, in which a qualitatively similar signal from the $\ell=4$ modes is found using a simulation length equal to the length of analyzed \textit{Kepler} data. Under the assumption that our simulations accurately describe the observed $SC$-spectrum of 16 Cyg B we can reject the null hypothesis at the $5\%$ significance level. Furthermore, we obtain a posterior odds ratio of $\mathcal{O}_{1,0}=6.2$, judged as ``substantial" evidence in favor of $H_1$ over $H_0$. Both tests are in favor of a detection of additional power from $\ell=4$ modes. From our simulations also find that if additional power is indeed present there will only be a ${\sim}21\%$ chance of actually being able reject $H_0$ at the $5\%$ significance level.
Our simulations further suggest that in any case, a solid detection of the modes should be possible with four times the amount of data currently available. Also, we see that in using only subsets of the solar data with a length equal to the time series of 16 Cyg A and B, we are still able to see an excess signal from the $\ell=4$ modes.
We find at this time no indications for the $\ell=5$ modes in the \textit{Kepler} data and a solid detection of these modes for 16 Cyg A and B will be very difficult even with very long time series, mainly due to the strength of the $\ell=0$ modes. A detection of these modes might have been possible at the end of the nominal length of the extended \textit{Kepler} mission if continued observations could have been made.

The validity of our simulations is of course conditioned by the ingredients used being correct. 
The major uncertainty in the simulations concerns the mode visibilities, where a discrepancy was found between values estimated from the power spectra and from theory. This issue clearly deserves some extra attention, not only due to the fact that it affects the reliability of our simulations, especially for the $\ell=4$ and $\ell=5$ modes where direct measurements are very difficult even for the Sun, but also because these values are most often fixed a priori in the process of peak-bagging which clearly will affect the validity of the returned results if these are in fact fixed to wrong values.

Another less important aspect is the uncertainty of the temperature dependence on the mode linewidths, which is currently a property with little consensus among different groups. However, we see the scaling of linewidths from the solar values as solid due to the similarity of 16 Cyg A and B to the Sun. 

The shot noise added in the simulations is set from data with an observing length of $643$ days; we note that there is no guarantee that the instrumental noise will remain at the same level if the \textit{Kepler} mission progresses. Additionally, colored noise sources, other than the fitted stellar noise, were neglected in the simulations.
   
We have shown the ability of the introduced method, \ie, the \textit{SC}-spectrum, to modify the power spectrum in a way that should increase the signal in the collapsed spectrum. It is clear that the gain from using the \textit{SC}-spectrum depends mainly on the amount of deviation from the general asymptotic description given by Equation~\ref{eq:asym}. Nonetheless, the simplicity of the method and the fact that it provides a well defined position in frequency in the collapsed spectrum for the potential excess power makes it worthwhile to implement and use the \textit{SC}-spectrum.     

The method will be applied again once more data becomes available, to test the validity of both our findings and the simulations. Here possibly also the inclinations and rotational splittings of the 16 Cyg A and B stars will be better constrained, and models better fitting the low degree modes are likely available.

\section*{Acknowledgements} 

The authors would like to thank the anonymous referee for the many suggestions that helped to improve the final version of this paper, Christoffer Karoff for supplying the solar data used in the study, Frank Grundahl for many useful discussions, G{\"u}nter Houdek for a useful discussion on the linewidths used in our simulations, William J. Chaplin for supplying the solar linewidths obtained with BiSON, Douglas Gough for useful discussions, and Travis Metcalfe for pointing us to the model results from AMP and for patiently aiding us in understand the details of the AMP output.

For the 16 Cyg A and B models, computational resources were provided by XSEDE allocation TG-AST090107 through the Asteroseismic Modeling Portal (AMP).

Funding for the Stellar Astrophysics Centre is provided by The Danish National Research Foundation (Grant agreement no.: DNRF106). The research is supported by the ASTERISK project (ASTERoseismic Investigations with SONG and Kepler) funded by the European Research Council (Grant agreement no.: 267864).

Funding for the \emph{Kepler} mission is provided by NASA's Science Mission Directorate.


\bibliography{MasterBIB.bib}


\end{document}